\documentclass[preprint,showpacs,preprintnumbers,amsmath,amssymb]{revtex4}

\usepackage{graphicx}
\usepackage{bm}

\begin{document}

\title{Spin polarized current and shot noise in carbon nanotube quantum dot in the Kondo regime}

\author{Stanis{\l}aw Lipi\'{n}ski}
\author{Damian Krychowski}
\affiliation{%
Institute of Molecular Physics, Polish Academy of Sciences\\M. Smoluchowskiego 17,
60-179 Poznan, Poland
}%
\date{\today}

\begin{abstract}
Using nonequilibrium Green functions and several complementary many-body approximations we calculate shot
noise and spin dependent conductance in carbon nanotube semiconducting quantum dot in spin-orbital Kondo regime. We point out on the possibility of reaching giant values of tunnel magnetoresistance in this range and discuss a prospect of its gate control. We also analyze the influence of
symmetry breaking perturbations on the shot noise with special emphasis on spin dependent effects. The gate and bias
dependencies of noise Fano factors influenced by magnetic field, polarization of electrodes and spin-flip processes are presented.
\end{abstract}

\pacs{73.22.-f; 73.23.-b; 73.63.Fg}
\maketitle

\section{\label{sec:level1}Introduction}

Within the last  decade a growing interest in studying of noise of mesoscopic systems is observed ~\cite{1,2,3,4}. As the dimension of the device goes on scaling down and only few conducting channels are saved giving a total conductance of order of  conductance quantum $e^{2}/h$  the system becomes more sensitive to the noise.  Due to the  random  processes governing transport current is fluctuating in time even under dc bias. These  fluctuations originate from granularity of carriers (shot noise) and from thermal disturbance (thermal noise) ~\cite{1}.  The thermal noise  reflects the fluctuations in the occupations of the leads due to thermal excitation, and vanishes at zero temperature. It is related to the linear conductance via the fluctuation-dissipation theorem and thus does not carry extra information other than that obtained from ordinary conductance measurements.  The shot  noise is a purely non-equilibrium property and it results from the fact that current is not a continuous flow, but a sum of discrete pulses in time, each corresponding to  the transfer of electron through the system.  The latter fluctuations do not manifest in macroscopic systems since the inelastic scattering e.g.  electron-phonon smoothes them, leaving only thermal noise.   In contrast to thermal noise, shot noise cannot be eliminated by lowering the temperature.   The shot noise reveals information of transport properties which are not accessible by conductance measurements alone  for example about the kinetics of electrons and about the correlations of electronic wave functions. For uncorrelated carriers, shot noise is Poissonian. Deviations from the Poissonian noise appear to be due to correlations between electrons. Fermi-Dirac statistics or the way carriers scatter and interact within a sample strongly affect shot. Usually Coulomb repulsion and Fermi statistics both tend to smooth electron flow, thereby reducing shot noise below the uncorrelated Poissonian limit, but under certain conditions the interplay of Fermi statistics and interactions can lead to electron bunching i.e. to super-Poissonian correlations ~\cite{5}.
The advances in nanofabrication techniques opened  a new path in studying correlation effects. Of special interest in this respect is Kondo effect - a formation of  the many body dynamical singlet between a localized spin and delocalized conduction electrons ~\cite{6}, observed in semiconductor-based quantum dots (QDs) ~\cite{7,8,9,10} and molecular nanostructures ~\cite{11,12,13,14}. The tunability of nanostructures and their amenability to electrical measurements have allowed studies of Kondo effect in a controlled way, in particular in nonequilibrium. Spin Kondo effect in nanostructures is well understood ~\cite{15,16,17,18,19,20,21,22,23,24,25,26,27} and recently this knowledge has been enriched by analysis of current correlations ~\cite{28,29,30,31,32,33,34,35,36,37}. It has been theoretically predicted ~\cite{35} and experimentally verified ~\cite{32} that in the unitary Kondo regime the shot noise of SU(2) QDs is enhanced via back-scattering processes and a universal effective charge $5/3e$ has been measured. An important issue for spintronics is how Kondo physics is affected by magnetic field ~\cite{38,39,40,41,42,43} or what is the impact of ferromagnetic electrodes ~\cite{44,45,46,47,48,49,50,51,52,53,54,55,56}. The Kondo effect can also occur, replacing the spin by orbital ~\cite{57} or charge ~\cite{58,59} degrees of freedom. Spin and orbital degeneracies can also occur simultaneously leading to highly symmetric Kondo Fermi liquid ground state. SU(4) group is the minimal group allowing orbital-spin entanglement, which guarantees rotational invariance in spin and orbital spaces.  The four states entanglement is interesting for quantum computing, because each 4-state bit is equivalent to two 2-state bits, so the 4-state bits double the storage density. The unambiguous consequence of SU(4) symmetry is a ``halfed" zero bias conductance reflecting a shift of Kondo peak away from the Fermi level to $\hbar\omega \sim k_{B}T_{K}^{SU(4)}$. This can be understood from Friedel sum rule, which in this case gives  corresponding phase shift  at the Fermi level  $\delta = \pi/4$.  The resulting broadening of Kondo peak, as compared to SU(2) case means exponential enhancement of Kondo temperature, which makes these structures interesting  for practical  applications. The observation of SU(4) Kondo effect has been reported in vertical quantum dots ~\cite{57} and in carbon nanotubes (CNTs) ~\cite{60,61,62,63}.
The problem  of simultaneous screening of charge or orbital degrees of freedom and spin   has been  widely discussed also from  the theoretical point of  view ~\cite{64,65,66,67,68,69,70,71,72,73,74,75,76,77,78,79}, but there are only very few attempts to discuss noise in this range ~\cite{79,80,81,82}.  References ~\cite{80,81}, which are based on the phenomenological Fermi liquid description ($F-L$) discussed some general aspects of the noise in the strong coupling fixed point regime of SU(4) and predicted enhanced shot noise with universal charge $e^{*}=0.3e$. This suggestion has not been verified yet, because, due to the residual symmetry breaking perturbations this range is not easily accessible in experiment.  The sole experiment on the noise in spin-orbital Kondo range of CNT-QD has been  carried by Dellatre et al. ~\cite{82} in the range  $T\sim T_{K}/3$ and $eV\leq 3k_{B}T_{K}$ . It has been shown for the first time in this paper that unitary Kondo regime remains noisy in SU(4) limit.   These authors also  pointed  out  on scaling properties of the noise and have shown  that their experimental results were  reasonably described by temperature-dependent  slave boson mean field theory.

The aim of the present paper is to examine an impact of the symmetry breaking perturbations on the noise of spin-orbital Kondo systems. In particular we focus on the spin-dependent shot noise. This topic has received very little attention up to now. With exception of a very recent single result on the magnetic field dependence of noise in SU(4) Kondo regime ~\cite{79}, all other publications on the noise in Coulomb blockade and Kondo ranges concern SU(2) symmetry (see e.g. ~\cite{83,84,85,86,87,88,89,90,91}). Although our considerations are general we address our discussion to CNT-QDs because these systems exhibit high Kondo temperature and are well suited for the examination of the noise ~\cite{82}. The low energy band structure of semiconducting carbon nanotubes is orbitally doubly degenerate at zero magnetic field ~\cite{92,93}. This degeneracy corresponds to clockwise and counterclockwise symmetry of the wrapping modes in CNTs ~\cite{92}. Field perpendicular to the nanotube axis breaks the spin degeneracy, whereas parallel field breaks both spin and orbital degeneracy. For perpendicular orientation apart from the central orbital Kondo peak also satellites reflecting spin fluctuations occur in the density of states (DOS), whereas for parallel field both, spin and spin-orbital fluctuation satellites are observed. Their occurrence is reflected by a depression of the shot noise for the fields or voltages corresponding to the energy of their position.

Currently there is a widespread interest in developing new types of spintronic devices and carbon nanotubes deserve special attention in this respect due to their long spin lifetimes ~\cite{94}. In this paper we discuss field induced  spin filtering and giant tunnel magnetoresistance (${\cal{TMR}}$). The role of polarization of electrodes is twofold, it makes the tunneling processes spin-dependent and it introduces an effective exchange field  via spin-dependent charge fluctuations ~\cite{48,49,50,54}. The calculated  Fano factors, which quantify the deviation from the Poissonian noise for both spin orientations are sub-Poissonian, but for large polarization the minority spin  Fano has a value close to Poissonian,  whereas majority shot noise is almost completely suppressed and corresponding Fano  factor takes the values  close to zero. An important question for spintronic application is the influence of any possible spin-flip processes on transport. In the absence of spin-flip scattering the currents of spin-up electrons and spin-down electrons are independent. Spin-flip scattering mixes the spin currents and induces current opposite-spin correlations. These correlations  provide additional information about spin-dependent scattering processes and spin accumulation, which is important for spintronic applications.

Our considerations are based on non-equilibrium Keldysh Green functions ~\cite{95}  and    we use  different complementary many-body methods that span the physically relevant regimes. The slave boson mean field approaches (SBMFA) ~\cite{96,97} well describe systems close to the Kondo fixed point i.e.  for the case of fully degenerate deep dot level at low temperatures and low voltages. Two other methods used by us: equation of motion method (EOM) ~\cite{98,99,100} and noncrossing approximation (NCA) ~\cite{101,102} are better adopted for higher temperatures and voltages.   Compared with the conductance of a system, shot noise is more difficult to investigate experimentally. There are several methods to detect noise such as cross correlation ~\cite{103} or SQUID-based resistance bridge ~\cite{104}. Shot noise occurs when the sample is biased and it can be only detected if the frequency is high enough to overcome extrinsic $1/f$ noise caused by fluctuations in the physical environment and measurement equipment. Typically these experiments are performed  in the kHz to MHz range, sometimes higher ~\cite{5,82,105,106}. In the present paper we focus on   the noise power spectrum in the zero-frequency limit, but to gain an understanding of the range of validity of zero-frequency results we compare them with finite frequency calculations.

The paper is organized as follows. In Sec. II we describe the single level model of CNT-QD, next we briefly review the many-body techniques used in the discussion and  we introduce the expressions for the current and shot noise  together with a short comment on the applicability of the latter. Numerical results and their analysis are given in Sec. III, where we first discuss frequency range of validity of  the calculations and  compare   shot noise for the  full spin-orbit degenerate SU(4) Kondo  obtained in different approximations and set the results together   with analogous for SU(2) symmetry. Next we discuss the influence on the noise of different symmetry breaking perturbations including parallel and perpendicular magnetic fields, polarization of electrodes and the effect of spin-flip scattering. Finally, we give the conclusions in Sec. IV.

\section{Formulation}

\subsection{Model Hamiltonian}

We consider a single-wall semiconducting carbon nanotube quantum dot coupled to two electrodes which can be either nonmagnetic or ferromagnetic. CNT exhibits fourfold shell structure in the low energy spectrum ~\cite{14,107}. This electronic behavior originates from a particular combination of the symmetry of the graphene band structure and the quantization of momentum imposed by periodic boundary conditions along the nanotube circumference ~\cite{108}. The orbital degeneracy can be intuitively viewed to originate from two equivalent ways electrons can circle the graphene cylinder, that is clockwise and anti-clockwise. The rotational motion results in additional to spin orbital magnetic moments of electrons, typically an order of magnitude larger than Bohr magneton.  In the following we use the single energy shell model, which corresponds to the case of a short CNT-QD, but it gives also a qualitative  insight into the many level SU(4) Kondo problem for large separations between the levels ~\cite{109,110}. The Hamiltonian of the system, which includes different symmetry breaking perturbations discussed in this paper takes the general form:

\begin{eqnarray}
{\cal H}={\cal H}_{L}+{\cal H}_{R}+{\cal H}_{QD}+{\cal H}_{T}
\end{eqnarray}

The first two terms describe noninteracting itinerant electrons in the leads:
\begin{eqnarray}
{\cal H}_{\alpha}(P_{\alpha})=\sum_{k m\sigma}\epsilon_{k\alpha m\sigma}c_{k \alpha m\sigma}^{+}c_{k \alpha m\sigma}
\end{eqnarray}

($\alpha=L$) for the left electrode and ($\alpha=R$) for the right, $\epsilon_{k \alpha m\sigma}$ is the energy of an electron in the lead $\alpha$ with wave vector $k$, spin $\sigma$ ($\sigma=\pm1$) and orbital number $m$ ($m = \pm1$). The spin polarization of the leads $P_{\alpha}$ is defined by spin dependent densities of the state $\rho_{\alpha\sigma}$ as $P_{\alpha}=(\rho_{\alpha +}-\rho_{\alpha -})/(\rho_{\alpha +}+\rho_{\alpha -})$, ($P=P_{\alpha}$). In the following the wide conduction band approximation with the rectangular density of states is used $\rho_{\alpha m\sigma}(\epsilon)=\rho_{\alpha\sigma}=1/(2D_{\alpha\sigma})$ for $|\epsilon|<D_{\alpha\sigma}$, $D_{\alpha\sigma}$ is the half  bandwidth.
The dot Hamiltonian is given by:
\begin{eqnarray}
&&{\cal H}_{QD}(\Delta_{orb},V_{g},h,{\cal R})=\sum_{m\sigma}\epsilon_{m\sigma}d_{m\sigma}^{+}d_{m\sigma}+\sum_{m}{\cal U}n_{m+}n_{m-}\nonumber\\
&&+\sum_{\sigma\sigma'}{\cal U'}n_{1\sigma}n_{-1\sigma'}+\sum_{m}{\cal R}(d_{m+}^{+}d_{m-}+h.c.)
\end{eqnarray}
We set $|e|=g=\mu_{B}=k_{B}=\hbar=1$. The first term represents the field ($h$) and gate voltage ($V_{g}$) dependent dot energies:
\begin{eqnarray}
\epsilon_{m\sigma}=\epsilon(V_{g})+m \mu_{orb}h\cos(\theta)+\sigma g\mu_{B}h-m \Delta_{orb}
\end{eqnarray}

$\epsilon(V_{g})=\epsilon_{0}+V_{g}$, $\theta$ specifies the orientation of magnetic field  relative to the nanotube axis, $\mu_{orb}$ is the orbital moment, $\Delta_{orb}$ is orbital level mismatch. According to our model assumptions the magnetic field enters only diagonal elements of the Hamiltonian. For the considered low magnetic fields it is assumed that off diagonal elements are not affected (no Peierls substitution). The next two terms in (3) describe intra (${\cal U}$) and interorbital (${\cal U'}$) Coulomb interactions and the last term denotes the possible spin-flip scattering in the dot. Finally, the tunneling Hamiltonian ${{\cal H}_{T}}$, in Eq. (1) takes the form:

\begin{eqnarray}
{\cal H}_{T}(\gamma)=\sum_{k\alpha m\sigma}t_{\alpha m}(c_{k\alpha m\sigma}^{+}d_{m\sigma}+h.c)
\end{eqnarray}

The spin-dependent coupling strength to the lead $\alpha$ is described by $\Gamma_{\alpha m\sigma}=2\pi\sum_{k}|t_{\alpha m}|^{2}\delta_{\alpha m\sigma}(\epsilon-\epsilon_{k\alpha m\sigma})$. It is assumed that the tunneling amplitude $t_{\alpha m}$ is independent of the spin  and only the spin-dependent density of states accounts for the ferromagnetic properties of the leads. We assume in the following equal coupling for both orbitals $t_{\alpha 1} = t_{\alpha -1}$, but in general allow for the left-right asymmetry  $t_{L m} =\gamma t_{R m}$. One can express coupling strengths for the spin-majority (spin-minority) electron bands using polarization parameter as $\Gamma_{\alpha m\sigma} = \Gamma_{\alpha m}(1+\sigma P_{\alpha})$  with $\Gamma_{\alpha m} = (\Gamma_{\alpha m+} + \Gamma_{\alpha m-})/2$ and  $\Gamma=\sum_{\alpha m\sigma}\Gamma_{\alpha m\sigma}$. The unperturbed Hamiltonian of the full spin-orbital symmetry ${\cal H}^{SU(4)}$ corresponds  in the present formulation to the case of paramagnetic electrodes ($P_{\alpha}=0$), vanishing magnetic field ($h =0$), full chiral symmetry (orbital degeneracy $\Delta_{orb}= 0$) and lack of the spin-flip  processes at the dot (${\cal R} =0$).

\subsection{Many-body aproximations and their limitations}

We briefly review now the different complementary methods employed by us to find the single particle Green functions. Concerning EOM approach we present the calculations for both the infinite Coulomb interaction limit and for  Coulomb parameters corresponding to the typical charging energies of CNT-QDs. We compare the  results   with the slave boson calculations performed in infinite Coulomb interaction limit. Such a simplifying approach is justified since  charging energy in the discussed CNT-QDs is substantially larger than all other energy scales.\\

\noindent \textbf{1) The slave boson mean field approximations}\\

For  ${\cal U}, {\cal U'}\rightarrow \infty$ the only allowed states are  empty and single occupied states.  In the simplest single boson representation of Coleman ~\cite{96} the electron annihilation operator of state $|ms\rangle$ is decomposed into slave boson creation operator $b^{+}$ which creates empty state at the dot and pseudofermion  $f_{m\sigma}$, $d_{m\sigma}\rightarrow b^{+}f_{m\sigma}$. In this representation, which is best adopted to the full spin-orbital degenerate case,  the effective Hamiltonian ${\cal H}^{SU(4)}$  reads:
\begin{eqnarray}
&&{\cal H}^{SU(4)}_{{\cal C}}=\sum_{k\alpha m\sigma}\epsilon_{k\alpha m\sigma}c^{+}_{k\alpha m\sigma}c_{k\alpha m\sigma}+
\sum_{k\alpha ms}\epsilon_{m\sigma}f^{+}_{m\sigma}f_{m\sigma}\nonumber\\&&
+\sum_{k\alpha m\sigma}t_{\alpha m}(c^{+}_{k\alpha m\sigma}b^{+}f_{m\sigma}+h.c.)\nonumber\\&&+ \lambda(\sum_{m\sigma}f^{+}_{m\sigma}f_{m\sigma}+b^{+}b-1)\end{eqnarray}

The last term in Eq.(6) with the Lagrange multiplier $\lambda$ is the constraint, which assures the single occupancy at the dot. In the mean field approximation (MFA) the slave boson operator is replaced by its expectation value $b=\langle b\rangle$, thereby neglecting charge fluctuations  and the problem is formally reduced  to the free-particle model with the renormalized hopping integral  $\widetilde{t}_{\alpha m} = bt_{\alpha m}$ and site energy $\widetilde{\epsilon}_{m\sigma} =\epsilon_{m\sigma} +\lambda$. The stable solution is found from the saddle point of the partition function i.e., from the minimum of the free energy with respect to the variables $b$ and $\lambda$.  The corresponding self-consistent SBMFA equations relating the MFA parameters with nonequilibrium Green functions (NGF) read:

\begin{eqnarray}
&&\sum_{ms}G^{<}_{m\sigma m\sigma}(t-t')+|b|^{2}-1=0\nonumber\\
&&\sum_{k\alpha m\sigma}t'_{\alpha m}G^{<}_{k\alpha m\sigma m\sigma}(t-t')+2\lambda|b|=0
\end{eqnarray}

where the nonequilibrium lesser Green functions  are defined as:
\begin{eqnarray}
&&G^{<}_{m\sigma m\sigma}(t-t')=-i\langle f^{+}_{m\sigma}(t')f_{m\sigma}(t)\rangle\nonumber\\
&&G^{<}_{k\alpha m\sigma m\sigma}(t-t')=-i\langle c^{+}_{k\alpha m\sigma}(t')f_{m\sigma}(t)\rangle
\end{eqnarray}

With some caution we extend the Coleman approach  also  to the case of weakly broken spin symmetry (weak perpendicular magnetic field or ferromagnetic leads) and compare the results with more reliable in this case many-boson representation of  Kotliar and Ruckenstein (${\cal K}-{\cal R}$) ~\cite{97}. In ${\cal K}-{\cal R}$ picture   different auxiliary bosons are used to project onto different orbital or spin states. For the infinite ${\cal U}$ case it is enough to use  slave bosons projecting only onto the empty ($e$) and single occupied states ($p_{m\sigma}$). We introduce again additional constraints to eliminate unphysical states. The completeness relation for the slave bosons  $e^{+}e+\sum_{m\sigma}p^{+}_{m\sigma}p_{m\sigma}=1$, and the condition for the correspondence between fermions and bosons  $Q_{m\sigma}=d^{+}_{m\sigma}d_{m\sigma}=p^{+}_{m\sigma}p_{m\sigma}$  with corresponding Lagrange multipliers $\lambda$, $\lambda_{m\sigma}$ are represented by the last terms in  Hamiltonian ($9$):

\begin{eqnarray}
&&{\cal H}^{SU(4)}_{{\cal K}-{\cal R}}=\sum_{k\alpha m\sigma}\epsilon_{k\alpha m\sigma}c^{+}_{k\alpha m\sigma}c_{k\alpha ms}+\sum_{k\alpha m\sigma}\epsilon_{m\sigma}f^{+}_{m\sigma}f_{m\sigma}\nonumber\\
&&+\sum_{k\alpha m\sigma}t_{\alpha m}(c^{+}_{k\alpha m\sigma}z_{m\sigma}f_{m\sigma}+h.c.)\nonumber\\
&&+\lambda(\sum_{m\sigma}Q_{m\sigma}+e^{+}e-1)\nonumber\\
&&+\sum_{m\sigma}\lambda_{m\sigma}(f^{+}_{m\sigma}f_{m\sigma}-Q_{m\sigma})
\end{eqnarray}

The effective hopping term in Eq.(9)  is expressed  by $z^{+}_{m\sigma}f^{+}_{m\sigma}$ ($z_{m\sigma}f_{m\sigma}$) with $z_{m\sigma}=e^{+}p_{m\sigma}/(\sqrt{Q_{m\sigma}}\sqrt{1-Q_{m\sigma}})$. The  parameters  $\lambda$, $\lambda_{m\sigma}$, $e$, $p_{m\sigma}$   are obtained in a similar way as in Eqs. (7) by minimization of  the MF free energy  of ${\cal H}^{SU(4)}_{{\cal K}-{\cal R}}$.  MFA is exact in the limit of infinite degeneracy ${\cal N}\rightarrow\infty$, for finite ${\cal N}$ it only furnishes the starting point for possible controlled $1/{\cal N}$ expansion. Mean field approach is correct for describing spin and orbital fluctuations in the unitary Kondo regime and it leads to a local Fermi-liquid behavior at zero temperature. The disadvantage of this approximation is that it breaks the required gauge invariance symmetry (break of the phase symmetry of slave bosons) which is associated with charge conservation. The related artifact of  MFA is a sharp, spurious transition to the state with vanishing expectation values of boson fields making these approaches unreliable for higher temperatures. For low temperatures $T<<T_{K}$ and low voltages $eV<<k_{B}T_{K}$ a neglect of fluctuations of boson fields and fluctuations of the renormalized levels is justified. To discuss the higher temperatures and bias voltages as well as  wider gate voltage range (charge fluctuations) we employ NCA and EOM.\\

\noindent \textbf{2) The non-crossing approximation}\\

One can  avoid the earlier mentioned high temperature   drawback of SBMFA  performing the $1/{\cal{N}}$ expansion around the mean-field solution ~\cite{6,101,102}. NCA is the lowest order self-consistent approximation which includes such corrections. Since in the present paper we only use this method marginally for comparison of approximations we do  not cite here the explicit expressions for coupled NCA equations which determine boson and fermion self energies and propagators. The interested reader is referred to ~\cite{101}. Let us only mention that NCA is valid for a wide range of voltages and temperatures, including the  region close to $T_{K}$  and it gives reliable results down to a fraction of  $T_{K}$, but  it fails to describe the low-energy Fermi liquid fixed point correctly.  It is also well known for  SU(2) symmetry that  NCA  introduces spurious peaks at chemical potentials for systems perturbed by magnetic field ~\cite{102}. This is a consequence of neglect of vertex corrections. The same drawback is expected for perturbations of SU(4) Kondo systems that lift the orbital degeneracy.\\

\noindent \textbf{3) Equation of motion method}\\

This method can work in the whole parameter space except  the close vicinity of Kondo fixed point. It breaks down at low temperatures.  We  apply this approach for the all types of the discussed perturbations.  EOM method consists of differentiating the Green functions with respect to time which generates the   hierarchy of equations with higher-order Green functions (GFs). In order to truncate the series of equations we use  the self-consistent procedure proposed by Lacroix ~\cite{98} which approximates the GFs involving two conduction-electron operators by:
\begin{eqnarray}
\langle\langle c^{+}_{k\alpha m'\sigma'}d_{m'\sigma'}c_{k\alpha m\sigma}|d^{+}_{m\sigma}\rangle\rangle\simeq\langle c^{+}_{k\alpha m'\sigma'}d_{m'\sigma'}\rangle\langle\langle c_{k\alpha m\sigma}|d^{+}_{m\sigma}\rangle\rangle\nonumber\\
\langle\langle c^{+}_{k\alpha m'\sigma'}c_{k\alpha m'\sigma'}d_{m\sigma}|d^{+}_{m\sigma}\rangle\rangle \simeq\langle c^{+}_{k\alpha m'\sigma'}c_{k\alpha m'\sigma'}\rangle\langle\langle d_{m\sigma}|d^{+}_{m\sigma}\rangle\rangle\nonumber\\
\langle\langle d^{+}_{m'\sigma'}c_{k\alpha m'\sigma'}c_{k\alpha m\sigma}|d^{+}_{m\sigma}\rangle\rangle\ \simeq\langle d^{+}_{m'\sigma'}c_{k\alpha m'\sigma'}\rangle\langle\langle c_{k\alpha m\sigma}|d^{+}_{m\sigma}\rangle\rangle
\end{eqnarray}

The correlations $\langle c^{+}_{k\alpha m'\sigma'}d_{m'\sigma'}\rangle$,$\langle c^{+}_{k\alpha m'\sigma'}c_{k\alpha m's'}\rangle$, and $\langle d^{+}_{m'\sigma'}c_{k\alpha m'\sigma'}\rangle$ occurring in Eq.(10)  play the leading role in Kondo effect. For detailed analysis of EOM hierarchy and decoupling schemes see e.g. ~\cite{99,100}. It has been theoretically predicted ~\cite{49} and experimentally confirmed ~\cite{54}  that ferromagnetic electrodes induce a local exchange field which polarizes the localized spin in the absence of any external fields and the Kondo resonance splits. This splitting originates in spin-dependent charge fluctuations. NCA cannot be used for description of these processes due to the mentioned drawbacks in analysis of effect  of magnetic field or polarization.
The equation of motion method  for finite Coulomb interactions is a tool which in principle can  handle this problem, but   getting a  consistent picture  requires going beyond Lacroix decoupling (Eq.(10)).  Instead of a tedious task of dealing with higher order GFs we adopt in our numerical calculations, following ~\cite{49}, an expression on the spin splitting based on perturbative scaling analysis ~\cite{111}.

\subsection{Current and shot noise}

Current flowing through CNT-QD in the $|m\sigma\rangle$ channel ${\cal I}_{m\sigma}=({\cal I}_{Lm\sigma}-{\cal I}_{Rm\sigma})/2$ can be expressed in terms of the lesser Green functions as follows ~\cite{95}:
\begin{eqnarray}
{\cal{I}}_{\alpha \sigma}(t)=\sum_{m}{\cal{I}}_{\alpha m\sigma}(t)=\sum_{km}t_{\alpha m}[G^{<}_{m\sigma,k\alpha m\sigma}(t)-G^{<}_{k\alpha m\sigma,m\sigma}(t)]
\end{eqnarray}
The corresponding conductances are defined as ${\cal G}_{m\sigma}=d{\cal I}_{m\sigma}/dV$, ${\cal G}_{s}={\cal G}_{1\sigma}+{\cal G}_{-1\sigma}$
and ${\cal G}=\sum_{m\sigma}{\cal G}_{m\sigma}$. The useful quantities  characterizing the spin dependent transport are polarization of conductance ${\cal PC}=({\cal G}_{+}-{\cal G}_{-})/({\cal G}_{+}+{\cal G}_{-})$ and   tunnel magnetoresistance ${\cal TMR}=({\cal G}^{{\cal{P}}}-{\cal G}^{{\cal{AP}}})/{\cal G}^{{\cal{AP}}}$ defined as the relative difference of conductances for different spin orientations $({\cal PC})$ or for parallel and antiparallel orientations of polarizations of the leads $({\cal TMR})$.
The temporal fluctuations of the currents are defined as:
\begin{eqnarray}
{\cal S}_{\alpha m\sigma\nu m'\sigma'}(t-t')=\langle[\Delta {\cal \hat{I}}_{\alpha m\sigma}(t),\Delta {\cal \hat{I}}_{\nu m'\sigma'}(t')]_{+}\rangle=\nonumber\\ \langle[{\cal \hat{I}}_{\alpha m\sigma}(t),{\cal \hat{I}}_{\nu m'\sigma'}(t')]_{+}\rangle-2\cdot{\cal I}_{\alpha m\sigma}(t){\cal I}_{\nu m'\sigma'}(t')
\end{eqnarray}
where $\Delta {\cal \hat{I}}_{\alpha m\sigma}(t)$ is the fluctuation of the current operator around its average value $\Delta {\cal \hat{I}}_{\alpha m\sigma}(t)={\cal \hat{I}}_{\alpha  m\sigma}(t) -{\cal I}_{\alpha  m\sigma}(t)$. The Fourier transform of the current noise called noise power is:
\begin{eqnarray}
{\cal S}_{\alpha m\sigma\nu m'\sigma'}(\omega)=2\int^{+\infty}_{-\infty}d\tau e^{\imath\omega\tau}{\cal S}_{\alpha m\sigma\nu m'\sigma'}(\tau)
\end{eqnarray}
More explicitly the current noise can be expressed by correlators, which involve two Fermi operators of the leads and two operators of the dot as follows ~\cite{32,87}:
\begin{eqnarray}
&&{\cal S}_{\alpha m\sigma\nu m'\sigma'}(t-t')=
t_{\alpha m}t_{\nu m'}[G^{<}_{1}(t,t')-G^{<}_{2}(t,t')-G^{<}_{3}(t,t')+\nonumber\\&&G^{<}_{4}(t,t')+h.c.]-2\cdot{\cal I}_{\alpha m\sigma}(t){\cal I}_{\nu m'\sigma'}(t')
\end{eqnarray}
where:
\begin{eqnarray}
G^{<}_{1}(t,t')=(i)^{2}\sum_{kq}\langle c^{+}_{k\alpha m\sigma}(t)d_{m\sigma}(t)c^{+}_{q\nu m'\sigma'}(t')d_{m'\sigma'}(t')\rangle\nonumber\\
G^{<}_{2}(t,t')=(i)^{2}\sum_{kq}\langle c^{+}_{k\alpha m\sigma}(t)d_{m\sigma}(t)d^{+}_{m'\sigma'}(t')c_{q\nu m'\sigma'}(t')\rangle\nonumber\\
G^{<}_{3}(t,t')=(i)^{2}\sum_{kq}\langle d^{+}_{m\sigma}(t)c_{k\alpha m\sigma}(t)c^{+}_{q\nu m'\sigma'}(t')d_{m'\sigma'}(t')\rangle\nonumber\\
G^{<}_{4}(t,t')=(i)^{2}\sum_{kq}\langle d^{+}_{m\sigma}(t)c_{k\alpha m\sigma}(t)d^{+}_{m'\sigma'}(t')c_{q\nu m'\sigma'}(t')\rangle
\end{eqnarray}

Finding an accurate expression for the shot is a formidable task since it involves not only the usual many body expansion, but also the analytical continuation of two and more particles GFs. Following many papers ~\cite{32,33,34,87,112}, we introduce a  crude approximation   which decouples the correlator (15):
\begin{eqnarray}
\langle c^{+}_{k\alpha m\sigma}(t)d_{m\sigma}(t)c^{+}_{q\nu m'\sigma'}(t')d_{m'\sigma'}(t')\rangle\simeq\nonumber\\
\langle c^{+}_{k\alpha m\sigma}(t)d_{m\sigma}(t)\rangle \langle c^{+}_{q\nu m'\sigma'}(t')d_{m'\sigma'}(t')\rangle-\nonumber\\
\langle c^{+}_{k\alpha m\sigma}(t)d_{m\sigma}(t')\rangle \langle c^{+}_{q\nu m'\sigma'}(t')d_{m'\sigma'}(t)\rangle
\end{eqnarray}
and  in a similar fashion the rest of correlation functions (15).
Decoupling  (16) is exact in the case of independent particles  and clearly it also applies for the  slave boson  MF Hamiltonians (6,9). The approximation (16) is consistent with Lacroix decoupling we use in the single particle problem (10).    It is believed that  the more conduction electrons a two-body Green function involves, the less correlation effects it includes ~\cite{34}. For the single particle properties however the omitted correlations  are only  small correction, whereas for the shot noise   these correlations might be of importance due to  the possible interaction-induced scattering. Inelastic scattering is out of importance for the current in SU(4) Kondo systems because backscattered charges vanish ~\cite{80,81}. But it has been shown that it had an impact on the noise ~\cite{81}.
 A more sophisticated treatment beyond approximation (16) is thus required if one wants  to take into account finite frequencies and properties far from equilibrium, but it is outside the scope of this paper. In the present discussion correlations  are implicitly included in formula (14) by Kondo correlations build up in the single particle Green functions and by non-equilibrium form of the Green function which is  influenced by correlations. Applying the operational rules as given by Langreth ~\cite{113} to the Dyson equation for the contour-ordered Green functions one gets the following  Keldysh equation for the lesser and greater functions  $\mathbf{G^{<(>)}(\omega)}=\mathbf{G^{r}(\omega)}\mathbf{\Sigma^{<(>)}(\omega)}\mathbf{G^{a}(\omega)}$
~\cite{95}, where $G^{r(a)}(\omega)$ denotes retarded(advanced) GF. We use Ng ansatz  ~\cite{114}  to approximate the lesser self energy $\mathbf{\Sigma^{<(>)}(\omega)}=\mathbf{\Lambda (\omega)}\mathbf{\Sigma^{(0)<(>)}(\omega)}$, where $\mathbf{\Sigma^{(0)<}(\omega)}=\sum_{\alpha}2f_{\alpha}(\omega)\mathbf{\Sigma^{(0)a}_{\alpha}(\omega)}=
\sum_{\alpha}2i f_{\alpha}(\omega)\mathbf{\Gamma_{\alpha}(\omega)}$ and $\mathbf{\Sigma^{(0)>}(\omega)}=\sum_{\alpha}2(1-f_{\alpha}(\omega))\mathbf{\Sigma^{(0)r}_{\alpha}(\omega)}=
-\sum_{\alpha}2i (1-f_{\alpha}(\omega))\mathbf{\Gamma_{\alpha}(\omega)}$ are noninteracting lesser and greater self-energies coming from the tunneling of electrons from the dot to outside leads and $f_{\alpha}(\omega)$ is Fermi-Dirac distribution function for the $\alpha$ electrode. $\mathbf{\Lambda}(\omega)$   is a matrix which can be determined by the Keldysh requirement $\mathbf{\Sigma^{>}(\omega)}-\mathbf{\Sigma^{<}(\omega)}=\mathbf{\Sigma^{r}(\omega)}-\mathbf{\Sigma^{a}(\omega)}$, where $\mathbf{\Sigma^{r(a)}(\omega)}$ are retarded (advanced ) self-energies for the interacting QD.  As a result one obtains $\mathbf{\Sigma^{<}(\omega)}=(\mathbf{\Sigma^{r}(\omega)}-\mathbf{\Sigma^{a}(\omega)})(\mathbf{\Sigma^{(0)r}(\omega)}-\mathbf{\Sigma^{(0)a}(\omega)})^{-1}\mathbf{\Sigma^{(0)<}(\omega)}$, . The advantage of Ng approximation is that it  is exact in equilibrium limit, is exact in nonequilibrium  for noninteracting particles and  it preserves continuity of current condition in the steady state limit ~\cite{114}.
Using decoupling (16) and performing Fourier transform of (14) one gets the noise power spectrum in Ng approximation in the form:
\begin{eqnarray}
&&{\cal S}_{\alpha m\sigma\nu m'\sigma'}(\omega)=\int^{+\infty}_{-\infty}d\epsilon
[\delta_{\alpha\nu}\Sigma^{(0)}_{\alpha m\sigma}(\epsilon)+\Sigma^{(0)}_{\alpha m\sigma}(\epsilon)G_{m\sigma m'\sigma'}(\epsilon)\Sigma^{(0)}_{\nu m'\sigma'}(\epsilon)]^{>}G^{<}_{m'\sigma'm\sigma}(\epsilon+\omega)
\nonumber\\&&+G^{>}_{m\sigma m'\sigma'}(\epsilon)[\delta_{\nu\alpha}\Sigma^{(0)}_{\nu m'\sigma'}(\epsilon+\omega)+\Sigma^{(0)}_{\nu m'\sigma'}(\epsilon+\omega)G_{m'\sigma' m\sigma}(\epsilon+\omega)\Sigma^{(0)}_{\alpha m\sigma}(\epsilon+\omega)]^{<}\nonumber\\&&-[G_{m\sigma m'\sigma'}(\epsilon)\Sigma^{(0)}_{\nu m'\sigma'}(\epsilon)]^{>}[G_{m'\sigma' m\sigma}(\epsilon+\omega)\Sigma^{(0)}_{\alpha m\sigma}(\epsilon+\omega)]^{<}-[\Sigma^{(0)}_{\alpha m\sigma}(\epsilon)G_{m\sigma m'\sigma'}(\epsilon)]^{>}\nonumber\\&&\times[\Sigma^{(0)}_{\nu m'\sigma'}(\epsilon+\omega)G_{m'\sigma' m\sigma}(\epsilon+\omega)]^{<}+h.c.(\omega\rightarrow-\omega)
\end{eqnarray}
To express the noise in a more compact way let us introduce, following Ref ~\cite{89,112}, the generalized transmission matrix ${\cal{\mathbf{T}}_{\alpha\beta}}$, which incorporates the nonequilibrium effects of Coulomb interaction:
\begin{eqnarray}
{\cal{\mathbf{T}}_{\alpha\beta}}
=4\mathbf{\Gamma}_{\alpha}\mathbf{G}^{r}\mathbf{\Lambda}
\mathbf{\Gamma}_{\beta}\mathbf{G}^{a}
\end{eqnarray}
The nonequilibrium GFs in Ng approximation can now be written as $G^{<}_{m\sigma m'\sigma'}(\epsilon)=\sum_{\alpha}i f_{\alpha}(\epsilon){\cal{T}}^{m\sigma m'\sigma'}_{L\alpha}(\epsilon)/2\Gamma_{Lm\sigma}$ and $G^{>}_{m\sigma m'\sigma'}(\epsilon)=\sum_{\alpha}-i (1-f_{\alpha}(\epsilon)){\cal{T}}^{m\sigma m'\sigma'}_{L\alpha}(\epsilon)/2\Gamma_{Lm\sigma}$. It is easy to show using the identity $G^{r}_{m\sigma m'\sigma'}G^{r}_{m'\sigma' m\sigma}+G^{a}_{m\sigma m'\sigma'}G^{a}_{m'\sigma' m\sigma}=G^{r}_{m\sigma m'\sigma'}G^{a}_{m'\sigma' m\sigma}+G^{r}_{m'\sigma' m\sigma}G^{a}_{m\sigma m'\sigma'}+(G^{>}_{m\sigma m'\sigma'}-G^{<}_{m\sigma m'\sigma'})(G^{>}_{m'\sigma' m\sigma}-G^{<}_{m'\sigma' m\sigma})$, that the spectral density of noise (17) in zero frequency limit can be written down as a sum of effective Landauer-Buttiker term with generalized transmission (the first two terms of (19) and the first term of (20)) and correction, the latter vanishes at zero temperature:
\begin{eqnarray}
&&{\cal{S}}_{Lm\sigma,Lm\sigma}(0)
=2\int^{+\infty}_{-\infty}d\epsilon
(f_{L}(\epsilon)-f_{R}(\epsilon))^2(1-{\cal{T}}^{m\sigma m\sigma}_{LR}(\epsilon))
{\cal{T}}^{m\sigma m\sigma}_{LR}(\epsilon)
+\nonumber\\&&{\cal{T}}^{m\sigma m\sigma}_{LR}(\epsilon)
[f_{L}(\epsilon)(1-f_{L}(\epsilon))+f_{R}(\epsilon)(1-f_{R}(\epsilon))]
+\nonumber\\&&2f_{L}(\epsilon)(1-f_{L}(\epsilon))[{\cal{T}}^{m\sigma m\sigma}_{LL}(\epsilon)
-4\Gamma_{Lm\sigma}G^{r}_{m\sigma m\sigma}(\epsilon)\Gamma_{Lm\sigma}G^{a}_{m\sigma m\sigma}(\epsilon)]\nonumber\\
\\
&&{\cal{S}}_{Lm\sigma,Lm-\sigma}(0)
=-2\int^{+\infty}_{-\infty}d\epsilon
(f_{L}(\epsilon)-f_{R}(\epsilon))^2{\cal{T}}^{m\sigma m-\sigma}_{LR}(\epsilon)
{\cal{T}}^{m-\sigma m\sigma}_{LR}(\epsilon)
+\nonumber\\&&f_{L}(\epsilon)(1-f_{L}(\epsilon))[
4\Gamma_{Lm\sigma}G^{r}_{m\sigma m-\sigma}(\epsilon)\Gamma_{Lm-\sigma}G^{a}_{m-\sigma m\sigma}(\epsilon)
+\nonumber\\&&+4\Gamma_{Lm-\sigma}G^{r}_{m-\sigma m\sigma}(\epsilon)\Gamma_{Lm\sigma}G^{a}_{m\sigma m-\sigma}(\epsilon)]
\end{eqnarray}
In a similar way current can be expressed as:
\begin{eqnarray}
{\cal{I}}_{L m\sigma}=\int^{+\infty}_{-\infty}d\epsilon
(f_{L}(\epsilon)-f_{R}(\epsilon)){\cal{T}}^{m\sigma m\sigma}_{LR}
\end{eqnarray}
It is seen, that for finite temperature and for finite frequency, noise even in the crude approximation (16) cannot be expressed solely in terms of effective transmission probabilities.
In the  SBMFA picture $\mathbf{\Lambda}(\omega)$ is the unit matrix $\mathbf{\Lambda}(\omega) = \mathbf{I}$ and effective noninteracting particles bias dependent transmission reads:
\begin{eqnarray}
\widetilde{\cal{\mathbf{T}}}_{\alpha\beta}(\epsilon,V)
=4\widetilde{\mathbf{\Gamma}}_{\alpha}(V)\mathbf{G}^{r}(\epsilon,V)
\widetilde{\mathbf{\Gamma}}_{\beta}(V)\mathbf{G}^{a}(\epsilon,V)
\end{eqnarray}
where $\widetilde{\mathbf{\Gamma}}_{\alpha}(V)=\mathbf{\Gamma}_{\alpha}|b|^2$ (Coleman) or $\widetilde{\Gamma}_{\alpha m\sigma}(V)=\Gamma_{\alpha m\sigma}|z_{m\sigma}|^2$ (${\cal{K}}-{\cal{R}}$). Due to the charge conservation the zero frequency auto and cross-correlations are related by ${\cal{S}}_{LL}(0)={\cal{S}}_{RR}(0)=-{\cal{S}}_{LR}(0)=-{\cal{S}}_{RL}(0)$. The total charge current noise is a sum of partial spin or spin-orbital contributions ${\cal{S}}^{c}_{LL}=\sum_{\sigma\nu}{\cal{S}}_{L\sigma L\nu}=\sum_{m\sigma\nu}{\cal{S}}_{Lm\sigma Lm\nu}$. A convenient means to assess how correlations affect shot noise is the Fano factor ${\cal F}$ defined as the ratio between the actual shot noise ${\cal S}$ and the Poissonian noise for uncorrelated carriers ${\cal F}={\cal S}/(2{\cal I})$.\\

\section{RESULTS}
\subsection{SU(4) Kondo system}

We parameterize the SU(4) Hamiltonian of  CNT-QD by   three parameters: charging energy  $\cal U$,  the tunnel rate between the QD and the reservoirs $\Gamma$ and the half bandwidth of the electrodes  $D$. The assumption of SU(4) symmetry is preserved if intra and interorbital Coulomb interactions are taken equal ${\cal{U}}={\cal{U'}}$, what is reasonable for ideal CNTs, because states $m=1$ and $m=-1$ have the same charge densities. We extend the above assumption also to the case of weakly broken symmetry. The value of ${\cal{U}}$ can be inferred from the  size of Coulomb diamonds. For semiconducting CNTs  the charging energy is of order of tens of meVs ~\cite{116}.  $\Gamma$  informs us about a quality of contacts. It can be estimated from the width of orbital or Coulomb peaks and for the interesting weak coupling regime $\Gamma$ is of order of several meVs ~\cite{60,61,117}.  Both charging energy and lead-dot coupling change with the gate voltage, which reflects variation  of tunnel barrier widths  with the number of electrons and related change of the source and drain capacitances.  A precise tuning of the coupling to the leads by energizing locally acting gate electrodes ~\cite{117} is yet  not possible in CNTs, but there are interesting trials along this line ~\cite{118}.   The value of  the orbital magnetic moment $\mu_{orb}$, which scales with CNT diameter, can be estimated from the experimental average slopes between the two Coulomb peaks that correspond to the addition energy of the electrons to the same orbital state ~\cite{60} and in our discussion we assume $\mu_{orb} = 10$, which corresponds to the diameter $d = 2.9$ nm. Kondo temperatures corresponding to the above intervals of parameters fall in the range of several K and the separation between fully degenerate energy states of  short CNT is of order of several meVs ~\cite{61,116}, and naturally is larger than $\Gamma$.   The Fermi energy is taken as zero energy  $E_{F} =0$. The dc bias voltage across the left and right leads is $V = \mu_{L} - \mu_{R}$. Here we choose  $\mu_{L}=-\mu_{R}=V/2$.
 Our discussion is based on the single shell model (1-5). It has been  shown ~\cite{109,110}, that such an oversimplified approach to the Kondo problem of multilevel systems is justified if the separation between the levels is large as compared to the Kondo temperature.  The position and coupling of the  effective  single level are then understood as renormalized by transitions  to higher levels. Charging energy is much larger than coupling to the leads and for illustrative purposes we compare in some cases, the large but finite $\cal U$ results with infinite $\cal U$ limit.  In the numerical calculations we concentrate  on quarter-filling i.e. it is assumed that one electron occupies the electronic shell of QD. The aim of the present paper is to discuss noise  in SU(4) Kondo systems tentatively omitting the detailed analysis of  the unitary limit for two reasons. First it is experimentally difficult to probe  the ultimate low-energy limit, and secondly a breakthrough in understanding of this region has been already achieved by publication of the  shot noise measurements and their SBMFA interpretation  for SU(4) Kondo CNT-QD ~\cite{82}.   Dellatre et al.  have shown   that in the unitary  SU(4) Kondo range  system remain noisy with Fano factor reaching value ${\cal{F}} =0.5$.    The same authors  also presented the scaling properties of the Kondo noise, what in addition to bias or temperature dependencies of conductance in this region  highlights  the fact that Kondo temperature is the only energy scale characterizing this range.  Based on the same conviction of governing  of low energy physics by $T_{K}$ alone,   but including both elastic and inelastic processes  Vitushinsky et al. ~\cite{80} and Mora et al. ~\cite{81} predicted, using the local Fermi liquid theory, enhanced shot noise  with universal charge $e^{*}= 0.3e$.  A difficult experimental confirmation of this interesting finding  is still missing.
Our calculations are addressed to CNT-QDs in the Kondo range. In the present Section  we compare in  test calculations     noise obtained   by different many-body approaches with the  predictions of the above mentioned articles ~\cite{80,81,82}.  The main analysis of the present work  focuses on  deviations  from strict unitary limit, where charge fluctuations are not negligible (this paragraph) or where perturbations break the spin-orbital  symmetry (next sections).  Naturally no strict scaling properties are expected in these  regions, and apart from Kondo temperature also other energy scales come into play connected with charge fluctuations or with the strength of symmetry breaking fields. Our task is to explain how the interplay of Kondo and other many-body correlations reflect in spin dependent current and  noise.

We focus in the following on the zero frequency shot noise, but it is not what is experimentally measured. In many experimental studies one finds that  the  low frequency spectrum is governed by $1/f$ noise. To eliminate this spurious noise high frequencies are used, typically from 10 kHz up to 1 GHz ~\cite{5,82,105,106}.
\begin{figure}
\includegraphics[width=6 cm,bb=0 0 741 725,clip]{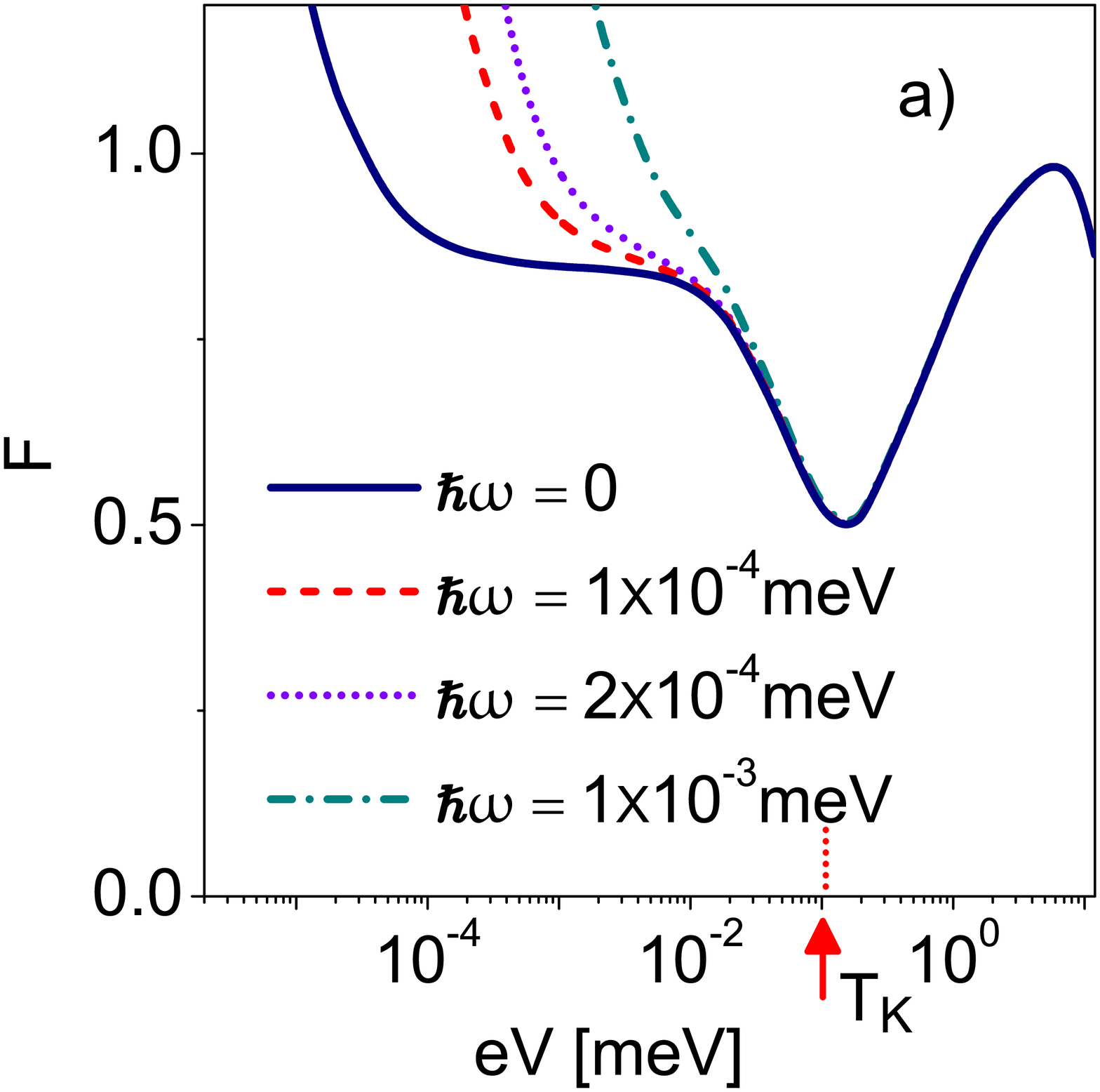}
\includegraphics[width=6 cm,bb=0 0 741 725,clip]{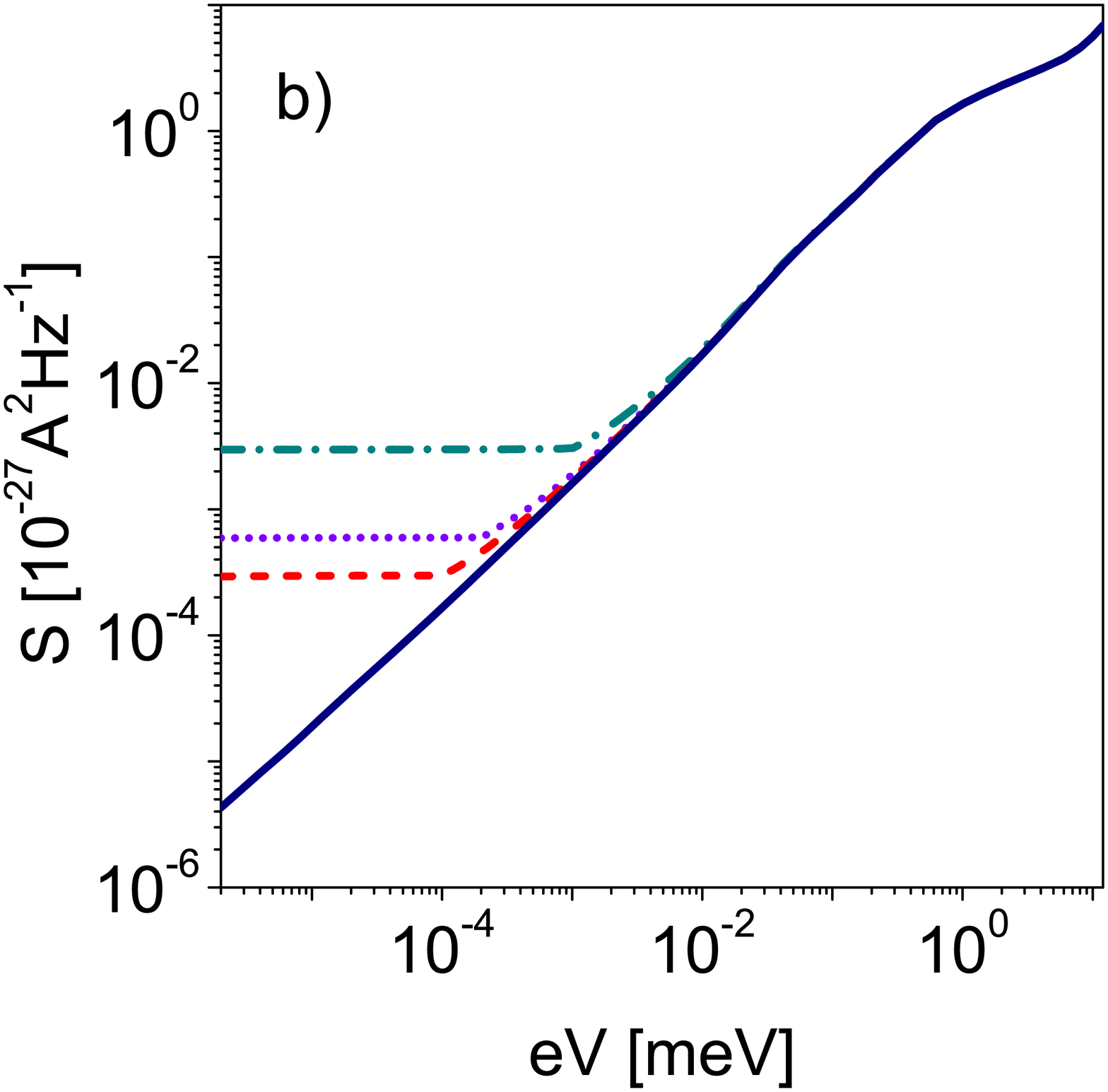}
\caption{\label{fig:epsart} (Color online) Zero frequency and finite frequency Fano factors (a) and shot noises (b) of SU(4) CNT-QD ($\epsilon=-6 \Gamma$, ${\cal{U}}=10\Gamma$, $\Gamma=2$ meV) calculated by means of EOM approach at $T=10^{-6}\Gamma$. Arrow marks Kondo energy $k_{B}T_{K}$.}
\end{figure}
To get an insight into the limitations of use of zero frequency results for analysis of experimental data it is useful to compare the zero frequency and finite frequency results in the voltage  range of interest.  Fig. 1a presents the noise Fano factor   plotted  versus bias voltage for several frequencies. For finite frequency $\omega$, like for finite temperature we do not have pure shot noise  and  for $|V| < \omega$, noise spectrum tends to the equilibrium value, which for $T = 0$ is determined by zero-point quantum fluctuations ${\cal S}(\omega)\rightarrow2\omega{\cal G}(V\rightarrow0)$, independent of the voltage (for the case presented at Fig. 1b ${\cal G}  = 1.5(e^{2}/h)$). The small deviations from this limit, observed in this region are the consequence of finite temperature.  For $T = 0$  the noise power spectrum has a discontinuous derivative at $\omega = V$ ~\cite{1} and a reminiscence of it is still visible in our finite temperature plots.  The low voltage dependence of ${\cal{F}}$ is mainly determined by Kondo correlations and the  minimum of ${\cal F}(V)$ occurs at $V \sim 2T_{K}$.   For higher voltages the influence of  Kondo correlations on transport dies off which is seen by an increase  of ${\cal F}$.  Maximum of ${\cal F}(V)$ for $V\gg T_{K}$  and the following decrease of ${\cal F}$ is due to Coulomb charge fluctuations. This high voltage range  should be taken with caution, since in this region the inelastic processes certainly play the role partially destroying the coherence and this is not taken into account in our calculations. The interesting part of the bias evolution of ${\cal F}(V)$ dictated by Kondo correlations for frequencies  in  MHz range does not differ significantly from  the zero frequency curve ($\hbar\omega=2\cdot10^{-4}$ meV corresponds to frequency $48$ MHz). The above statement can be safely applied to CNTs, because the observed Kondo temperatures are as high as $10-15 K$ ~\cite{60,61}, i.e. much higher than $T_{K}$ from Fig.1 ($T_{K}\simeq1.2$ K). At very small bias  noise is dominated by thermal noise and  the Fano factor is $2T/V$ due to fluctuation dissipation theorem and divergent at $V = 0$.  Henceforth we concentrate on the shot noise calculated for  $V > T$, but first let us elucidate the effect of many-body correlations on equilibrium noise. This is illustrated on Fig. 2. Kondo correlations, which are hardly visible in the temperature dependence of conductance clearly  appear, as has been shown in ~\cite{82}, as maximum of ${\cal{S}}(T)$ dependence. Solid curve on Fig. 2 has been calculated by SBMFA method with $T_{K}$  marking the position of Kondo resonance in DOS ~\cite{119}.
\begin{figure}
\includegraphics[width=6 cm,bb=0 0 741 725,clip]{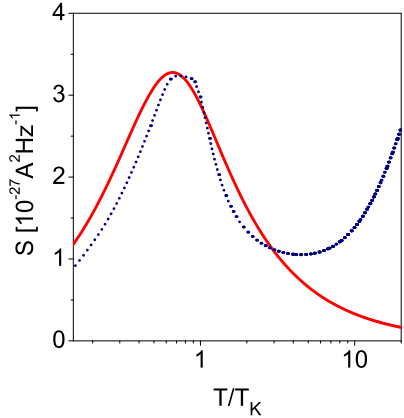}
\caption{\label{fig:epsart} (Color online) Equilibrium noise of CNT-QD ($\epsilon=-6\Gamma$) calculated by means of SBMFA approach (solid line) and EOM ($\epsilon=-6\Gamma$, ${\cal{U}}=12\Gamma$), $\Gamma=1$ meV.}
\end{figure}
Dotted curve denotes EOM temperature dependence of equilibrium noise.  Apart from Kondo maximum also charge fluctuation correlations reflect  in ${\cal{S}}(T)$ dependence  by  an upturn of the  curve for higher temperatures. Now let us test the applicability for the noise analysis of different complementary many-body techniques reviewed before. Concerning  slave boson calculations we present results both in the  single boson (Coleman) ~\cite{96} and double  boson representations (${\cal{K}}-{\cal{R}}$)~\cite{97}. The numerical solutions of self-consistent equations of the former method are identical in the deep dot level range  with analytical  temperature and bias  dependencies  proposed in ~\cite{82}.  Use of SBMFA beyond unitary limit is less justified and requires the full numerical solutions of equations, but we also use this technique in this region. Moreover we present  also some  SBMFA results  for mixed valence range, where in principle this approximation does not hold, but we show them only  to visualize the tendency.  One more comment is necessary. It is well known in literature ~\cite{32,82},that  for large bias ($V>2T_{K}$) SBMFA breaks down ($b\rightarrow0$), and the width of the resonance peaks decrease in an abrupt manner. This reflects  e.g. in  the appearance of artificial  negative magnetoresistance. To avoid this problem some authors introduce regularization procedures ~\cite{32,82}.  In our analysis we restrict in SB discussion only to the region $V<2T_{K}$.
In the following pictures, if not stated in a different manner, all the energies are given in units of $\Gamma$ and the half bandwidth is chosen $D=(D_{L+}+D_{L-})/2=50$.
Figure 3 presents a  comparison of different   many-body methods and approximations for non-equilibrium GFs used by us for calculation of the shot noise in the case of  ${\cal U} \rightarrow \infty$. Fig. 3a shows the low voltage value of Fano factor, which for a very  deep level at the dot and  chosen low temperature almost reaches the limiting value 0.5 corresponding to the ``halfed" zero bias single channel conductance of SU(4) Kondo systems. As expected  this result is numerically best reproduced by SBMFA calculations. EOM and NCA results show for the  deep levels  temperature induced upper deflection of ${\cal F}$  for energies much  higher than the SBMFA calculations because the former two methods overestimate Kondo temperature.
\begin{figure}
\includegraphics[width=6 cm,bb=0 0 741 725,clip]{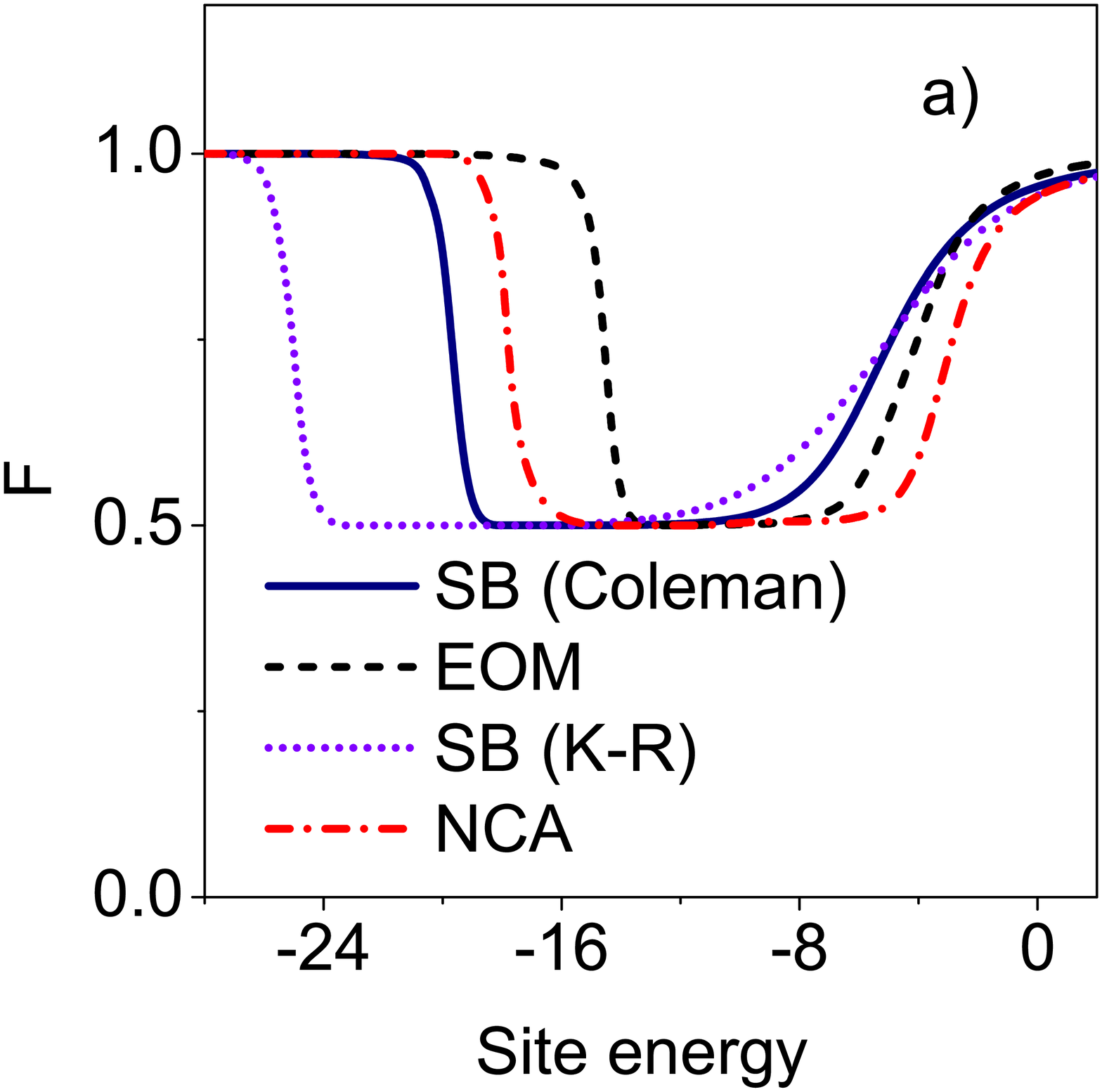}
\includegraphics[width=6 cm,bb=0 0 741 725,clip]{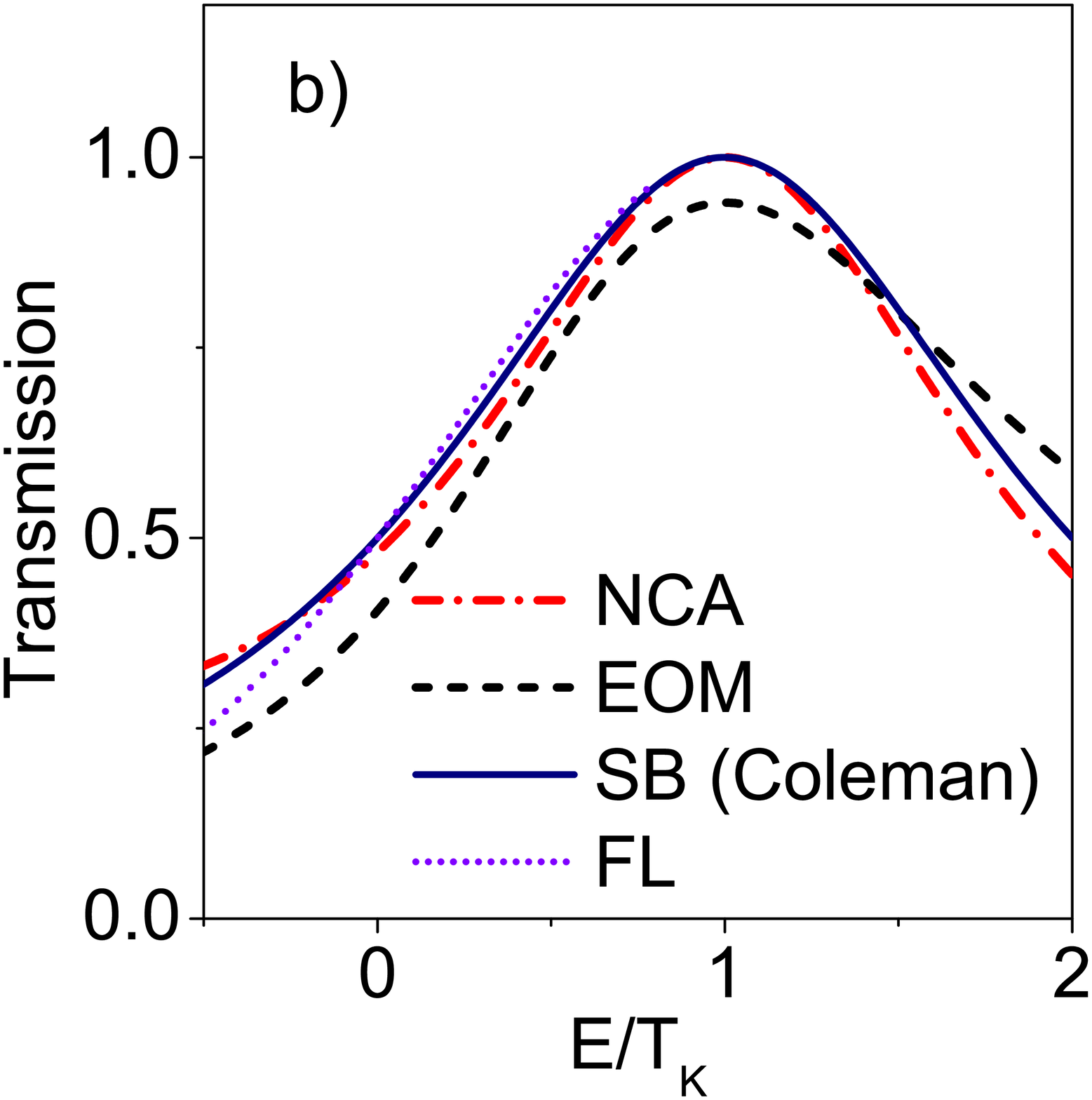}
\includegraphics[width=6 cm,bb=0 0 741 725,clip]{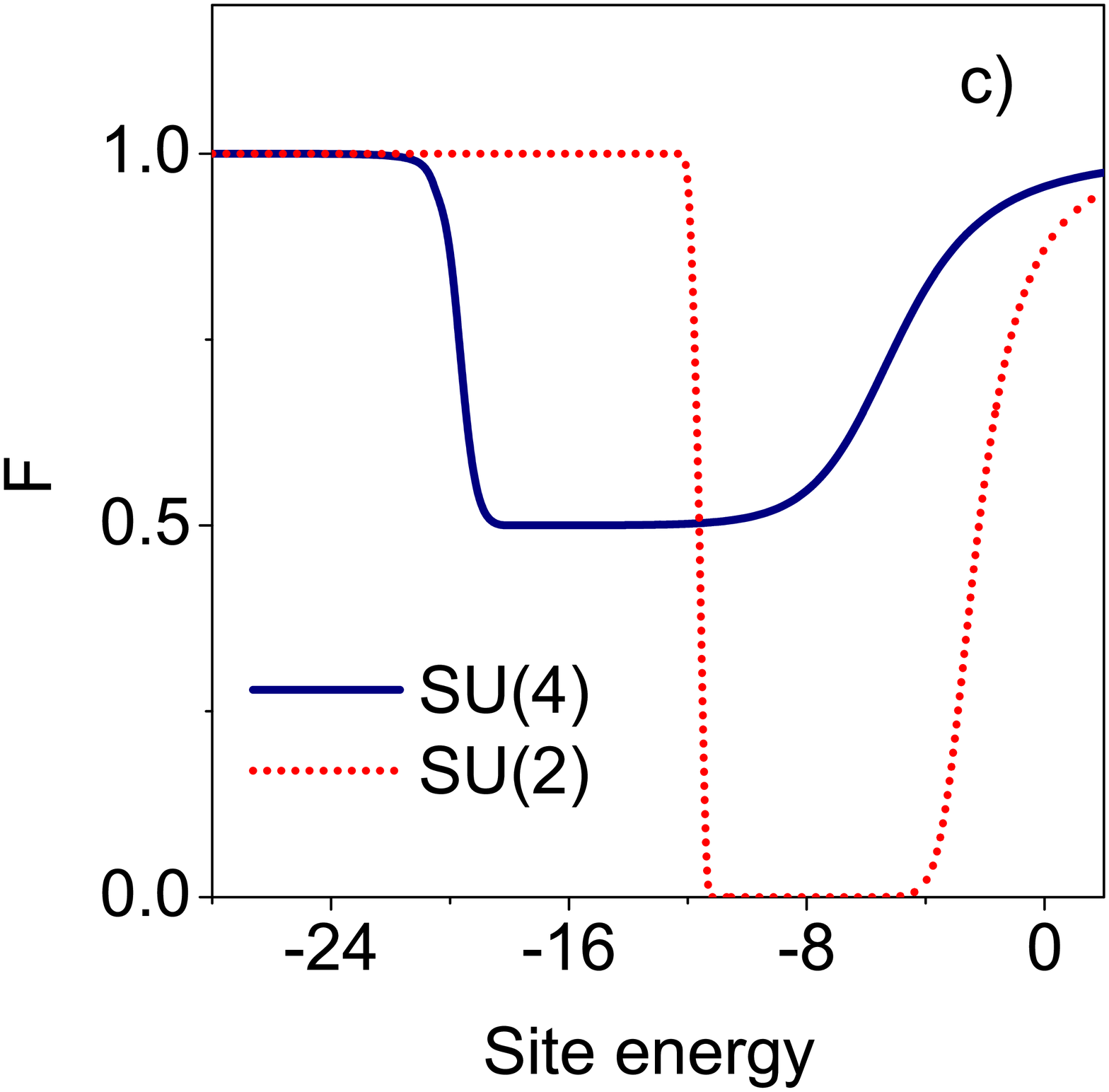}
\caption{\label{fig:epsart} (Color online) (a) Comparison of linear Fano factors of SU(4) CNT-QD (${\cal{U}}=\infty$) in the Kondo range
              calculated by  the slave boson methods  (Coleman, ${\cal{K-R}}$), NCA and EOM. (b) Zero-bias transmissions of CNT-QD ($\epsilon=-9$) calculated by means of  SBMFA, NCA (${\cal{U}}=\infty$) and EOM (${\cal{U}}=25$) compared with Fermi liquid predictions ~\cite{81}. FL curve is presented for energies close to $E_{F}$. (c) Comparison of linear Fano factors in the Kondo regime for SU(2) and SU(4) symmetries (SBMFA - Coleman).}
\end{figure}
This is a consequence of the  mentioned difficulties of these  approaches  in describing the region close to the Kondo fixed point. The estimated Kondo temperatures in ${\cal{K}}-{\cal{R}}$ approach are lower than in Coleman formalism and this is reflected in the difference of the corresponding deflection points. Since we assume  $V\gg T$ the discussed increase of ${\cal F}$ does not account for thermal noise, but  reflects  the role of temperature on the shot noise.   The correlations and the mean field parameters (SBMFA) are temperature dependent. For dot energies, where deflection is observed Kondo temperature is lower than the  chosen  temperature and the observed increase of ${\cal F}$ is a result of  weakening of Kondo correlations. The effect of  a broad dot level resonance peak in the mixed valence region on the  shot noise is represented by EOM and NCA curves and is most clearly visible in the latter. This can be understood by an insight into the  transmissions (Fig. 3b), where especially strong hybridization of atomic and Kondo peaks is observed in NCA case. The problem of large width of Kondo resonance of the exact NCA solution in the limit $T\rightarrow0$ is well known in literature ~\cite{101}. For comparison we have also plotted on Fig. 3b Fermi-liquid transmission, that includes both elastic and inelastic contributions as discussed in ~\cite{81}. SB and F-L curves exactly reproduce the ``halfed" value of transmission at the Fermi energy.
In Fig. 3c Fano factor for SU(4) symmetry is compared with ${\cal{F}}$ for SU(2). The SU(2) Hamiltonian can be straightforwardly written down by restricting  the index m in Hamiltonian ($1$) to $m = 1$ and truncating the corresponding sum. For the SU(2) case a complete suppression of Fano factor for the  deep dot levels is observed and for even deeper levels also the  temperature induced upper deflection of ${\cal F}$ is visible. Note that the ranges of dot energies with complete suppression of ${\cal F}$ (SU(2)) or half suppression (SU(4)) are different, which is a consequence of remarkable differences of Kondo temperatures for both symmetries. Fig. 4a  presents a comparison of bias dependencies of Fano factors ${\cal{F}}(V)$ calculated by different methods  in the ${\cal{U}}\rightarrow\infty$ limit for the deep dot level. The SBMFA value of ${\cal F}$ is almost bias independent and takes  value ${\cal F}=1/2$ in the whole presented range.  Curve EFL denotes ${\cal F}(V)$ calculated within F-L approach as presented in ~\cite{81}, but including only elastic terms and curve FL takes into account both elastic and inelastic contributions ~\cite{81}. As it is seen elastic scattering leads to suppression of the noise for higher voltages, whereas  inelastic processes enhance the noise. Charge fluctuations which are inherent  in EOM or NCA formalism cause deviation from the limiting value $0.5$ at small bias. Minima of EOM and NCA  ${\cal F}(V)$ curves are mainly determined by Kondo temperature,  but they are also  influenced by position of dot energy. The curve denoted as EOM$2$ illustrates the effect of more accurate treatment of correlations in noise beyond decoupling ($16$). In EOM$2$ approximation instead of using decoupling $(16)$ we have written down EOM for the corresponding two-electron GFs ($15$) and then decoupled the two-particle (conduction electron-dot electron) GFs occurring at the r.h.s. of the mentioned EOM in the spirit of Lacroix approximation and replaced occupation operators in the three-particle functions by their averages:
\begin{eqnarray}
 &&\langle\langle c^{+}_{k\alpha m\sigma}c_{k''\alpha'' m\sigma}|c^{+}_{k'\alpha' m'\sigma'}d_{m'\sigma'}\rangle\rangle\simeq\langle c^{+}_{k\alpha m\sigma}d_{m'\sigma'}\rangle\langle\langle c_{k''\alpha'' m\sigma}|c^{+}_{k'\alpha' m'\sigma'}\rangle\rangle\nonumber\\
 &&\langle\langle n_{m''\sigma''}c^{+}_{k\alpha m\sigma}d_{m\sigma}|c^{+}_{k'\alpha' m'\sigma'}d_{m'\sigma'}\rangle\rangle\simeq\langle n_{m''\sigma''}\rangle\langle\langle c^{+}_{k\alpha m\sigma}d_{m\sigma}|c^{+}_{k'\alpha' m'\sigma'}d_{m'\sigma'}\rangle\rangle
\end{eqnarray}
As it is seen on Fig. 4a taking into account the  two-particle correlations in this lowest approximation (EOM$2$) suppresses the shot noise. Many experimentalists use instead of traditional noise  Fano factor ${\cal{F}}={\cal{S}}/2e{\cal{I}}$, experimentally more relevant quantity,  so called invariant or generalized  Fano factor ${\cal{IF}}$ ~\cite{81,82}.  In the  three  subsequent pictures (Figs 4b,c,d) we use this quantity in order  to help the reader to see the validity of the mentioned  many-body techniques in the context of  the sole experimental data that illustrate  SU(4) Kondo noise  ~\cite{82}. The mentioned data have been  presented with the use of ${\cal{IF}}$.  We also elucidate the  deviations from noise scaling if charge fluctuations are included. To define invariant Fano factor  let us first introduce the limiting values of Kondo current and noise for $T_{K}\rightarrow\infty$, ${\cal{I}}_{0}=(2e^{2}/h)V$, ${\cal{S}}_{0} = (4e^{2}/h)(k_{B}T+eVcoth(eV/2k_{B}T)/2)$ and define following ~\cite{82},  the excess noises   ${\cal{S}}^{{\cal{I}}}_{exc}={\cal{S}}_{{\cal{I}}}(V)-{\cal{S}}_{{\cal{I}}}(0)$ and ${\cal{S}}^{0}_{exc}={\cal{S}}_{0}(V)-{\cal{S}}_{0}(0)$. Figures 4b,c,d present the noise deviation $\delta {\cal{S}}={\cal{S}}^{0}_{exc}-{\cal{S}}^{{\cal{I}}}_{exc}$ as a function of current deviation $\delta {\cal{I}}={\cal{I}}_{0}-{\cal{I}}$. The question of our interest is how different approximations reproduce the linear scaling law proposed in  ~\cite{82} and what deviations from linearity are expected away from the unitary limit. The value of the slope of $\delta {\cal{S}}$ curve versus $2e\delta {\cal{I}}$ defines invariant Fano factor ${\cal{IF}}$. The experimental data of Dellatre et al. for different CNT-QDs and different gate voltages ~\cite{82} give values of ${\cal{IF}}$ very close to $0.5$, which is the number predicted by SBMFA theory. This is a consequence of the fact that quasiparticles in this picture are scattered elastically on spin-orbital singlet resonance. Fig.4b  shows that also other approximations  trace this result in the deep dot level limit where charge fluctuations are of minor importance. Exception is Fermi liquid theory, which  gives ${\cal{IF}}$$=- 0.3$ and this is a consequence of the earlier mentioned  inelastic scattering resulting from  polarization effects of spin-orbital singlet ~\cite{81}. Figs 4c,d   illustrate the robustness of noise scaling ~\cite{82} on  the deviation from the unitary limit. The $\delta {\cal{S}}$ versus $\delta {\cal{I}}$ curves calculated by EOM method for different values of Coulomb interaction and dot energies for temperature of order of  $T_{K}/3$ are presented. These  pictures illustrate the case when   the effective  spin-orbital pseudospin is not totally quenched due to the interplay of spin-orbital fluctuations  with charge fluctuations and in consequence ${\cal{IF}}$ deviates from $0.5$.  We only announce here the interesting problem of the role of charge fluctuations on the noise of fully symmetric SU(4) Kondo   system leaving the detailed analysis for  the future publication and in the following we rather concentrate on the impact of  symmetry breaking perturbations. Not trying to do any qualitative predictions it is worth to observe the qualitative resemblance of some of the finite ${\cal{U}}$  curves  from Fig.4d  to the experimental data presented in ~\cite{82}. Apart form the oversimplified approximations used by us one has also to remember   that our calculations concern zero frequency whereas experiment ~\cite{82} has been done for finite frequency. In this case if $T\neq0$ it is not possible completely separate the shot noise from thermal noise (see eq. ($127$) from Ref. ~\cite{1}).

Let us close this paragraph by a comparison of ${\cal{F}}$  for finite ${\cal{U}}$ with infinite ${\cal{U}}$ limit (Fig. 5a). This will help to follow the analysis presented in the next section.
\begin{figure}
\includegraphics[width=6 cm,bb=0 0 741 725,clip]{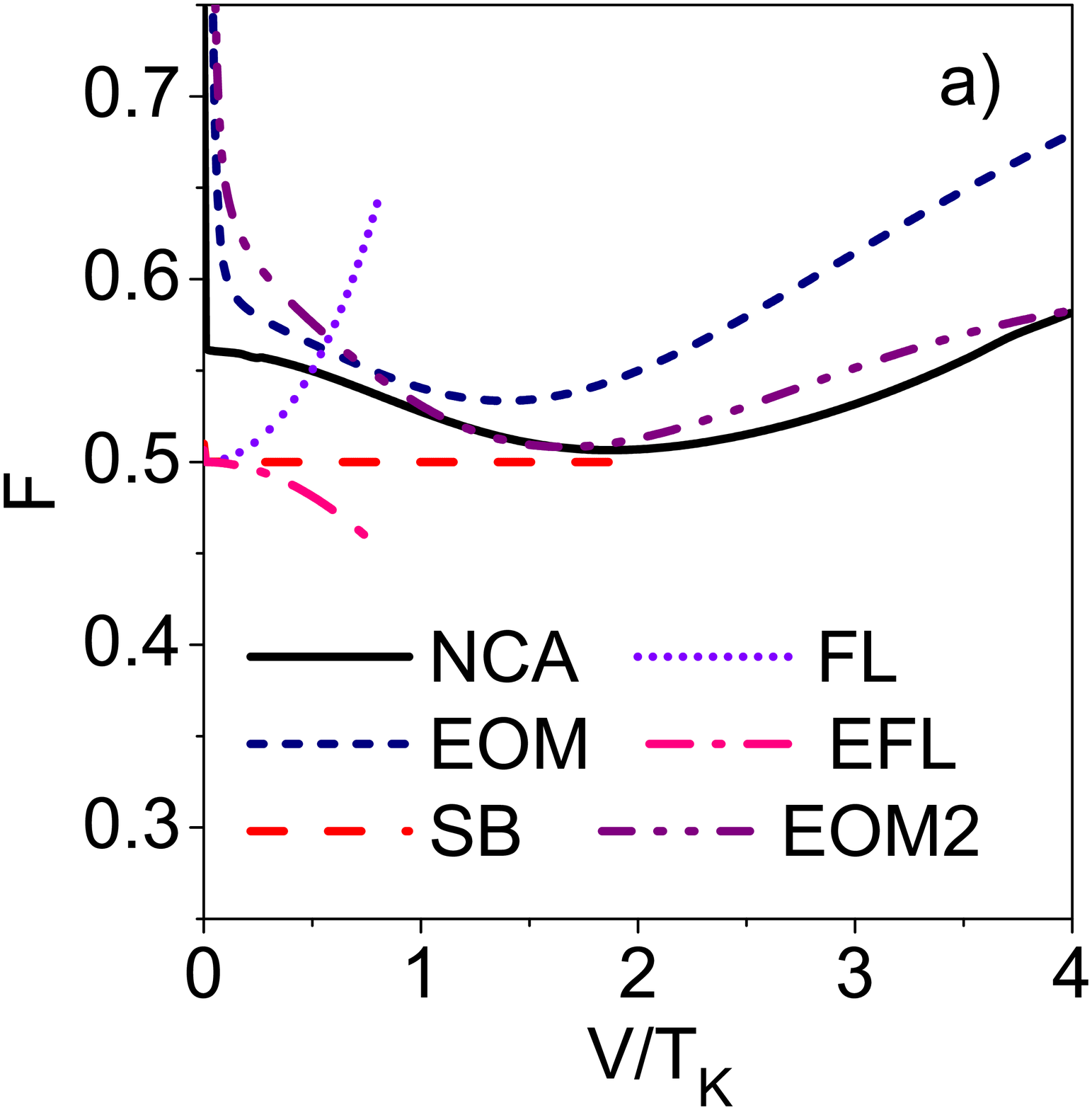}
\includegraphics[width=6 cm,bb=0 0 741 725,clip]{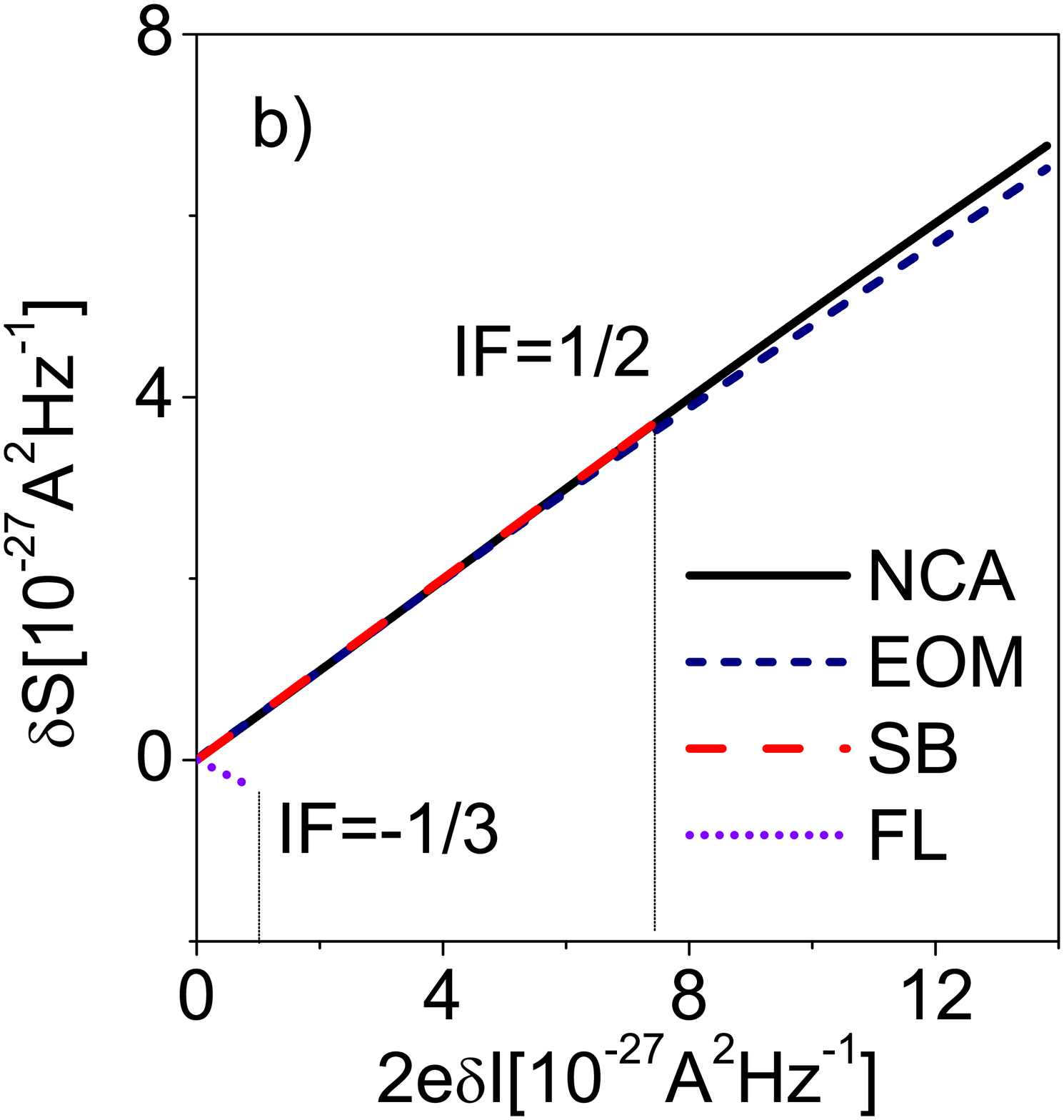}
\includegraphics[width=6 cm,bb=0 0 741 725,clip]{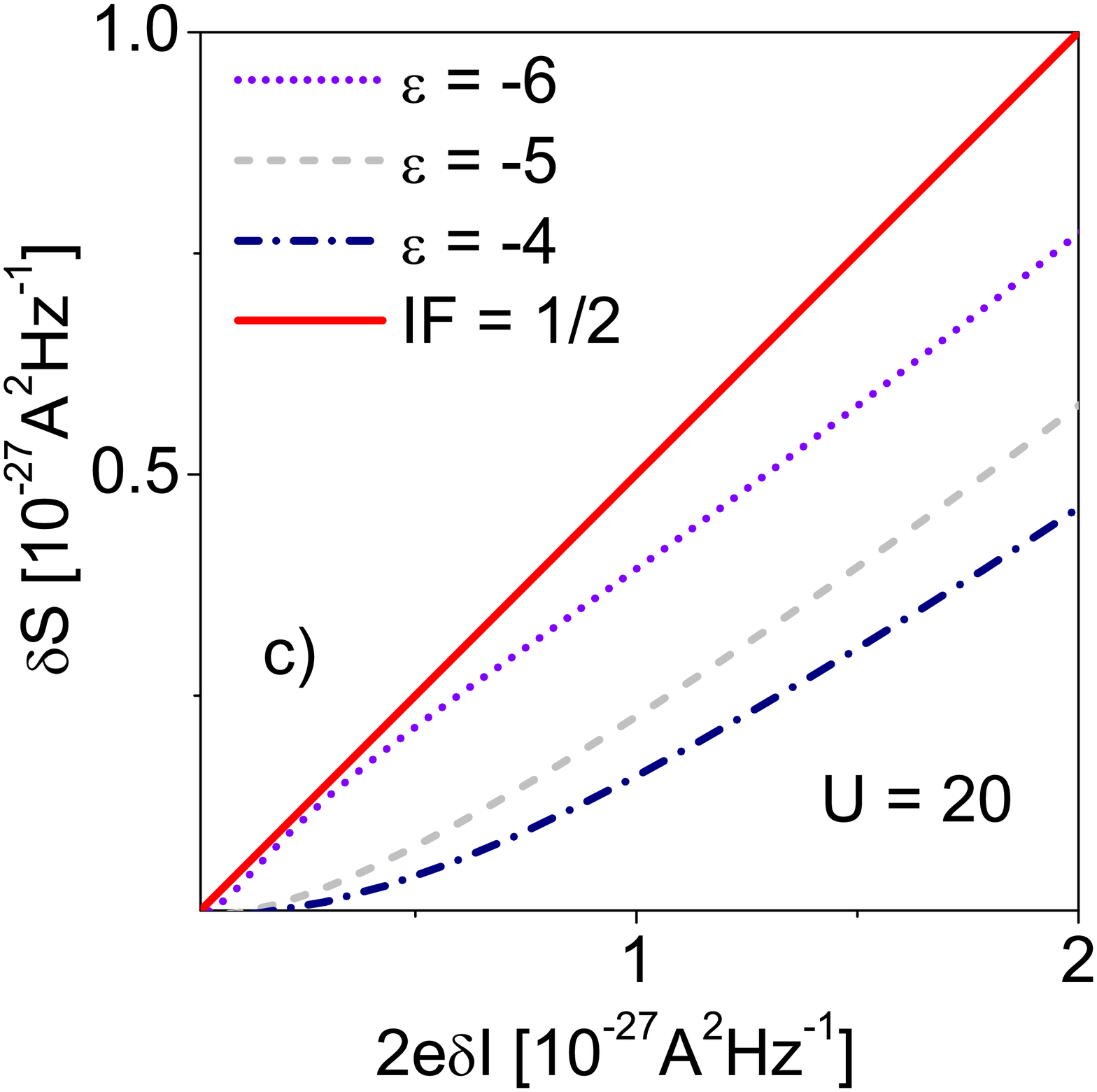}
\includegraphics[width=6 cm,bb=0 0 741 725,clip]{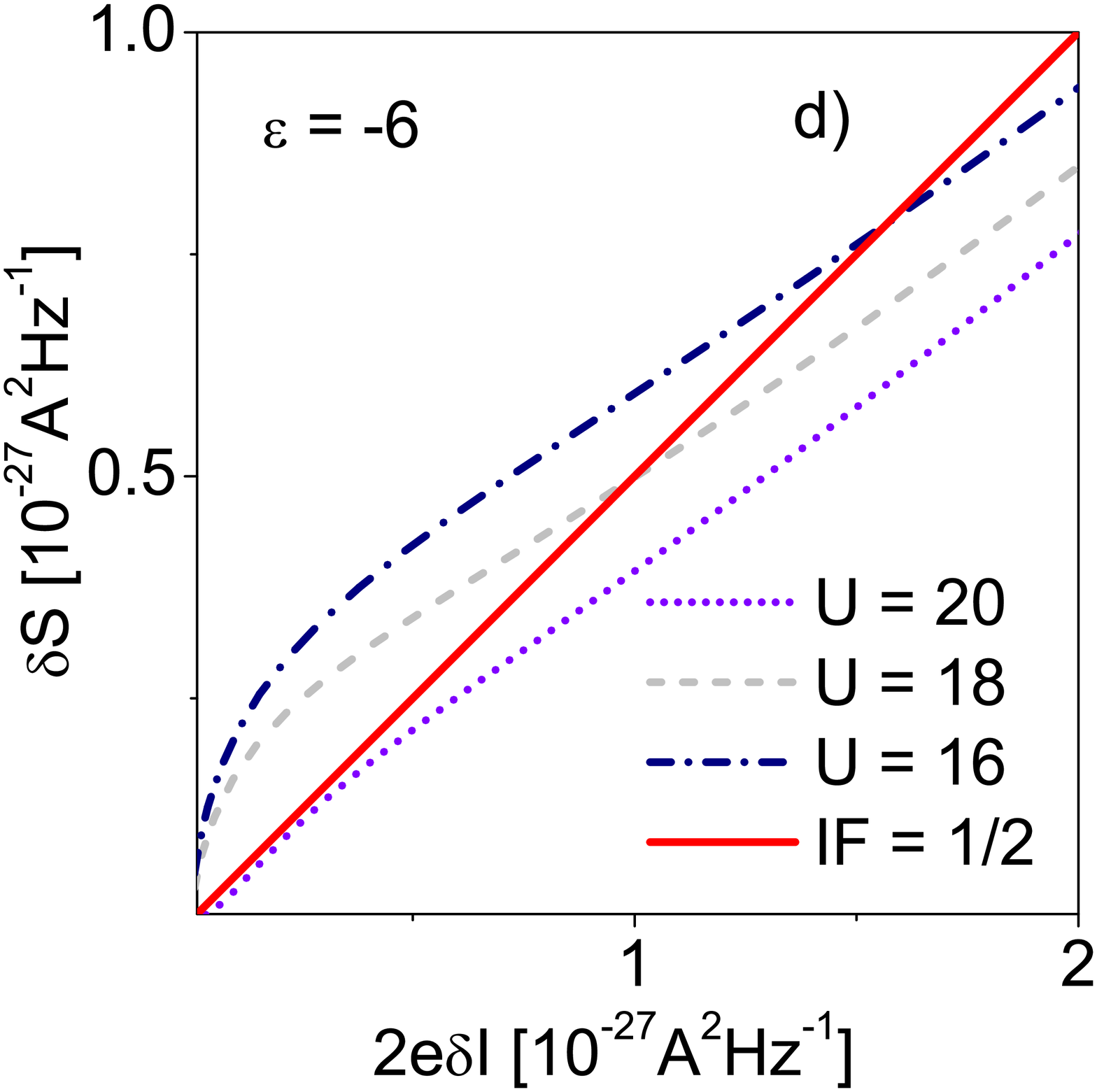}
\caption{\label{fig:epsart} (Color online) a) Bias dependencies of Fano factors ${\cal{F}}={\cal{S}}/2e{\cal{I}}$ of CNT-QD ($\epsilon=-9$) calculated by different methods compared with Fermi liquid predictions ~\cite{81}. SB, NCA (${\cal{U}}=\infty$), EOM and EOM$2$ (${\cal{U}}=25$). EOM and EOM$2$ curves have been calculated by the equation of motion method with decoupling ($16$)(EOM) or by taking into account correlations ($15$) in the lowest order (EOM$2$, see the text). Fermi liquid curves have been obtained using expressions for the current and noise given in ~\cite{81}. EFL denotes Fermi liquid Fano factor dependence that includes only elastic terms and FL curve takes into account both elastic and inelastic processes. b) Noise deviation $\delta{\cal{S}}={\cal{S}}^{0}_{exc}-{\cal{S}}^{{\cal{I}}}_{exc}$ as a function of current deviation $\delta{\cal{I}}={\cal{I}}_{0}-{\cal{I}}$ calculated by SBMFA, EOM and NCA methods compared with F-L predictions ~\cite{81} (parameters are the same as in Fig.4a). The slope of the lines determines the generalized Fano factors ${\cal{IF}}$. The dotted vertical lines mark the bias range limits for F-L and SBMFA from Fig. 4a. c) Deviations from the linear noise scaling ~\cite{82} - $\delta{\cal{S}}$ versus $\delta{\cal{I}}$ for different values of dot energy (EOM, ${\cal{U}}=20\Gamma$, $\Gamma=1$ meV). d) Deviations from linear scaling - $\delta{\cal{S}}$ versus $\delta{\cal{I}}$ for different values of Coulomb interaction ${\cal{U}}$ (EOM, $\epsilon=-6\Gamma$, $\Gamma=1$ meV).}
\end{figure}
Similarly as we mentioned in the discussion of ${\cal{IF}}$,  the low bias dependence of ${\cal{F}}$ outside  the unitary limit is not solely determined by Kondo correlations. For infinite ${\cal{U}}$ charge fluctuations ($n=0\leftrightarrows n=1$) play also the role and for finite ${\cal{U}}$ additional fluctuations ($n=1\leftrightarrows n=2$) come into play. The reader is referred to  an example of DOS (Fig. 5c).  Apart from Coulomb peak ($\epsilon\sim4$) corresponding to the fluctuation into double occupied state also a track of fluctuations into higher occupancy is visible for higher energies ($\epsilon\sim14$, $n=3$) , they are however not relevant for the low energy transport discussed here. The quantitative difference of the curves for ${\cal{U}}\rightarrow\infty$ and ${\cal{U}}=10$ is caused both by weakening of Kondo correlations with the decrease of ${\cal{U}}$ and by the increase of the role of charge fluctuations. Two features are clearly visible, deviations from $1/2$ limit for small bias and a shift of minimum of ${\cal{F}}(V)$ from $V\approx2T_{K}$ towards smaller voltages. The full symmetric  case discussed so far is not easily accessible experimentally.
\begin{figure}
\includegraphics[width=6 cm,bb=0 0 741 725,clip]{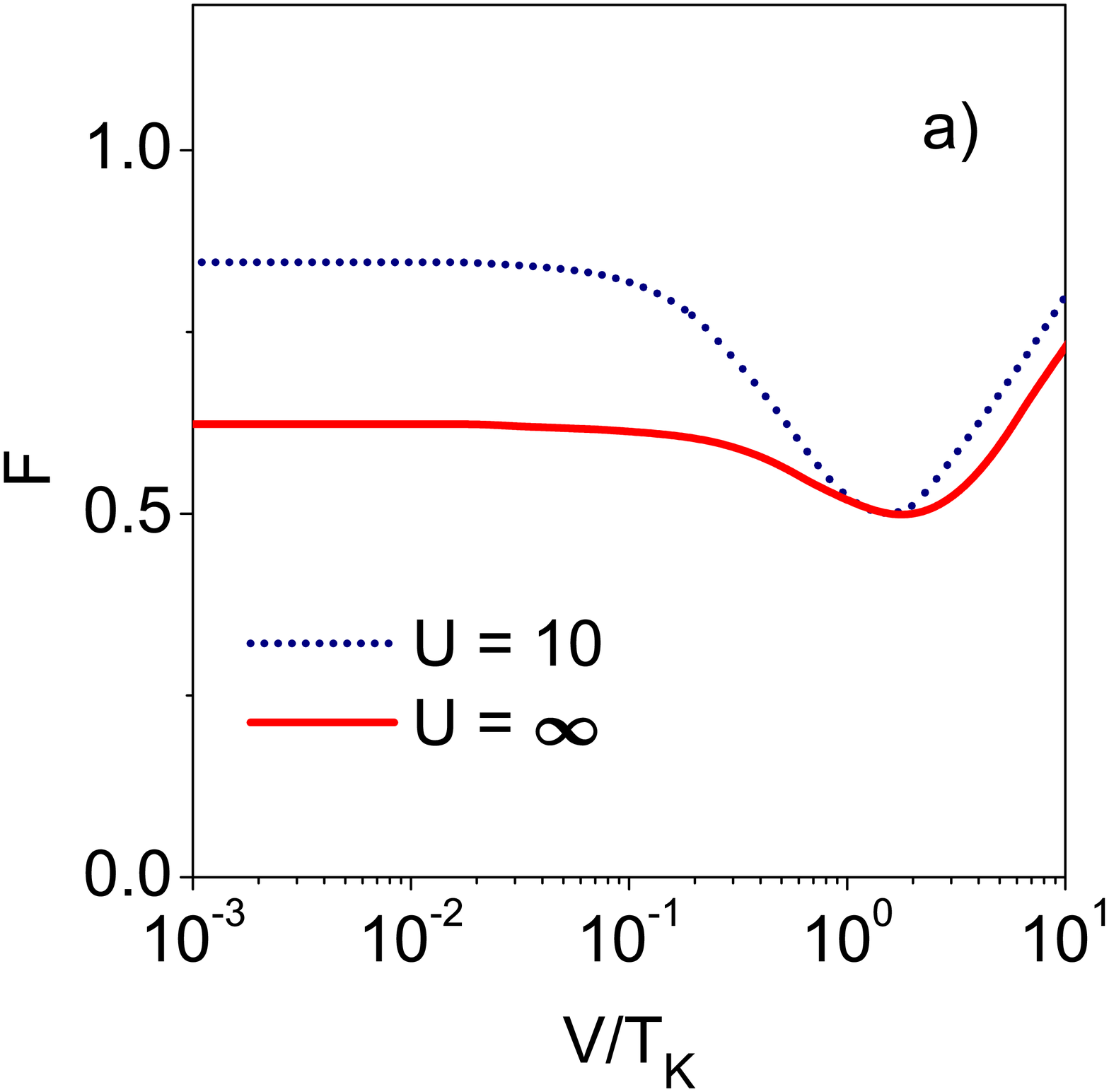}
\includegraphics[width=6 cm,bb=0 0 741 725,clip]{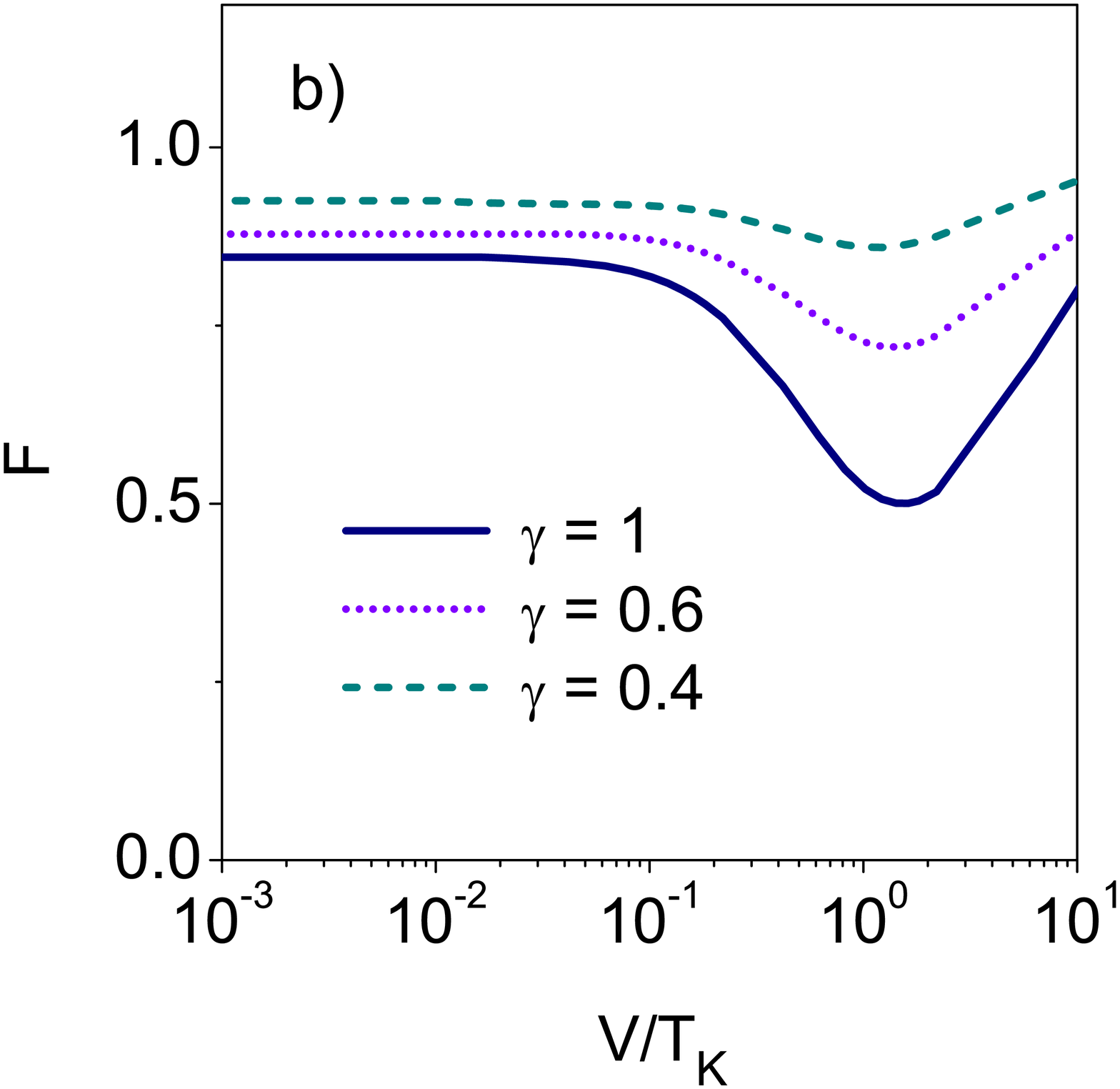}
\includegraphics[width=6 cm,bb=0 0 741 725,clip]{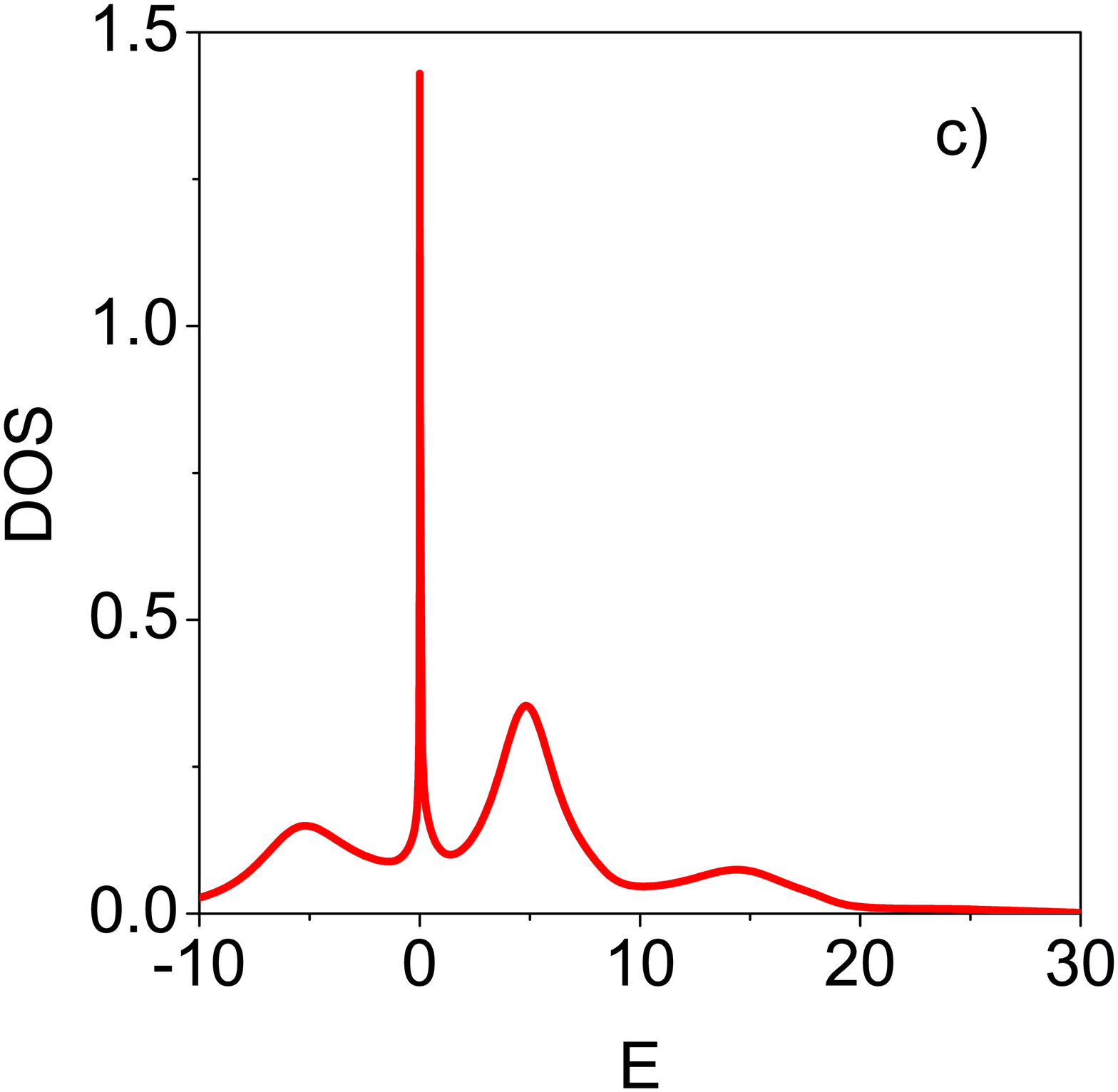}
\caption{\label{fig:epsart} (Color online) a) Bias dependence of  Fano factors of SU(4) CNT-QD ($\epsilon=-6$) in the Kondo range calculated by means of EOM approach for finite ${\cal{U}}$
and in the infinite ${\cal{U}}$ limit. (b) Fano factor for ${\cal{U}}=10$ and asymmetric coupling $\gamma\neq1$. (c) Density of states of SU(4) CNT-QD ($\epsilon=-6$, ${\cal{U}}=10$) (EOM).}
\end{figure}
On Fig. 5b we present as an example  the effect of one of the perturbations - the  left-right asymmetry.  This perturbation does not break the spin-orbital SU(4) symmetry. The asymmetry between the left and right barriers gives rise to asymmetry of  the shot noise and current with respect to the bias reversal. Asymmetry weakens Kondo correlations which is observed in a weaker suppression of Fano factor for $\gamma\neq1$  both in small bias limit and for $V\sim 2T_{K}$.

\subsection{Broken SU(4) symmnetry of the spin-orbital  CNT-QD Kondo system}
So far we have discussed SU(4) CNT-QD, where due to the entanglement the  spin and orbital degrees of freedom participate on the same footing. In this  Section we will analyze the systems where the role of  one of the degrees of freedom is suppressed and the system is left in an SU(2) Kondo state  stemming from the other degree of freedom or where both quantities are knocked out from the degeneracy, but the dot is still close enough to the degeneracy and  the richness of  cotunneling processes strongly influence the shot noise.\\

\noindent \textbf{1) Magnetic field}\\

Let us first discuss the influence of  field perpendicular to carbon nanotube axis, which breaks only the spin degeneracy. In Fig. 6a the field dependence of the Fano factors for both spin channels calculated within SBMFA (Coleman, ${\cal{K-R}}$) are presented. As it is seen in the limits of small fields $h\ll T_{K}$ and large fields $h\sim T_{K}$, both approximations give qualitatively similar results. The  monotonic increase of Fano factor for up spin channel ${\cal{F}}_{++}={\cal{F}}_{L1+L1+}={\cal {S}}_{L1+L1+}/(2{\cal{I}}_{L1+})={\cal{F}}_{L2+L2+}$ with the increase of  the field observed in SBMFA picture and the decrease of ${\cal{F}}_{--}={\cal{S}}_{L1-L1-}/(2{\cal {I}}_{L1-})$ are the consequence of the increasing splitting of the  Kondo peak of DOS (Fig. 6e). The  down spin peak ($\sigma= -$)  moves towards the Fermi level and up spin towards higher energies.  Certainly the charge fluctuations not included in MF approach would modify the picture, but for the deep dot level and infinite ${\cal U}$ this is of minor importance.  The low field region is believed to be well reproduced by SBMFA. Concerning the  moderate fields ($0 < h < T_{K}$), where a crossover from  spin-orbital SU(4) Kondo effect to two-level SU(2) orbital Kondo effect (TL SU(2)) starts, the SBMFA description is questionable and the plotted dependencies should be considered as an interpolation between low and high field range only. Spin and orbital degrees of freedom are already not fully entangled in this range  and  are not enough  detangled to induce perturbed  SU(2) type behavior (the regime where SBMFA formalism is  expected to give  again a reasonable description).
Before we present EOM results, which give  more detailed insight into the  richness of the many-body fluctuations in systems with broken symmetry, let us first comment on the large field limit of SBMFA calculations ($h > T_{K}$) ~\cite{120}.
The orbital fluctuations for each spin channel play the dominant role in this regime and their interplay between different channels decreases with the increase of magnetic field. The SBMFA spin resolved Fano factors approach in high fields  the limits $0$ or $1$.
Despite the crudeness of MF approach, which overestimates the weight of the up spin resonance shifted from the Fermi level by $2h$, the limiting values of Fano factors seem to be correct. This conviction is based on a comparison with the similar results for double dot systems obtained in the Numerical Renormalization Group (NRG) approach ~\cite{65}, where a crossover to a purely orbital Kondo state SU(2) for down-spin electrons has been predicted. Now let us analyze the EOM results presented on Fig. 6b.
\begin{figure}
\includegraphics[width=6 cm,bb=0 0 741 725,clip]{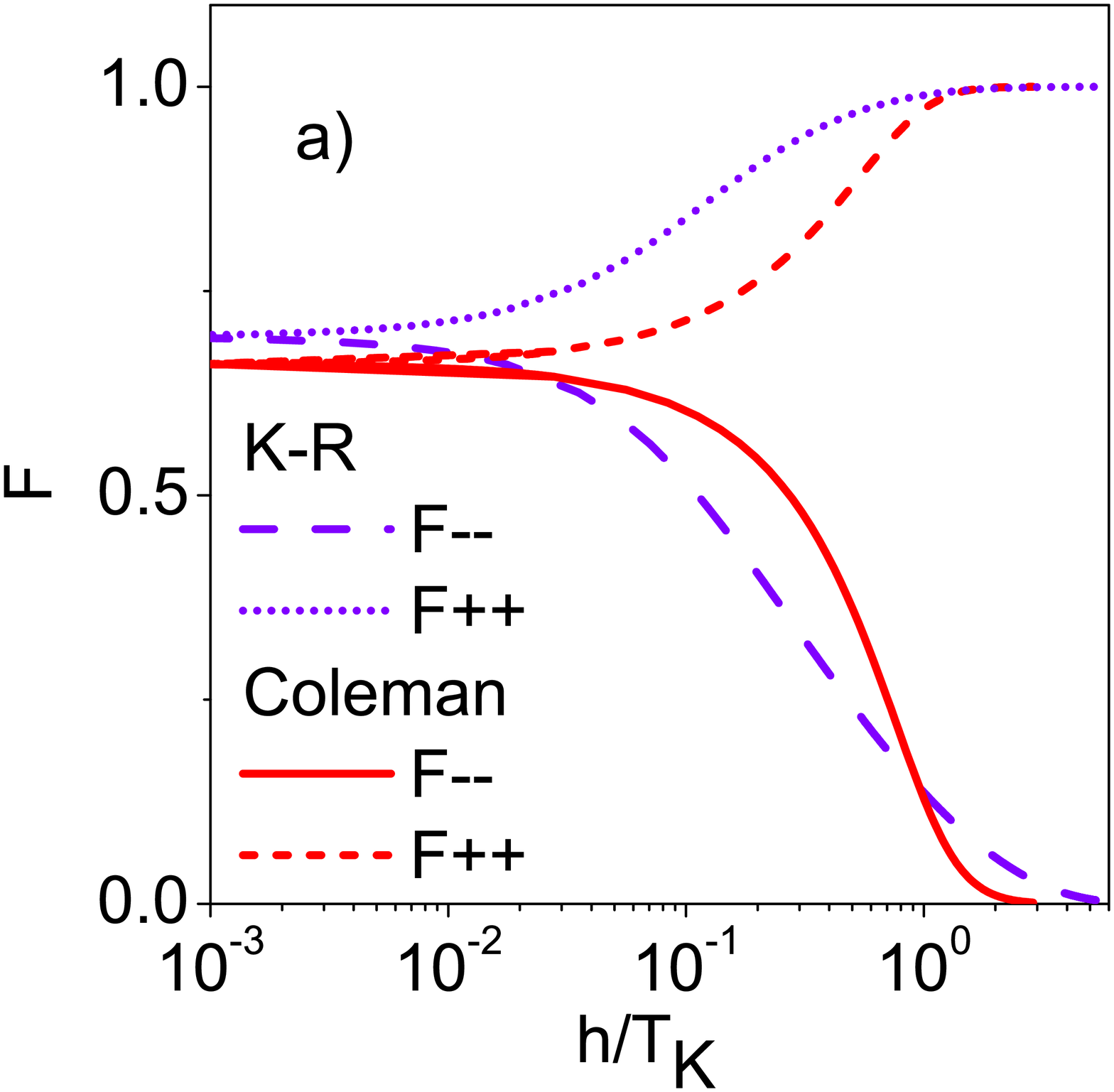}
\includegraphics[width=6 cm,bb=0 0 741 725,clip]{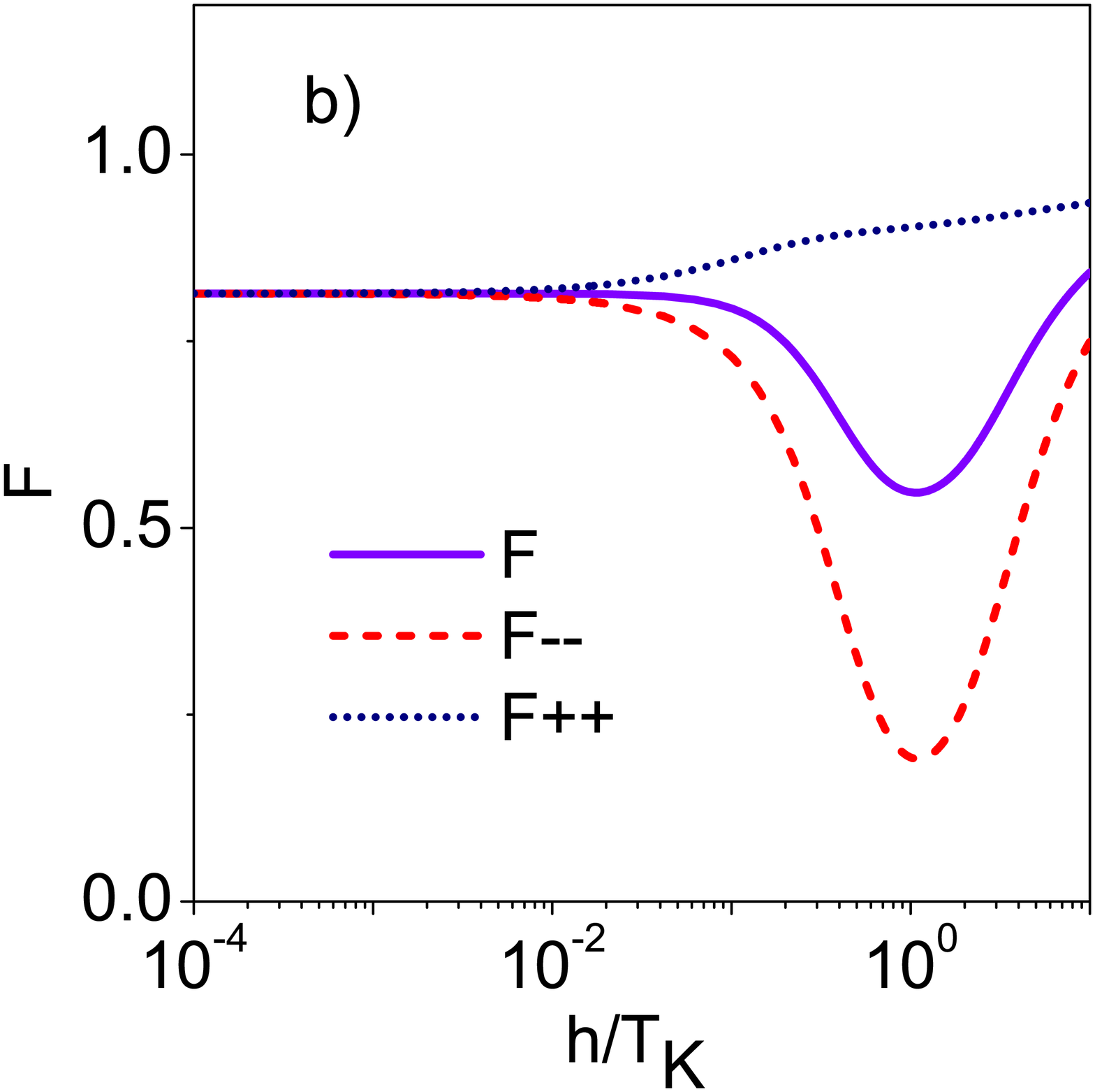}
\includegraphics[width=6 cm,bb=0 0 741 725,clip]{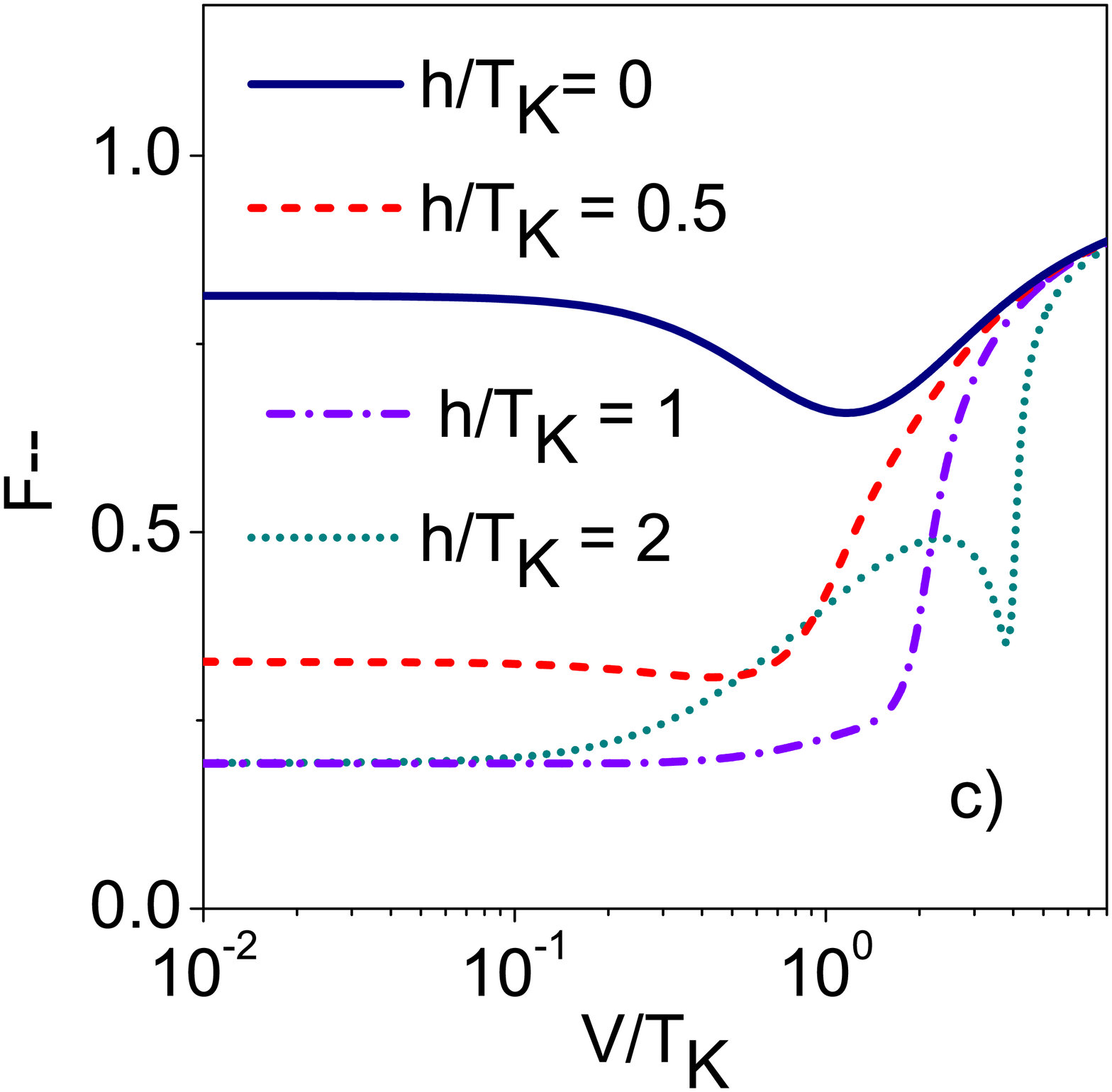}
\includegraphics[width=6 cm,bb=0 0 741 725,clip]{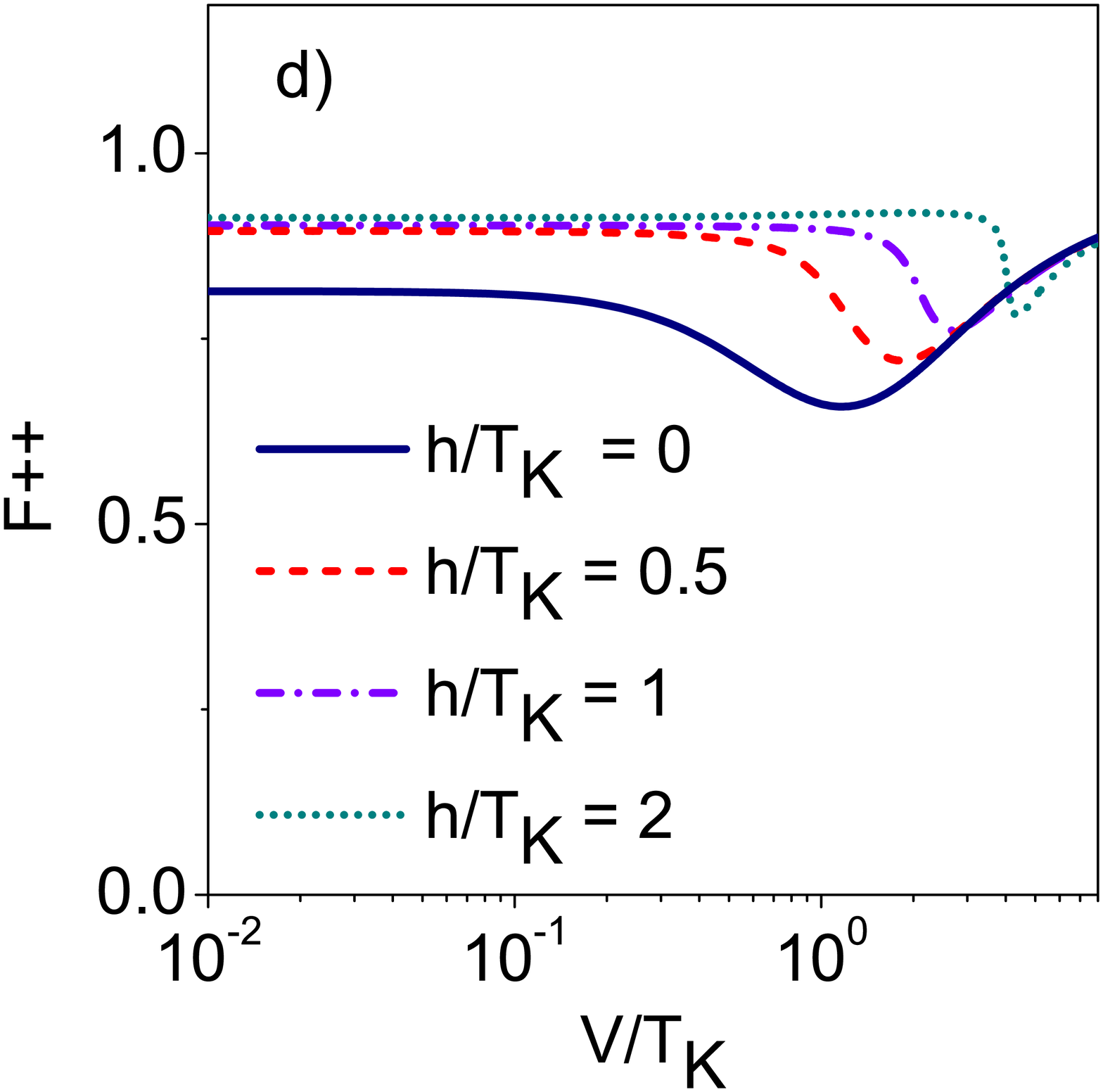}
\includegraphics[width=6 cm,bb=0 0 741 725,clip]{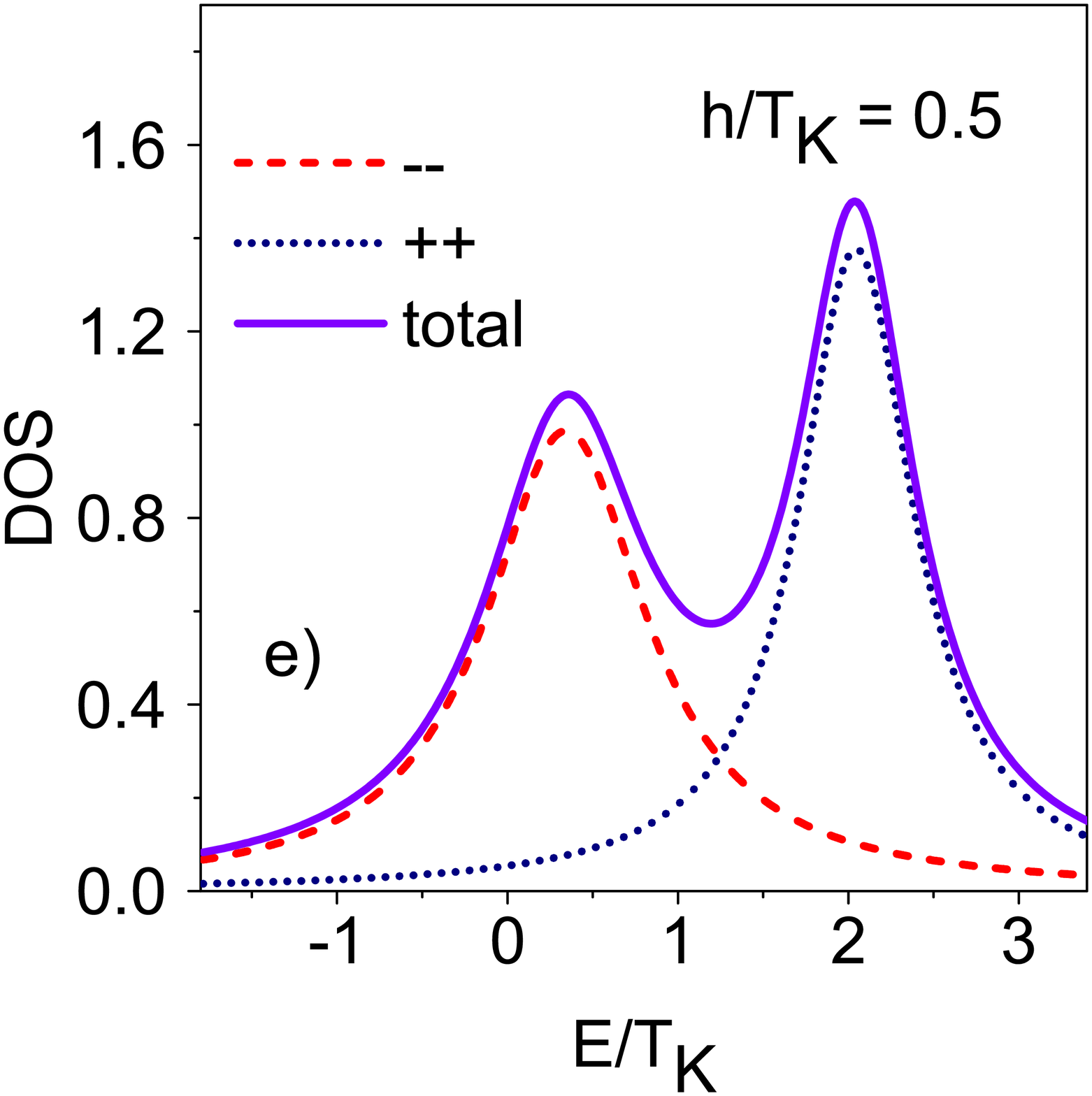}
\includegraphics[width=6 cm,bb=0 0 741 725,clip]{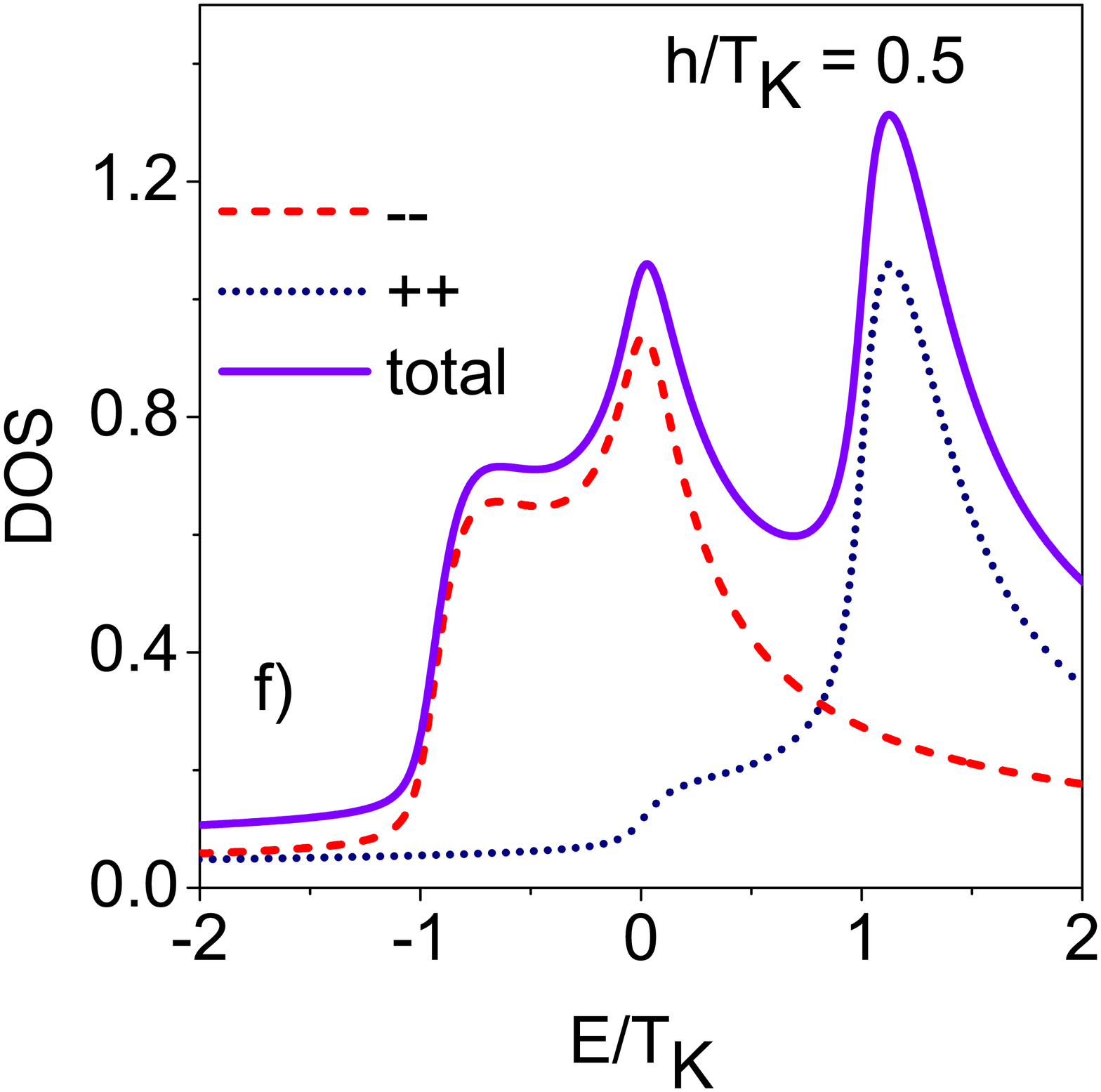}
\caption{\label{fig:epsart} (Color online) Magnetic field dependence of spin-resolved Fano factors of CNT-QD in the Kondo
               range ($\epsilon= -6$) for perpendicular orientation of the field ($\theta= \pi/2$). (a) Linear Fano factors (${\cal{U}}=\infty$, SBMFA-Coleman, {\cal{K}}-{\cal{R}}). (b) Linear Fano factors calculated for finite ${\cal{U}}=15$ (EOM). (c,d) Bias dependencies of  spin resolved Fano factors for different values of the field (EOM). (e,f) Total and partial densities of states of  CNT-QD ($\epsilon=-6$) in perpendicular magnetic field (e) ${\cal{U}}=\infty$ (SBMFA - Coleman) and (f) ${\cal{U}}=15$ (EOM).
}
\end{figure}
Breaking of SU(4) symmetry reflects in the decrease of total current Fano factor with the increase of the field. For more detailed discussion the reader is referred  to the illustration of corresponding DOS presented on Fig. 6f.   For small fields (not presented) the many body resonance has still a single peak structure, the orbital, spin and spin-orbital fluctuations are not well resolved in this range, but the partial spin densities of states shift in energy with the  increase of the field.
\begin{figure}
\includegraphics[width=6 cm,bb=0 0 741 725,clip]{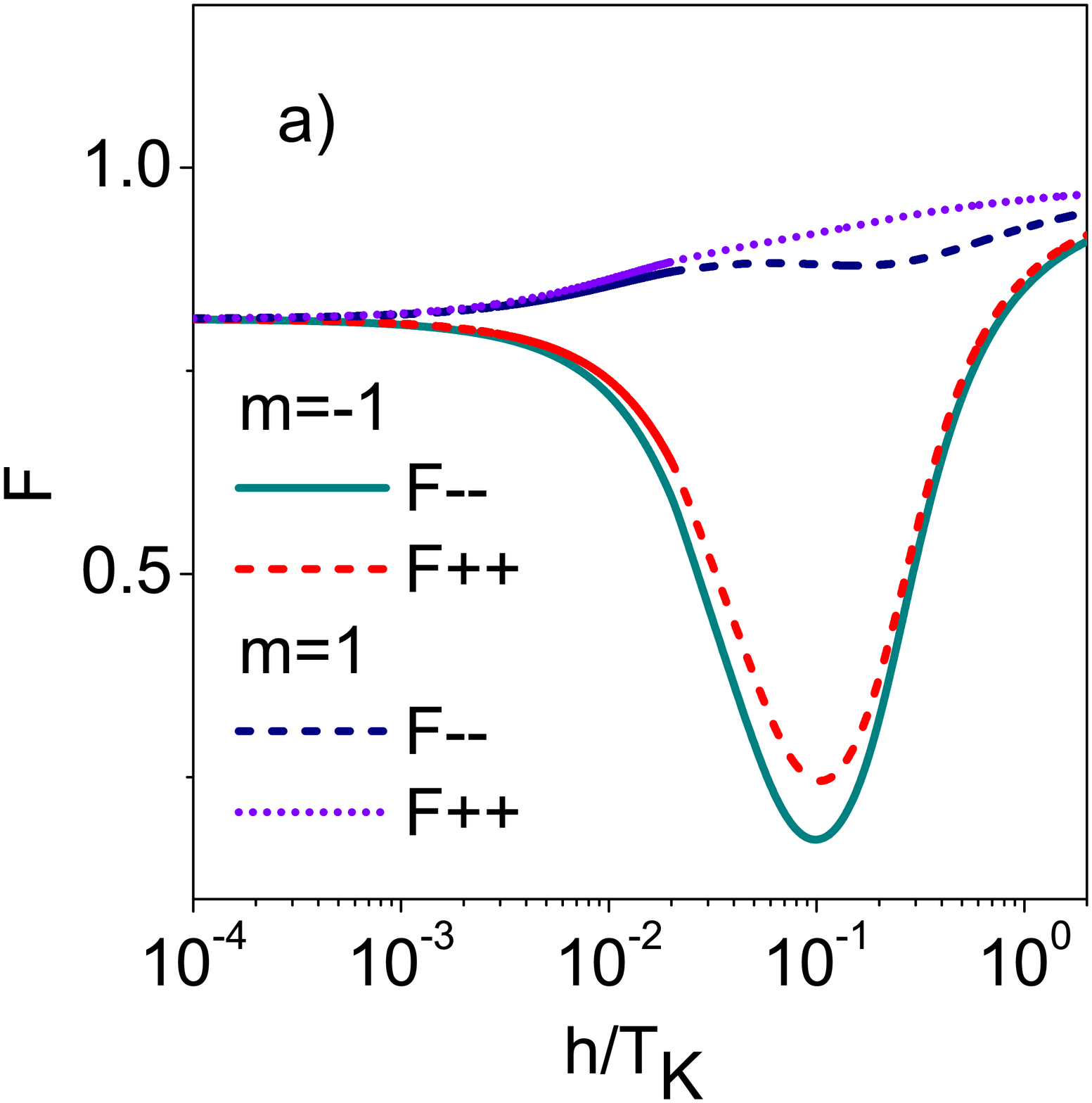}
\includegraphics[width=6 cm,bb=0 0 741 725,clip]{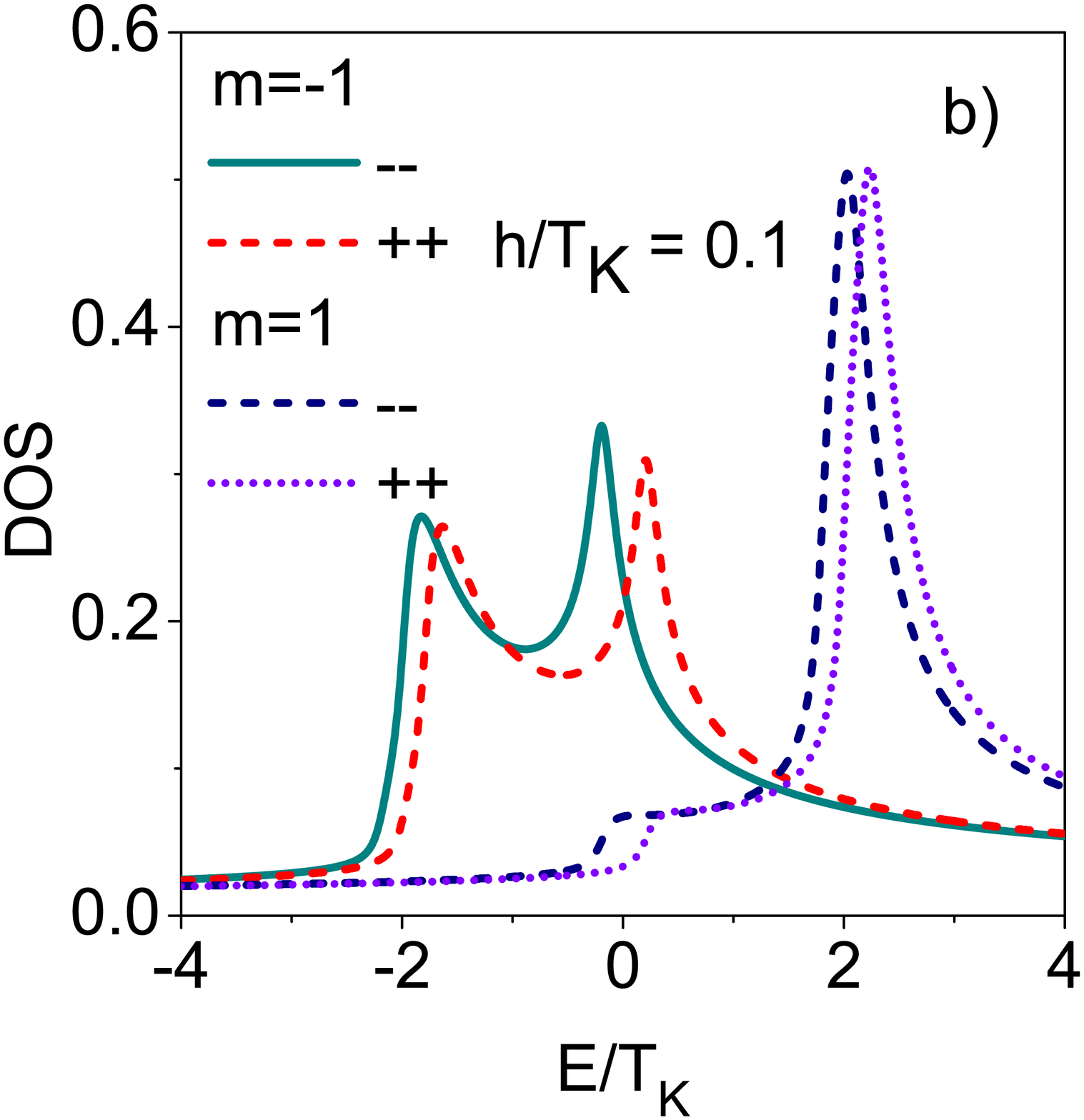}
\caption{\label{fig:epsart} (Color online) (a) Magnetic field dependence of spin and orbital resolved linear Fano factors of CNT-QD for
             axial field ($\theta= 0$) calculated by means of EOM approach. (b) Partial densities of states of CNT-QD in parallel field  ($\epsilon= -6$, ${\cal{U}}=15$) (EOM).}
\end{figure}
The down spin partial DOS dominates   in the neighborhood of $E_{F}$ in this range  which results in   a  strong suppression of linear Fano factor  for this spin orientation. For higher fields (Fig.6f) the three peak structure in DOS is observed, what is in contrast to SBMFA DOS (Fig. 6e). The central peak corresponds to the coupled orbital  fluctuations for both spin orientations and  a pair  of  the satellites accounts for  spin and simultaneous spin-orbital fluctuations. The higher energy fluctuations do not manifest in  the low  voltage shot noise, but their effect can be observed in the bias dependence of ${\cal F}$ (Figs. 6c,d). The opposite shift of minima of ${\cal F}_{++}(V)$ and ${\cal F}_{--}(V)$ for small fields ($h<T_{K}$)  reflects the earlier mentioned   spin dependent  energy  redistribution of DOS,  the sharp minima  seen for higher fields ($h>T_{K}$) account for fluctuations responsible for  the  satellites. Now let us turn to the parallel magnetic field case.  Parallel field breaks both spin and orbital degeneracies. The orbital pseudospin is more sensitive to the parallel magnetic field than the real spin  ($\mu_{orb}\gg \mu_{B}$) and therefore the similar effects as described for the perpendicular field occur in this case, but for much lower fields. Fig. 7 presents spin and orbital resolved Fano factors. Already for small axial fields a strong depression of ${\cal F}$ is observed and opposite tendency is seen  for different orbital channels. The spin resolution of the noise for these fields  is much weaker.  Breaking of SU(4) symmetry in this case is associated with a dramatic reconstruction of the many-body DOS (Fig. 7b).    The peaks in the center account for spin Kondo fluctuations for both orbitals and the satellites correspond to orbital isospin and simultaneous isospin and spin fluctuations.  The field evolution of the linear Fano factors is related to the changes of DOS close to $E_{F}$.\\

\noindent \textbf{2) Spin filtering}\\

We discuss now the spin dependent shot noise of CNT Kondo single shell spin filter ~\cite{121}.  This recently proposed by us    filtering mechanism  is based on the  field induced tuning of the spin-polarized nondegenerate states from the same shell  into orbital degeneracy. A similar idea was put forward earlier in ~\cite{60},  but the involved   orbital states were from different shells of CNT-QD, which implied   a use of high magnetic fields for filtering of order of several Teslas in contrast to the present proposition where fields of order of fraction of Tesla are sufficient. In our proposal it is assumed that  the considered CNT-QD are characterized by  an orbital level mismatch ($\Delta_{orb}\neq 0$), which is expected to occur e.g. in nanotubes with torsional deformation. Axial magnetic field might recover the orbital degeneracy either within the same spin sector ($\epsilon_{-1+} =\epsilon_{1+}$ and $\epsilon_{-1-} =\epsilon_{1-}$ for $h = \Delta_{orb}/2\mu_{orb}$) or with  the mixing of spin channels  ($\epsilon_{-1-}  = \epsilon_{1+}$ for $h = \Delta_{orb}/2(\mu_{orb} -1)$, or $\epsilon_{-1+}  = \epsilon_{1-}$  for $h = \Delta_{orb}/2(\mu_{orb} +1)$.
In the former case almost the same polarizations of conductance are observed for both orbital channels, whereas for the latter the spin polarizations of different orbital sectors have opposite signs.
\begin{figure}
\includegraphics[width=6 cm,bb=0 0 741 725,clip]{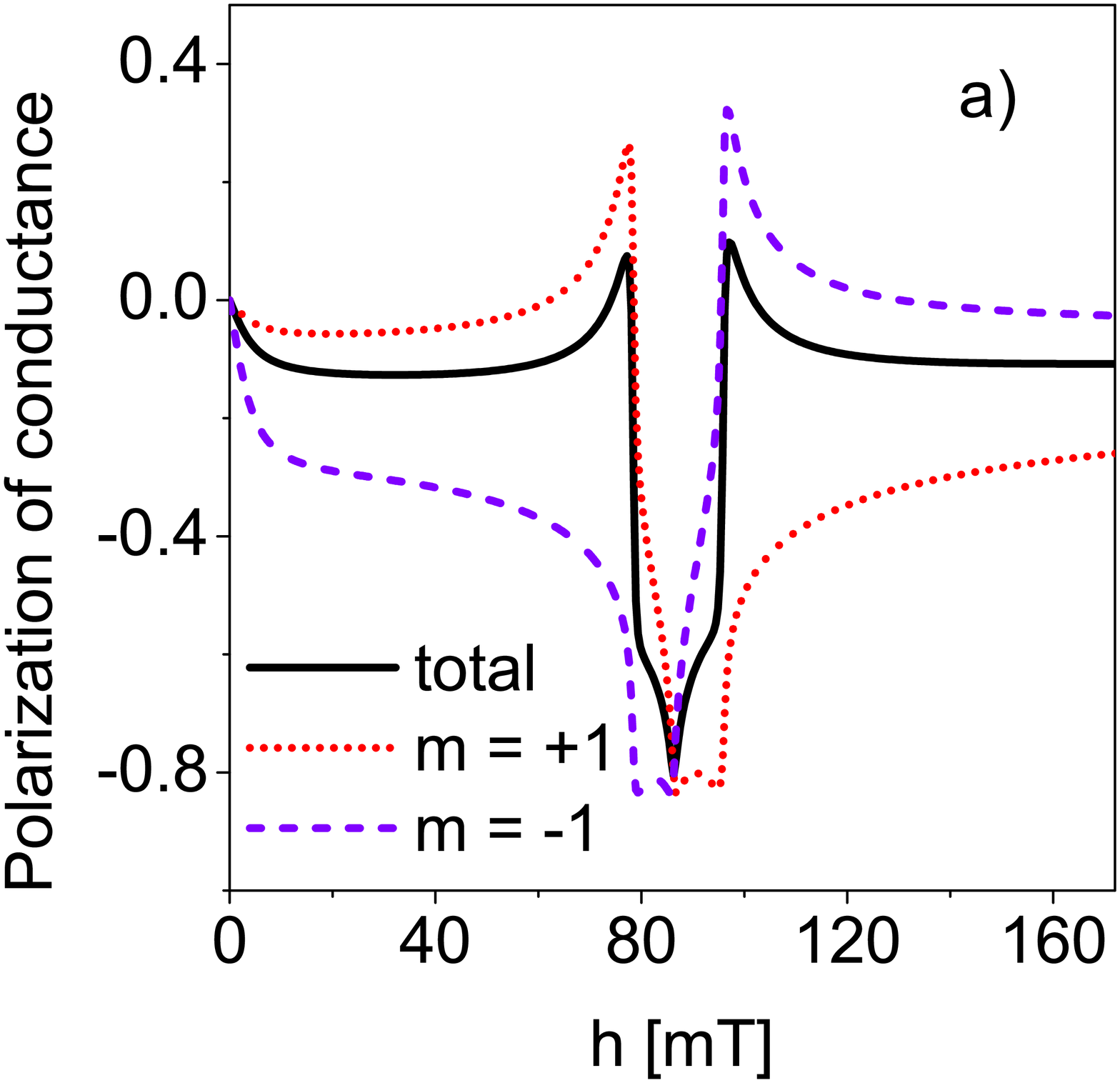}
\includegraphics[width=6 cm,bb=0 0 741 725,clip]{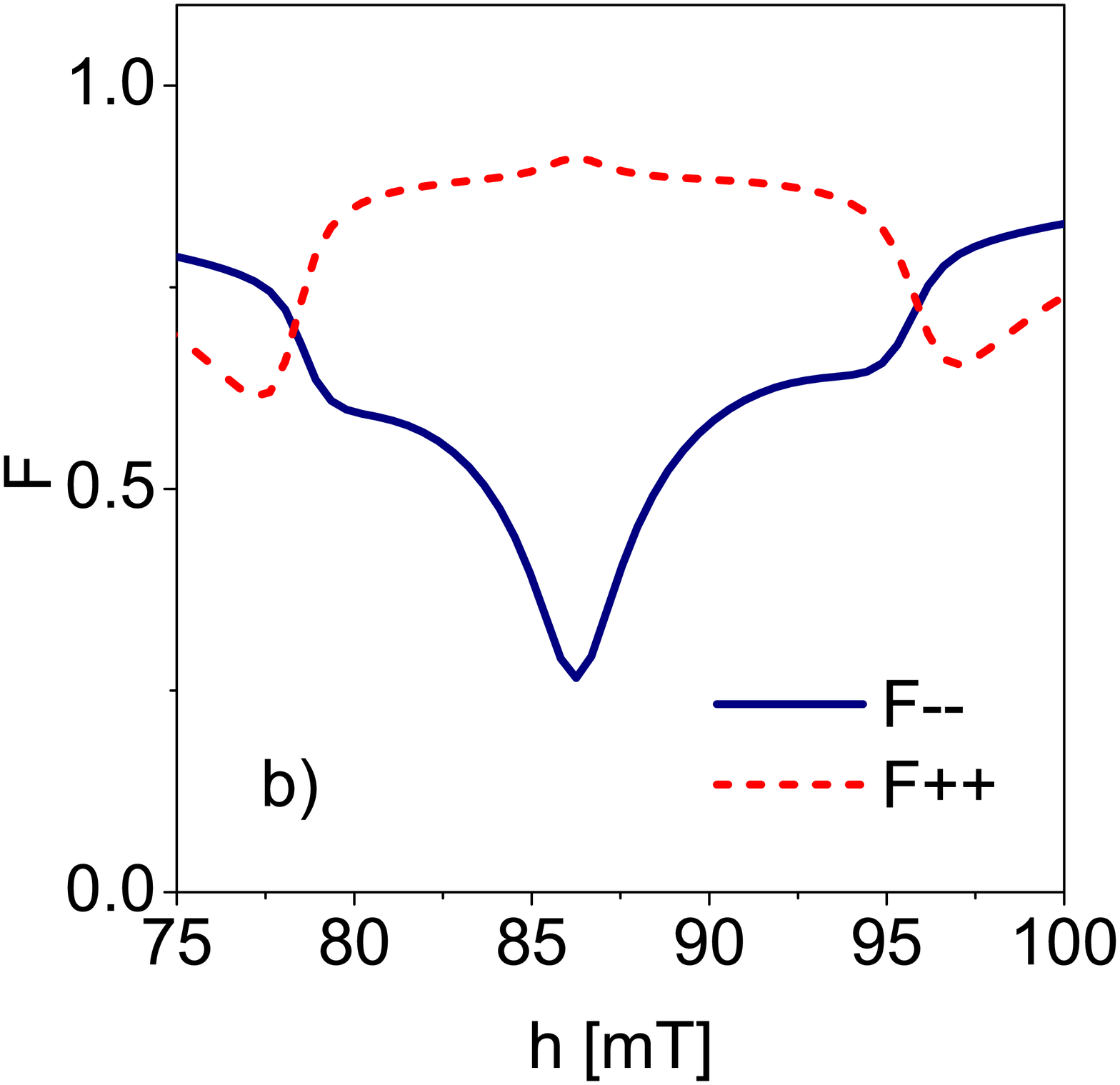}
\caption{\label{fig:epsart} (Color online) Spin filter (a) Total and orbital resolved polarization of conductance of CNT-QD in the Kondo
            regime ($\epsilon=-6$, $\Delta_{orb}=0.1$, ${\cal{U}}=15$, $\Gamma=1$ meV) (EOM). (b) Corresponding spin-resolved Fano factors.}
\end{figure}
The total polarization of conductance of CNT-QD spin filter  together with the orbital resolved polarizations are displayed on Fig. 8a. Fig. 8b presents the corresponding spin resolved Fano factors. The field induced restoring of orbital degeneracy allowing   the occurrence of orbital Kondo effect reflects also in a strong field suppression of Fano factor for one of the spin channels.\\

\noindent \textbf{3) CNT-QD coupled to ferromagnetic leads}\\

The Kondo effect in a quantum dot attached to ferromagnetic electrodes was widely discussed in literature, but only for SU(2) case ~\cite{44,45,46,47,48,49,50,51,52}. The same regards experimental investigations ~\cite{53,54,55,56}. Here we discuss the spin-orbital Kondo effect perturbed by polarization of the leads. The presence of  ferromagnetic electrodes breaks the spin degeneracy at the dot and  the  fluctuations of the real spin and orbital psudospin play in formation of the many-body resonance different role.   For the deep dot level far from charge degeneracy points  the spin  distinction reflects only in  the difference of the widths of the  many body resonances for the  opposite spin channels. Moving closer to the degeneracy points, but still remaining  in the Kondo regime, the spin-dependent charge fluctuations induce  an effective exchange field. To find the spin splitting we use, following ~\cite{49}  the perturbative scaling approach ~\cite{111}, where charge fluctuations are integrated out, but effectively introduce spin and orbital dependent renormalization of the effective dot energies.   Since in the following we present the results only for the symmetrically coupled CNT-QD with orbitally degenerate state we bring up below the formula for the exchange splitting of the dot levels characterized by single parameter $\Delta=\delta\epsilon_{1+}-\delta\epsilon_{1-}$:
\begin{eqnarray}
&&\Delta=-\int^{-\infty}_{+\infty}\frac{d\omega}{2\pi}\sum_{\alpha}
Re \left[\frac{\Gamma_{\alpha1+}(1-f_{\alpha}(\omega))}{\omega-\epsilon_{1+}+\imath0^{+}}
+\frac{\Gamma_{\alpha 1-}f_{\alpha}(\omega)}{-\omega+\epsilon_{1-}+{\cal{U}}+\imath0^{+}}- \right.\nonumber\\
&&\left.-\frac{\Gamma_{\alpha1-}(1-f_{\alpha}(\omega))}{\omega-\epsilon_{1-}+\imath0^{+}}
-\frac{\Gamma_{\alpha 1+}f_{\alpha}(\omega)}{-\omega+\epsilon_{1+}+{\cal{U}}+\imath0^{+}} \right]
\end{eqnarray}

The first term in (24) corresponds to electron-like processes and the second to  hole-like.
 The renormalizations can intuitively be understood as follows ~\cite{54}. In the  emptying processes of the dot  i.e. fluctuations between single occupied state $|m\sigma\rangle$ and empty  state $|0\rangle$  an electron with majority-spin   can tunnel between the QD and the leads easier than an electron with opposite orientation and this effectively  shifts down in energy the majority-spin state at the dot. Concerning filling processes there are two types of them intra and interorbital. Only the former are spin sensitive and induce spin splitting of the effective dot energies. This is a consequence of Pauli principle which allows   for the virtual tunneling of electron of the opposite spin to the electron that already resides at a given orbital ($|m\sigma\rangle \rightarrow|m\sigma\rangle|m-\sigma\rangle$). The intraorbital fluctuations cause  shift down of the renormalized dot energy of the minority spin electrons. Depending on the dot level position the dominant role in determining the exchange splitting is played either by electron or by hole processes. This gives the possibility of gate (electric field) control of spin-splitting in a quantum dot.

\begin{figure}
\includegraphics[width=6 cm,bb=0 0 741 725,clip]{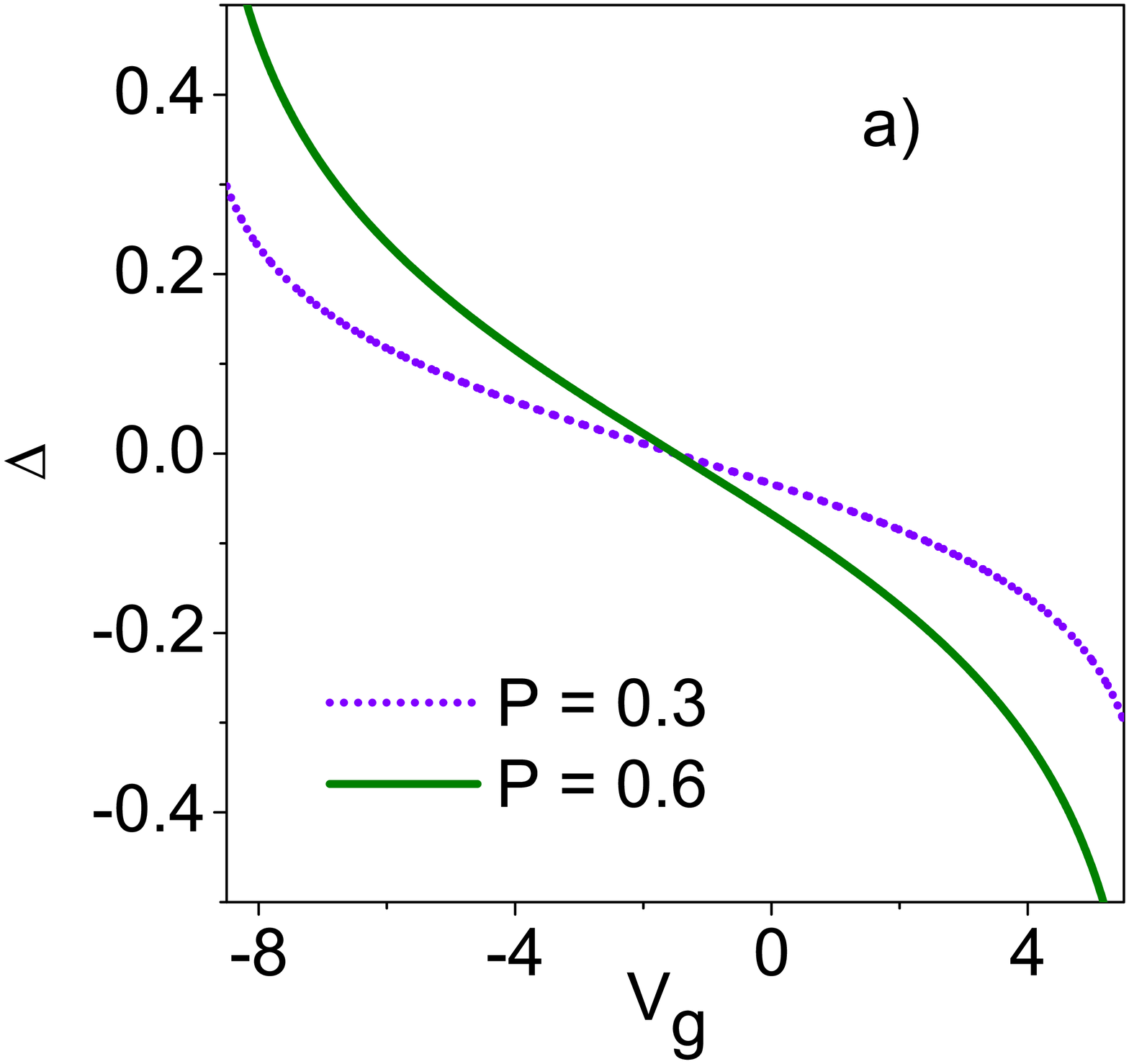}
\includegraphics[width=6 cm,bb=0 0 741 725,clip]{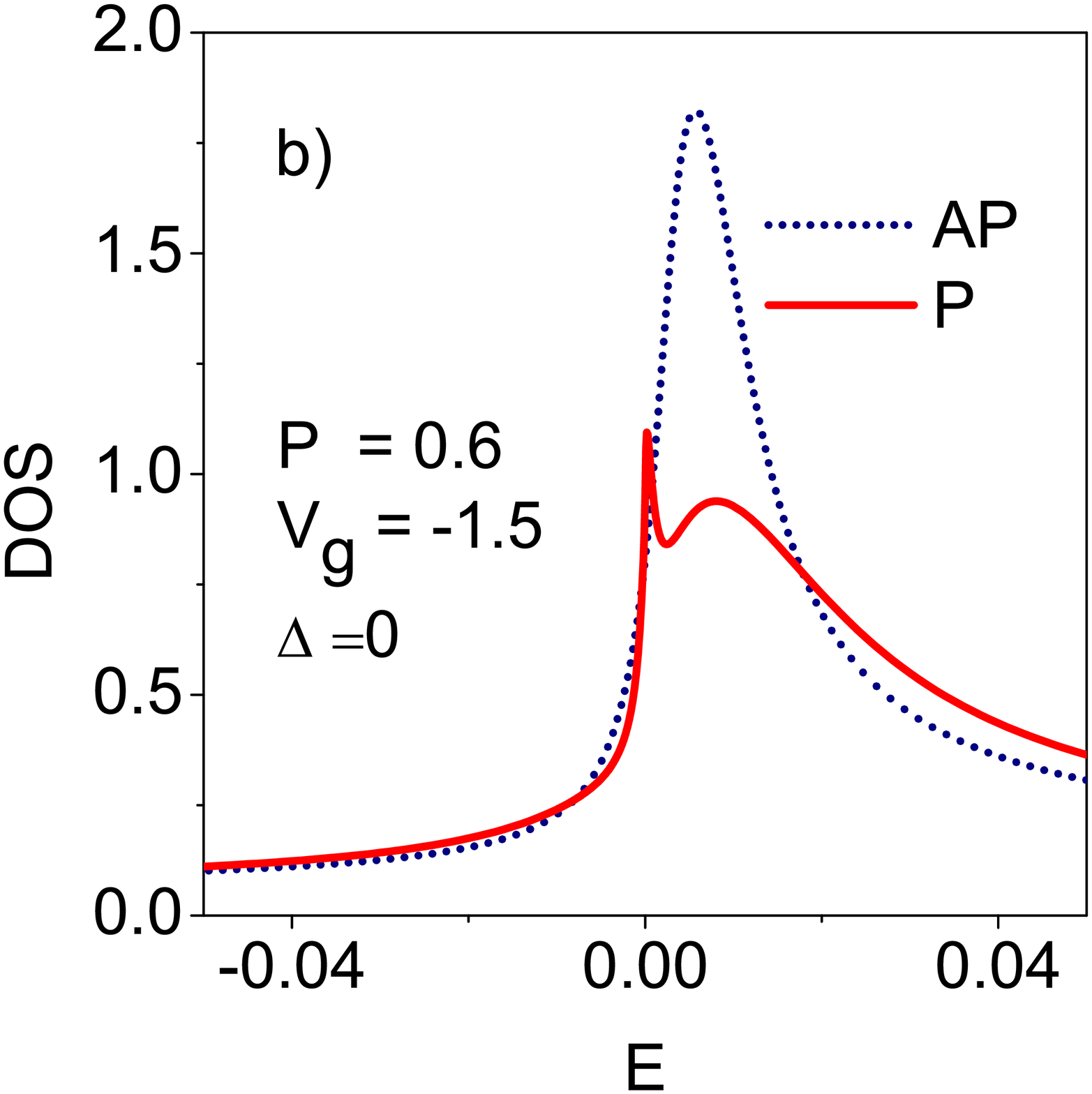}
\includegraphics[width=6 cm,bb=0 0 741 725,clip]{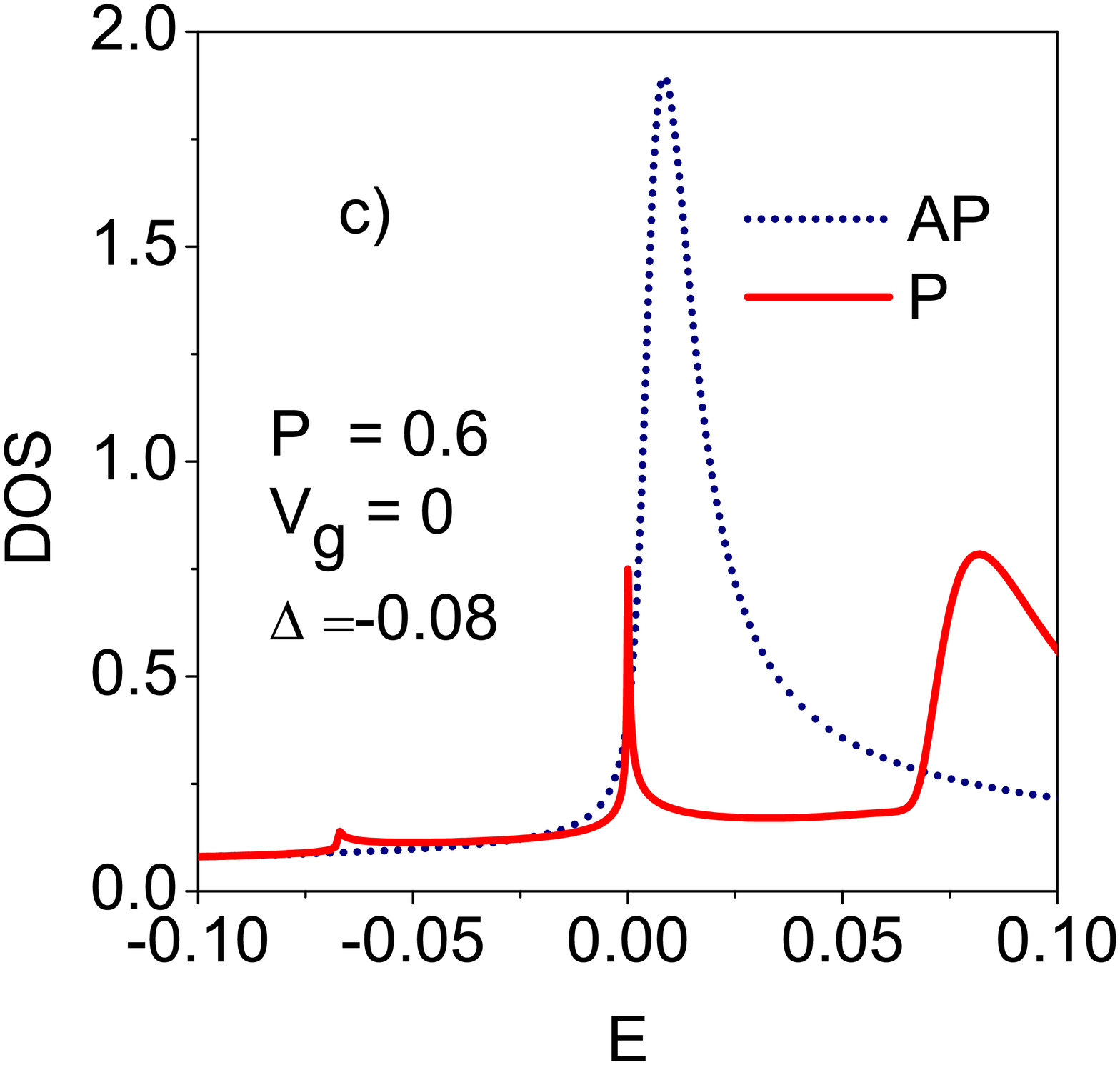}
\caption{\label{fig:epsart} (Color online) (a) Exchange splitting as a function of gate voltage ($\epsilon_{0}=-6$, ${\cal{U}} = 15$). (b,c) Examples of DOS for parallel  orientation of polarization of the leads (${\cal{P}}$) for the
             case of vanishing exchange splitting (b) and finite splitting (c).  In addition  DOS
             for ${\cal{AP}}$ configuration for the same values of gate voltage are  presented ( EOM).
}
\end{figure}

 As an  illustration  we show on Fig. 9a    the plots of exchange splitting as a function of gate voltage  for two values of polarization and  in  Fig. 9b the examples of DOS for vanishing exchange field and for finite exchange splitting. The influence of polarization on the  many-body DOS is twofold, first it introduces the difference in heights and widths of spin resolved peaks  due to the spin dependence of tunneling rates and second, it redistributes in energy  the spin partial contributions to DOS due to nonvanishing exchange field. As it is seen for $\Delta=0$   the spin-orbital fluctuations are not fully resolved in energy and only  a weak dip is marked in DOS,  for large enough exchange splitting ($|\Delta|\gg0$)    the three peak structure is observed, the central peak originates from orbital spin conserving processes and the satellites  reflect the spin and spin-orbital fluctuations. The relative position of majority  and minority spin satellites depend on the sign of $\Delta$. For   $\Delta>0$  lower satellite is characterized by minority spin orientation. The additional curves of DOS for $\cal{AP}$ configuration displayed also on Figs. 9b,c will be used later in the text for interpretation of dependencies presented on Fig. 11.  In the next picture (Fig. 10) we  show  polarization of conductance versus gate voltage and bias voltage for parallel  orientation of polarizations of electrodes. The crossing point of high polarization lines corresponds to $\Delta =0$.    The  vertical high polarization line of linear conductance, or alternatively line of high polarization of linear current,  is  a consequence of difference in  transmission for both spin orientations at the Fermi level. Interestingly not dependent on the sign of $\Delta$ the majority linear transmission dominates, despite the reverse of the role of the spin resolved densities of states for both signs of exchange field. Finite bias high conduction polarization  lines,  characterized by an opposite sign occur due to entering of the satellites into transport window and  the minority spin satellite transmission dominates for arbitrary  $\Delta$. The reverse of polarization of  conductance can be controlled by gate voltage or bias.  This point is more clearly visible by an inspection of the selected cross-sections of the map from Fig. 10, presented on Fig. 11d.

\begin{figure}
\includegraphics[width=6 cm,bb=0 0 741 725,clip]{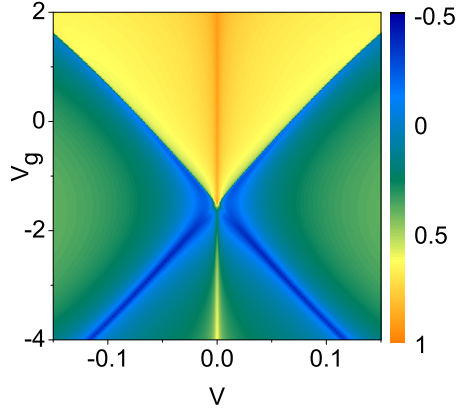}
\caption{\label{fig:epsart} (Color online) Polarization of conductance of CNT-QD  ($\epsilon_{0}=-6$, ${\cal{U}} = 15$) versus gate voltage $V_{g}$ and
            transport voltage $V$ for parallel orientation of polarization of the leads ($P=0.6$)(EOM).
}
\end{figure}

Let us focus now  on the problem   of possible  control of  transport of CNT-QD by the change of relative orientations of magnetic moments of the leads. Fig. 11a presents gate voltage dependence of linear ${\cal {TMR}}$ and Fig. 11b shows examples of bias dependencies of magnetoresistance for negative and positive exchange fields. Two features are most interesting, the giant values of linear ${\cal {TMR}}$ observed for   gate voltages corresponding to negative exchange fields, which reach the values of  several hundred percent and also huge values of ${\cal {TMR}}$ in nonlinear regime for voltages corresponding to the exchange splittings. To understand the presented dependencies we again  refer   to the picture of  the many-body contribution to the DOS of CNT-QD for both configurations of magnetizations of the leads (Fig. 9c).
The single peak for ${\cal {AP}}$ configuration reflects a compensation of left  and right electrode contributions to the exchange field for the symmetric case ($\gamma= 1$). Depending on the ratio of exchange splitting to the Kondo temperature ($|\Delta|/T_{K}$), either a single ($|\Delta|/T_{K}<1$) or triple peak structure ($|\Delta|/T_{K}>1$) is observed in DOS  for ${\cal {P}}$ configuration. The central  orbital fluctuation peak for  ${\cal {P}}$ configuration is sharper  and located closer to $E_{F}$ than the spin-orbital fluctuation peak for ${\cal {AP}}$ alignment and the linear transmission for $\Delta< 0$ case  is much higher for ${\cal {P}}$ configuration which leads to large positive ${\cal {TMR}}$.
The sharpness of the  central Kondo peak with   parallel ferromagnetic electrodes  causes in this case   a dramatic decrease of ${\cal {TMR}}$ and change of the sign for slightly increased voltage. For  $\Delta>0$  the low bias  evolution is less dramatic, for $\Delta>\Delta_{0}=0.04$ the ${\cal {AP}}$ transmission dominates over the ${\cal {P}}$ even  in the limit  $V\rightarrow0$ and therefore small, but negative linear ${\cal {TMR}}$ (inverse ${\cal {TMR}}$) is observed in this range. It is worth to underline  that the positive giant linear ${\cal {TMR}}$ for $\Delta<0$ is in contrast to  negative linear ${\cal {TMR}}$  observed for SU(2) symmetry ~\cite{27,50}. For SU(2) case however only spin is engaged and  the exchange field splits the Kondo peak for ${\cal {P}}$ configuration. For $V_{g}=-1.5$, where exchange splitting vanishes (Fig. 9a) linear ${\cal {TMR}}$ has a local maximum. In the neighborhood of this point exchange splitting is small and it does not play the important role in determining the dominance of Kondo transmission at $E_{F}$ for any configuration. The magnitude of linear TMR in this region is mainly determined by the difference of effective couplings of the dot to the leads for both configurations.
\begin{figure}[h]
\includegraphics[width=6 cm,bb=0 0 741 725,clip]{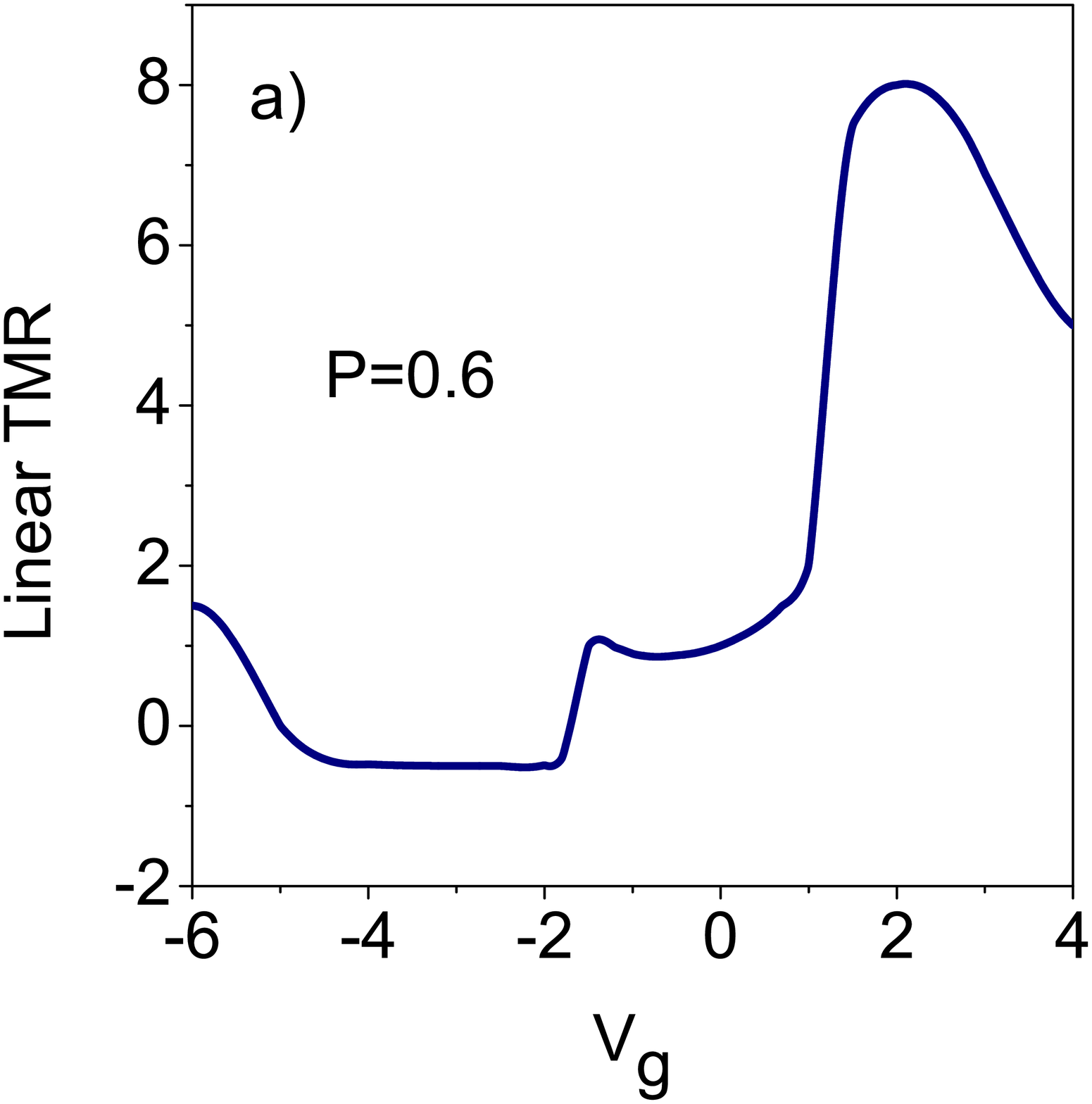}
\includegraphics[width=6 cm,bb=0 0 741 725,clip]{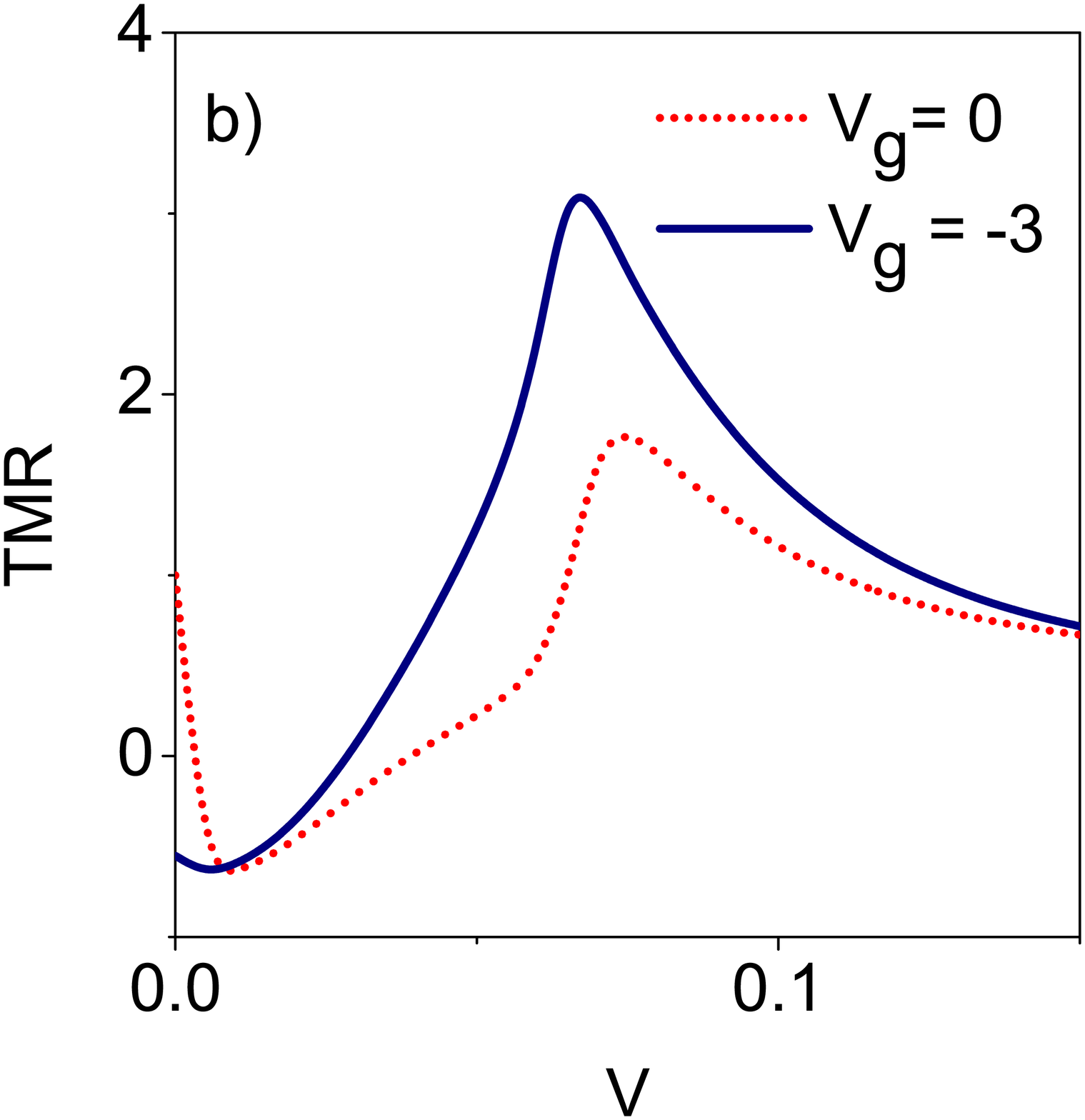}
\includegraphics[width=6 cm,bb=0 0 741 725,clip]{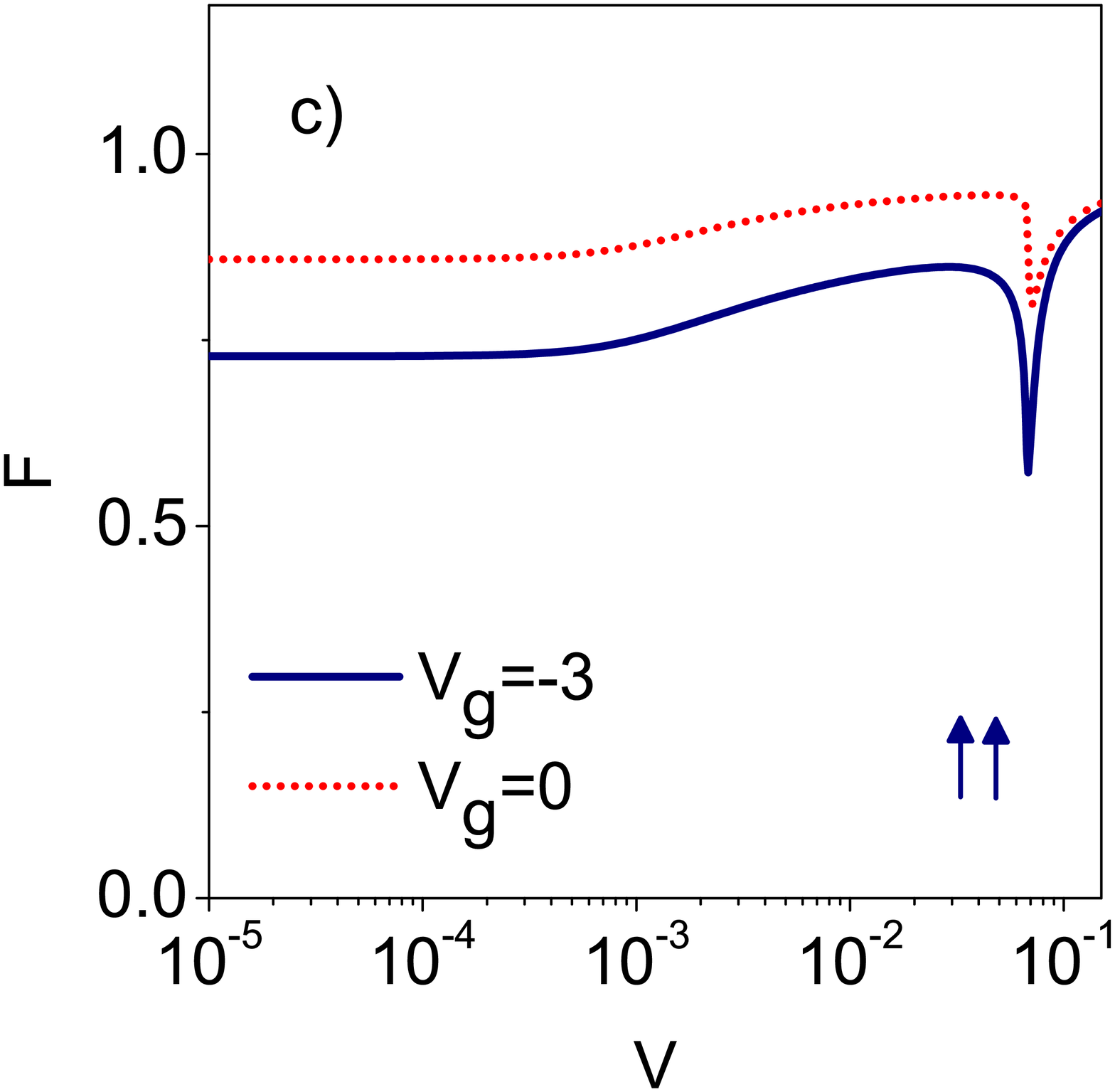}
\includegraphics[width=6 cm,bb=0 0 741 725,clip]{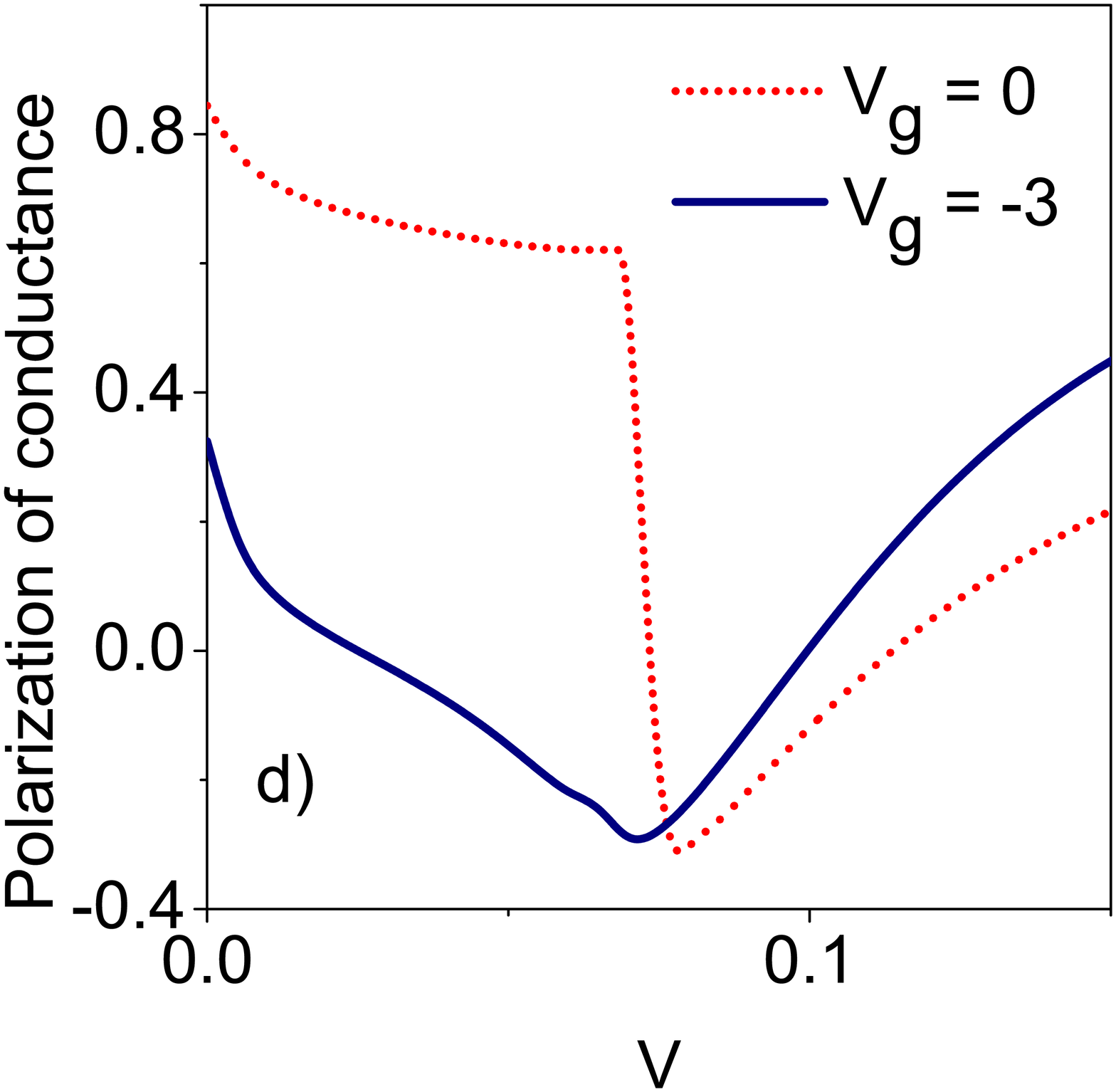}
\includegraphics[width=6 cm,bb=0 0 741 725,clip]{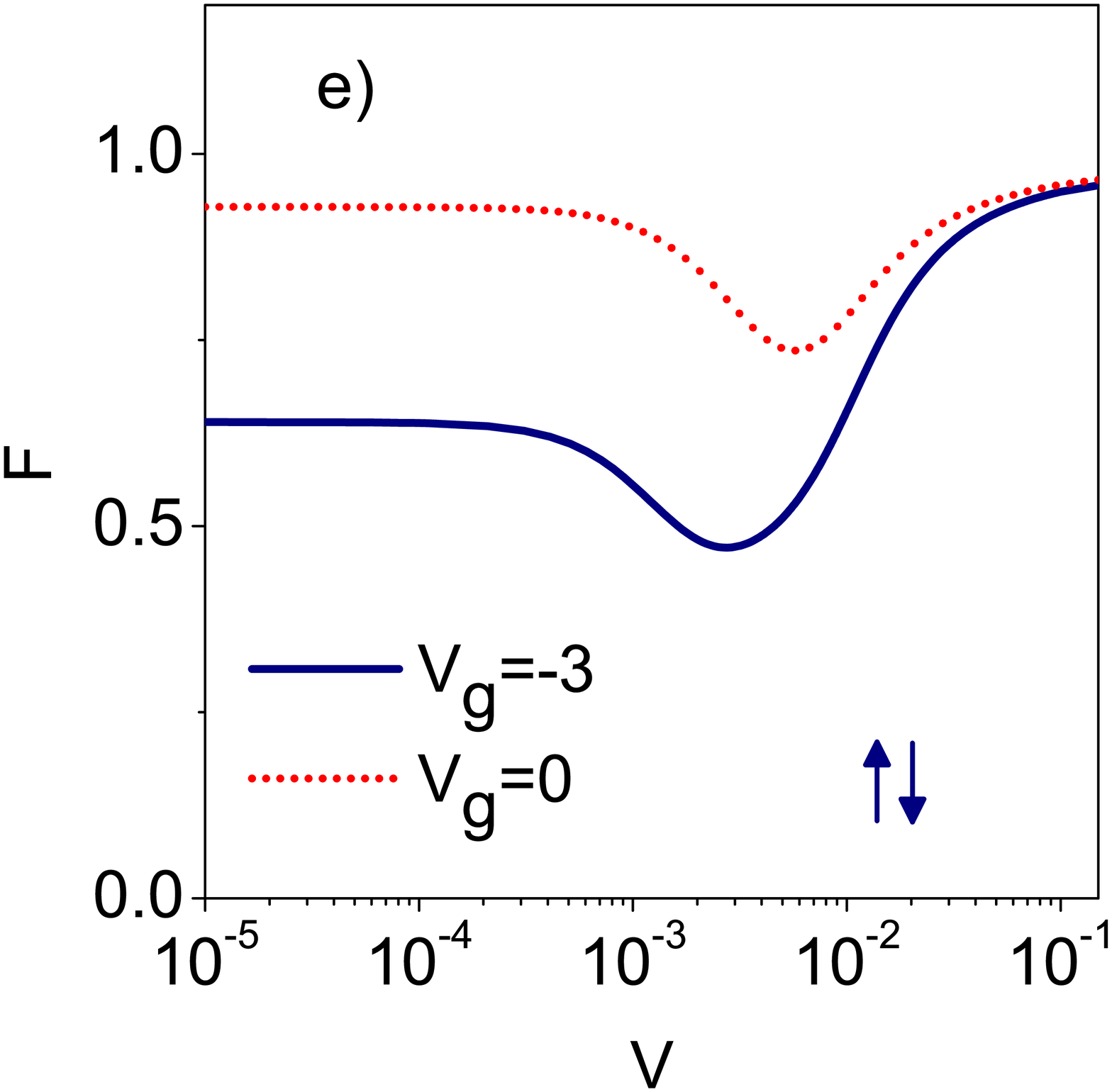}
\caption{\label{fig:epsart} (Color online) a) Gate dependence of linear ${\cal{TMR}}$ ($\epsilon_{0}=-6$, ${\cal{U}}=15$), $P=0.6$. b) Bias dependencies of ${\cal{TMR}}$ for $V_{g}=0$ ($\Delta=-0.08$), $V_{g}=-3$ ($\Delta=0.08$). c,e) Fano factors of CNT-QD in ${\cal{P}}$ and ${\cal{AP}}$ configurations respectively. d) Polarization of conductance for ${\cal{P}}$ configuration.}
\end{figure}
\clearpage
For higher voltages, as it is seen on Fig. 11b ${\cal {TMR}}$ increases and a smooth change of sign is observed. Nothing dramatic happens with differential conductance for ${\cal {AP}}$ configuration. For parallel  orientation conductance sharply increases close to bias voltage equal to exchange splitting $V\sim\Delta$ and  this reflects  in the occurrence of peaks in ${\cal {TMR}}$.  For still higher voltages ${\cal {TMR}}$ saturates,  the observed limit is slightly enhanced in comparison to the  Julliere prediction ~\cite{122} (for  $P=0.6$ the Julliere limit is $0.53$).  This is a consequence of the weak influence of high energy charge fluctuations.  It is instructive  to compare the discussed bias dependence of ${\cal {TMR}}$ with the bias dependence of polarization of conductance for parallel  magnetization configuration ${\cal {PC}}$  (Figs. 11d).   Since the bias evolution of both quantities  has to a larges extend the same source it is not surprising that maxima of ${\cal {TMR}}$ coincide with minima of polarization ${\cal {PC}}$.
With the change of the sign of $\Delta$ the spin up and spin down satellite change their role on the energy scale. This is reflected in the observed difference in the character of bias dependence of polarization reverse, from the mild transition for $\Delta>0$ to the sharp jump for $\Delta<0$.
Such a behavior is dictated by an asymmetric shape  of the satellites (Fig. 9c)
. Figs. 11c,e depict Fano factors for parallel and for antiparallel configuration. The sharp finite bias minima for $\cal{P}$ orientation located in the positions of $\cal{TMR}$ peaks occur because the exchange satellites put up in the transport window for these voltages. The gate dependence of linear Fano factor for ${\cal{P}}$ alignment of ferromagnetic electrodes ${\cal{F}}^{\uparrow\uparrow} (V=10^{-5})$ is nonmonotonic and takes the minimal value for $V_{g}=-1.5$ i.e. for the case of vanishing exchange splitting. The bias dependence of ${\cal{F}}^{\uparrow\downarrow}$ (${\cal{AP}}$ alignment) qualitatively resembles a similar behavior for unpolarized case, minima of ${\cal{F}}^{\uparrow\downarrow}$ correspond to polarization renormalized Kondo temperatures.\\

\noindent \textbf{4) Spin-flip noise}\\

Currents flowing through the  dot placed in a  magnetic field or coupled to ferromagnetic electrodes are spin polarized  ${\cal{I}}^{+}\neq{\cal{I}}^{-}$. Alternatively we can say that apart from the  charge current  ${\cal{I}}^{c}\equiv{\cal{I}}^{+}+{\cal{I}}^{-}$  also spin current  ${\cal{I}}^{s}\equiv{\cal{I}}^{+}-{\cal{I}}^{-}$ can be nonzero. The total (charge) current noise ${\cal{S}}^{c}\equiv{\cal{S}}_{++} + {\cal{S}}_{--} +  {\cal{S}}_{+-} +{\cal{S}}_{-+}$  and the spin  current noise  ${\cal{S}}^{s}\equiv\cal{S}_{++} + {\cal{S}}_{--} -  {\cal{S}}_{+-} -{\cal{S}}_{-+}$ are equal if the off-diagonal terms vanish.
\begin{figure}[h]
\includegraphics[width=6 cm,bb=0 0 741 725,clip]{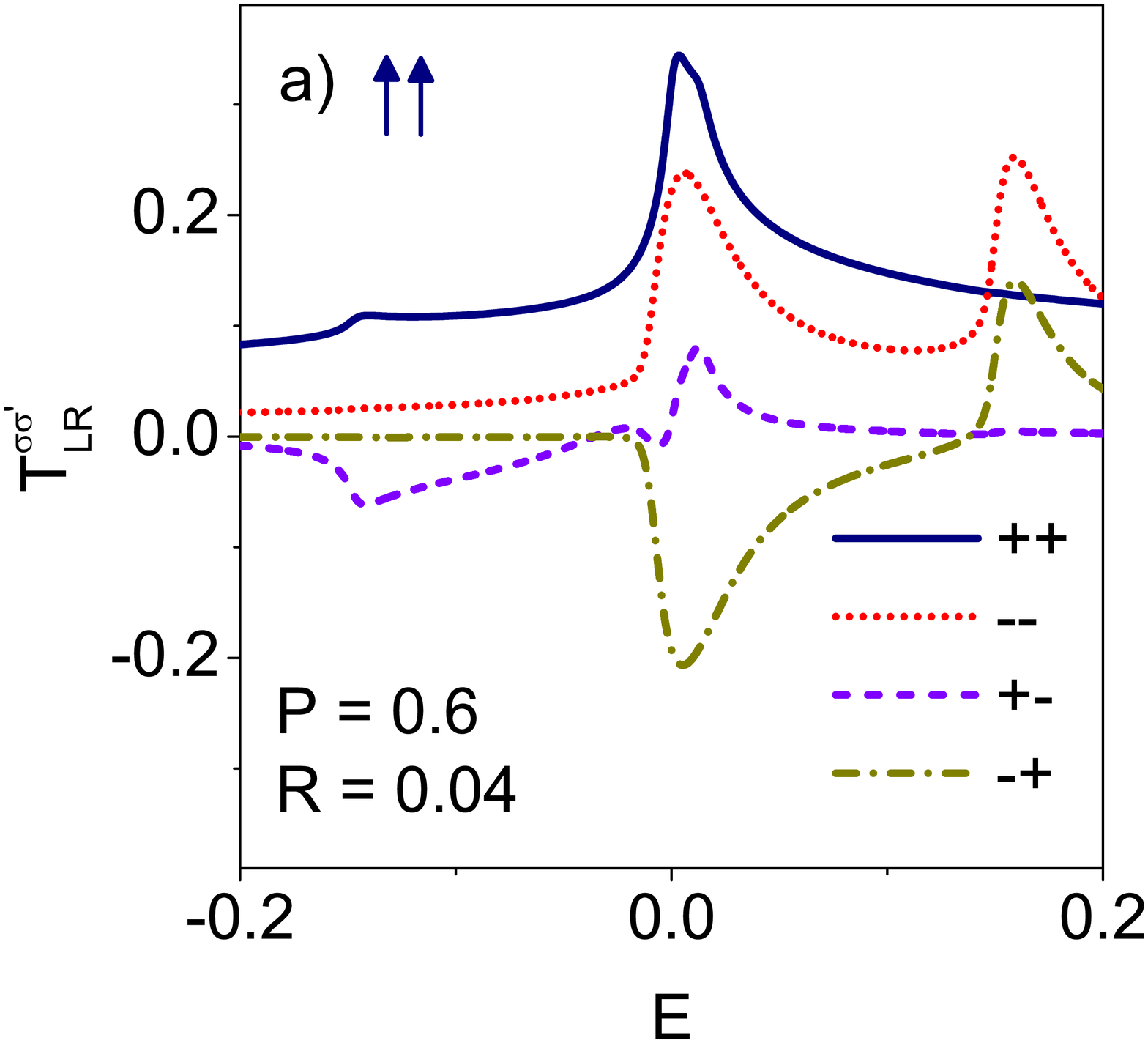}
\includegraphics[width=6 cm,bb=0 0 741 725,clip]{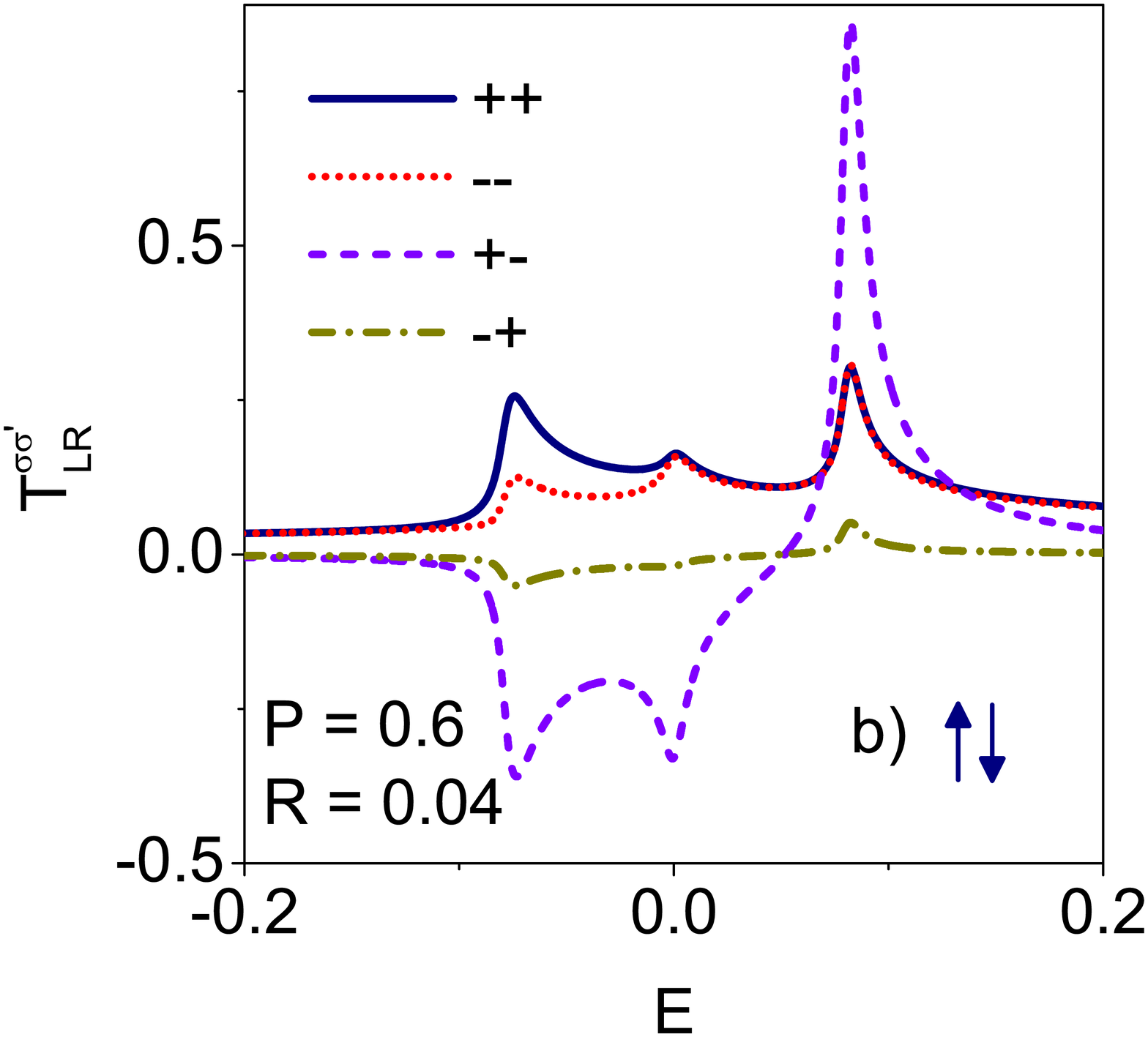}
\caption{\label{fig:epsart} (Color online) Generalized transmission coefficients ${\cal{T}}^{\sigma\sigma'}_{LR}$  (18)  of CNT-QD  ($\epsilon_{0}=-6$, ${\cal{U}} = 15$, $V_{g}=0$) coupled to polarized electrodes ($P = 0.6$) in the presence of spin-flip scattering  for parallel ($\Delta=-0.08$)(a) and antiparallel configuration  (b).
}
\end{figure}
In the presence of Coulomb interactions  it is not in general  the case   and correlations between spin-up and spin-down channels occur. To take them into account one has to go beyond the formalism we use e.g. by introducing higher order truncation beyond Lacroix's,  but this point is postponed for the future publication. Here we discuss the  case where, spin-opposite current noise ${\cal{S}}_{\sigma-\sigma}$  results from the real spin-flip scattering. The spin-flip term (3) is assumed to be coherent, in the sense that spin-flip strength ${\cal{R}}$ involves reversible transitions between up and down-spin states on the dot. These transitions may be caused e.g. by transverse component of a local magnetic field ~\cite{123}. Fig. 12 presents matrix elements of zero bias generalized transmissions ${\cal{T}}^{\sigma \sigma'}_{LR}(\omega)\equiv{\cal{T}}^{m\sigma m\sigma'}_{LR}(\omega,V=0)$ (18) for ${\cal{P}}$ and ${\cal{AP}}$ configurations plotted for intermediate spin-flip scattering  strength ${\cal{R}}/|\Delta| =0.6$.  ${\cal{T}}^{\sigma\sigma}_{LR}$ is transmission corresponding to Kondo processes associated with spin flips of even order and ${\cal{T}}^{\sigma-\sigma}_{LR}$  describes transmission in the Kondo range accompanied by spin-flips of odd order.     In the limit of weak (${\cal{R}}/|\Delta|\ll1$) or strong (${\cal{R}}/|\Delta|\gg1$) spin-flip scattering the expected positions of maxima of ${\cal{T}}^{\sigma\sigma'}_{LR}(\omega)$ for ${\cal{P}}$ orientations are located around $\widetilde{T_{K}}$, $\widetilde{T_{K}}\pm(\Delta\pm2{\cal{R}})$ leading to the five peak structure in the case of  well separation of the peaks ($\widetilde{T_{K}}$ denotes Kondo temperature in the presence of spin-flip scattering and exchange field $\widetilde{T_{K}}=T_{K}({\cal{R}},\Delta)$). In the intermediate scattering case some of the peaks overlap as is illustrated for example on Fig.12 a ($\widetilde{T_{K}}=0.01$, $\Delta=-0.08$, ${\cal{R}}=0.04$), where   the satellites placed  close to $\widetilde{T_{K}}-(\Delta+2{\cal{R}})$ and $\widetilde{T_{K}}+(\Delta+2{\cal{R}})$ are not well separated from the central peak and the low energy satellite $\widetilde{T_{K}}+\Delta-2{\cal{R}}$  is only  poorly visible. For ${\cal{AP}}$ configuration the splitting is caused by  spin-flips alone and  ${\cal{T}}^{\sigma\sigma}_{LR}$ has a  three peak structure  ($\omega\simeq \widetilde{T_{K}}$  and $\omega\simeq \widetilde{T_{K}} \pm{2\cal{R}}$) (Fig.12 b).
\begin{figure}[h]
\includegraphics[width=6 cm,bb=0 0 741 725,clip]{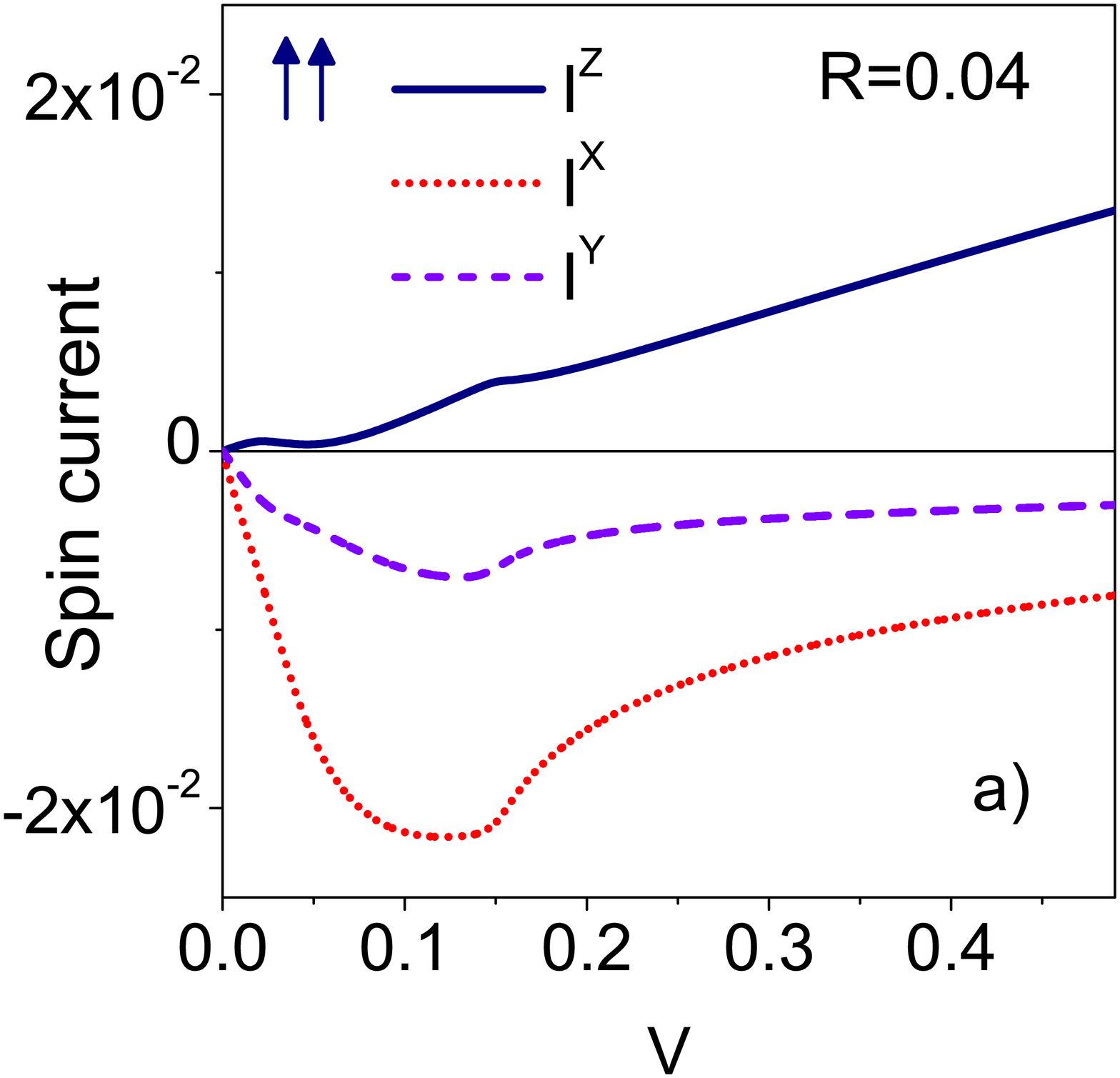}
\includegraphics[width=6 cm,bb=0 0 741 725,clip]{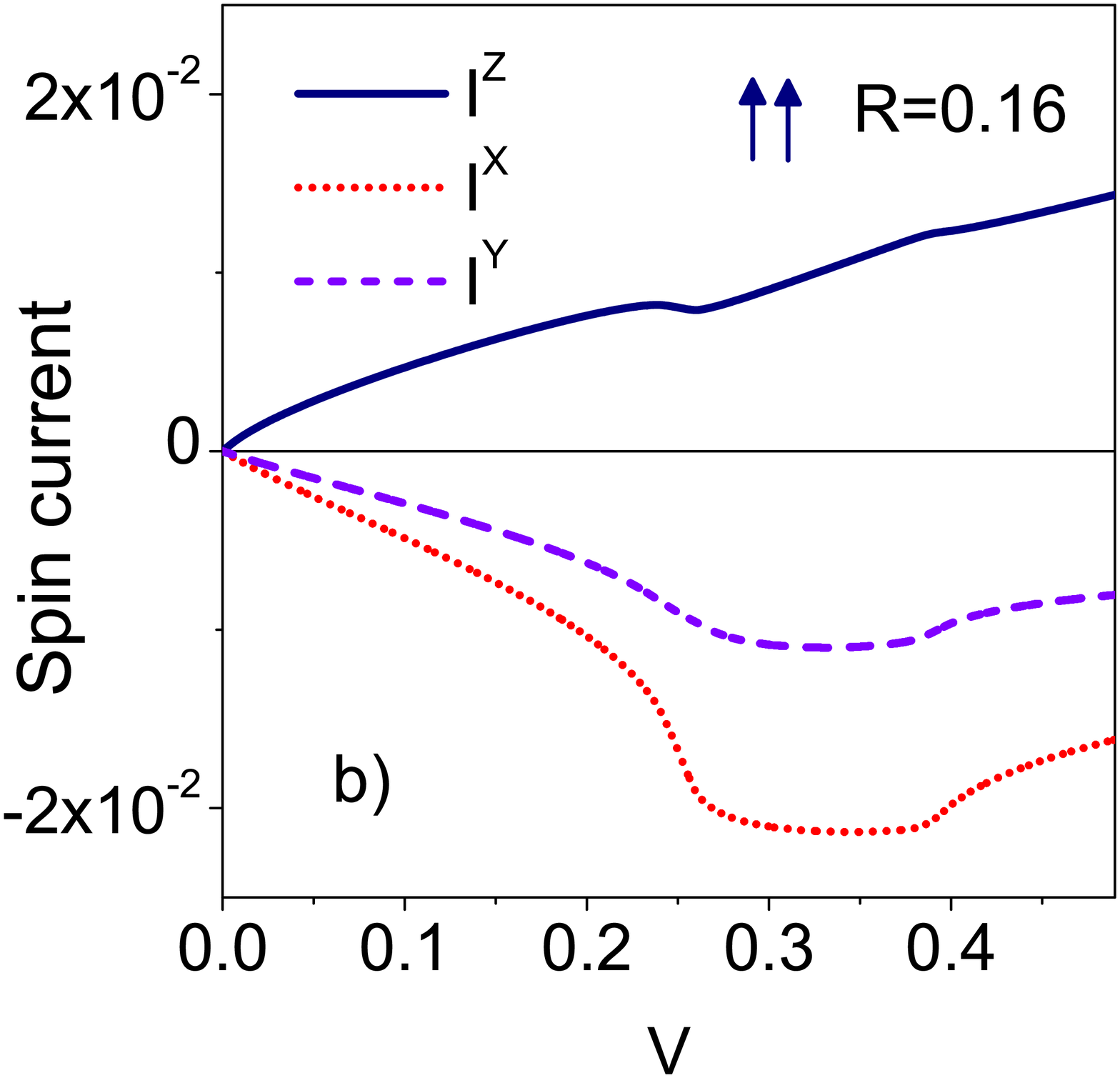}
\includegraphics[width=6 cm,bb=0 0 741 725,clip]{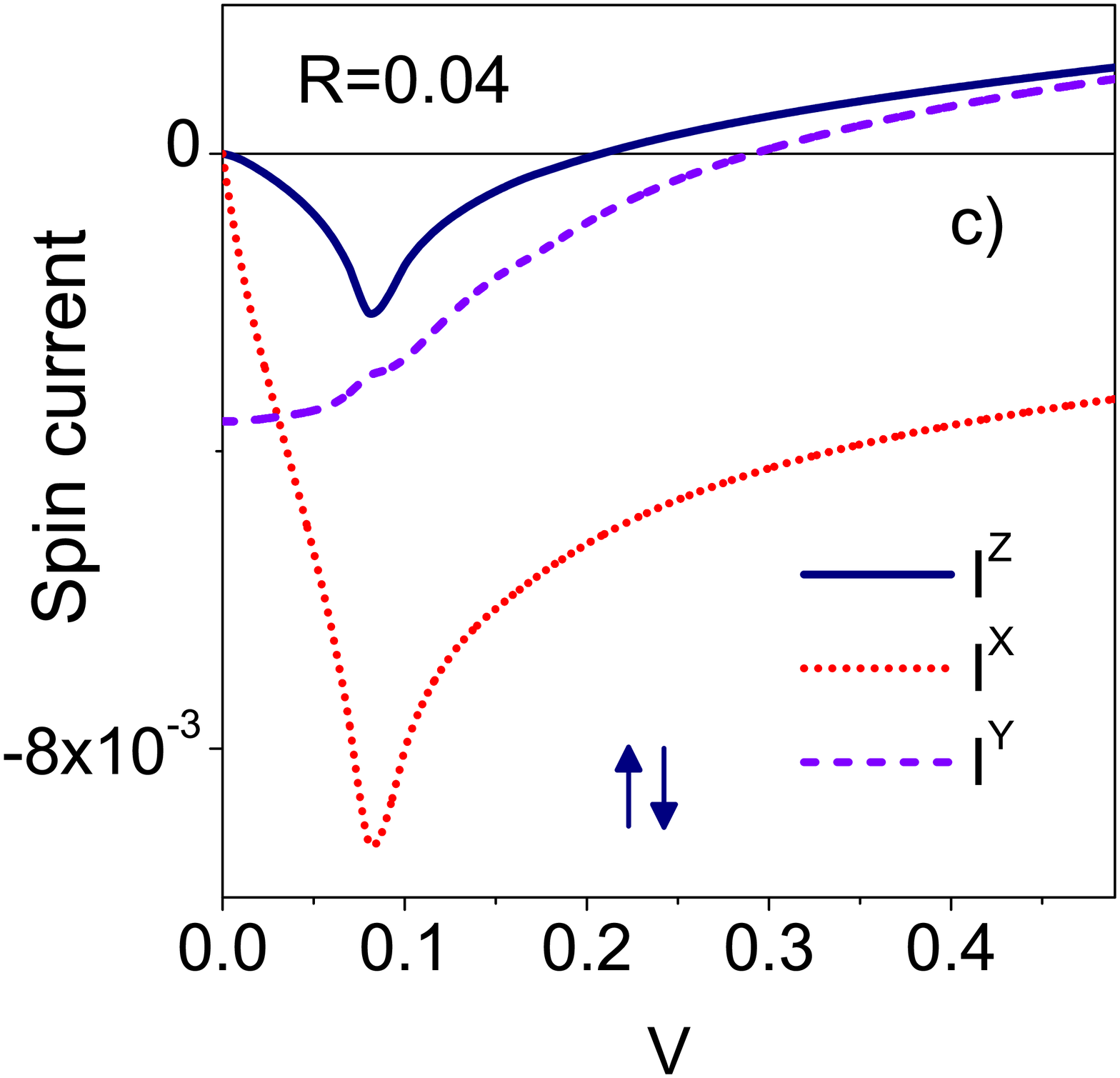}
\includegraphics[width=6 cm,bb=0 0 741 725,clip]{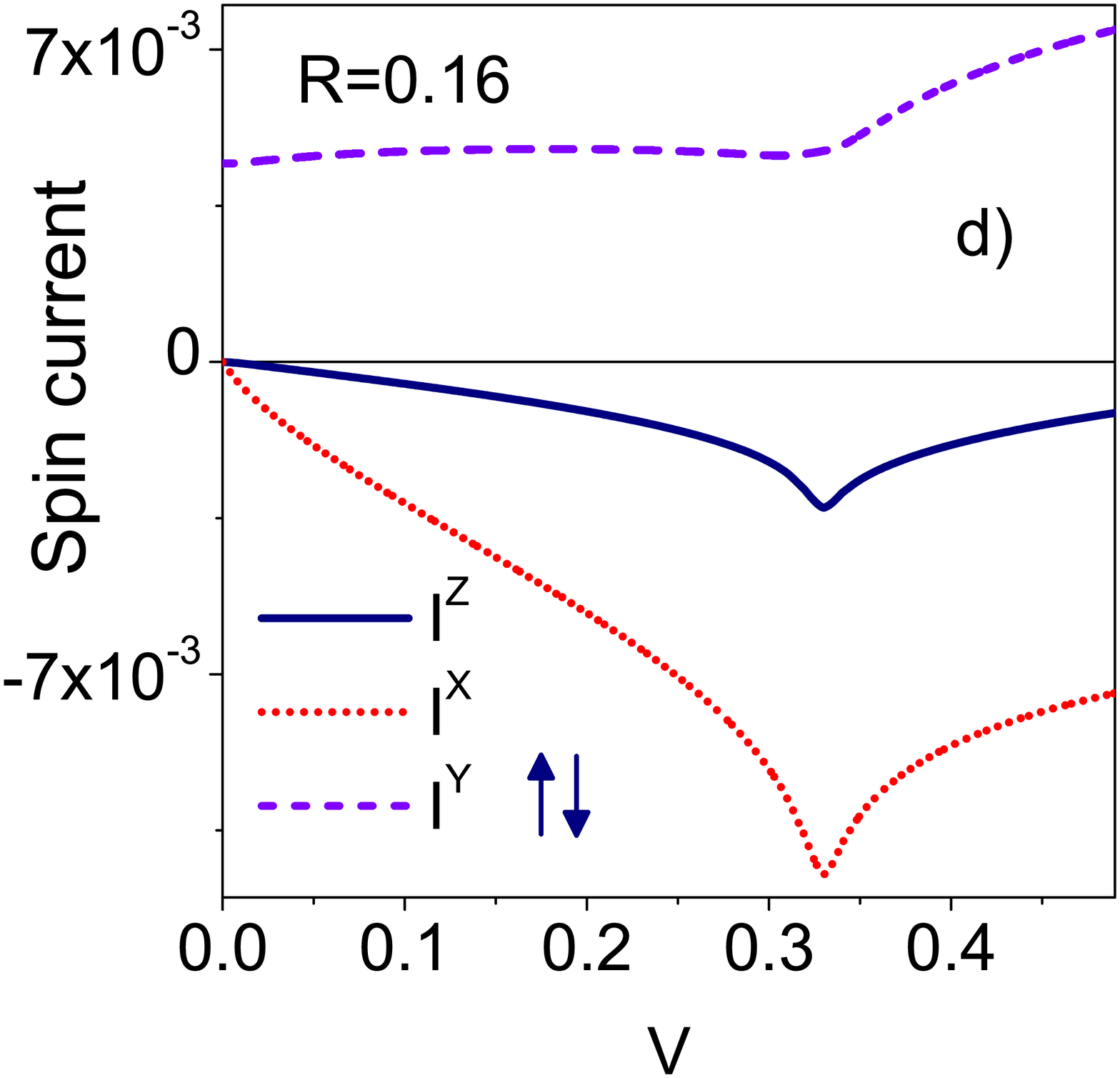}
\includegraphics[width=6 cm,bb=0 0 741 725,clip]{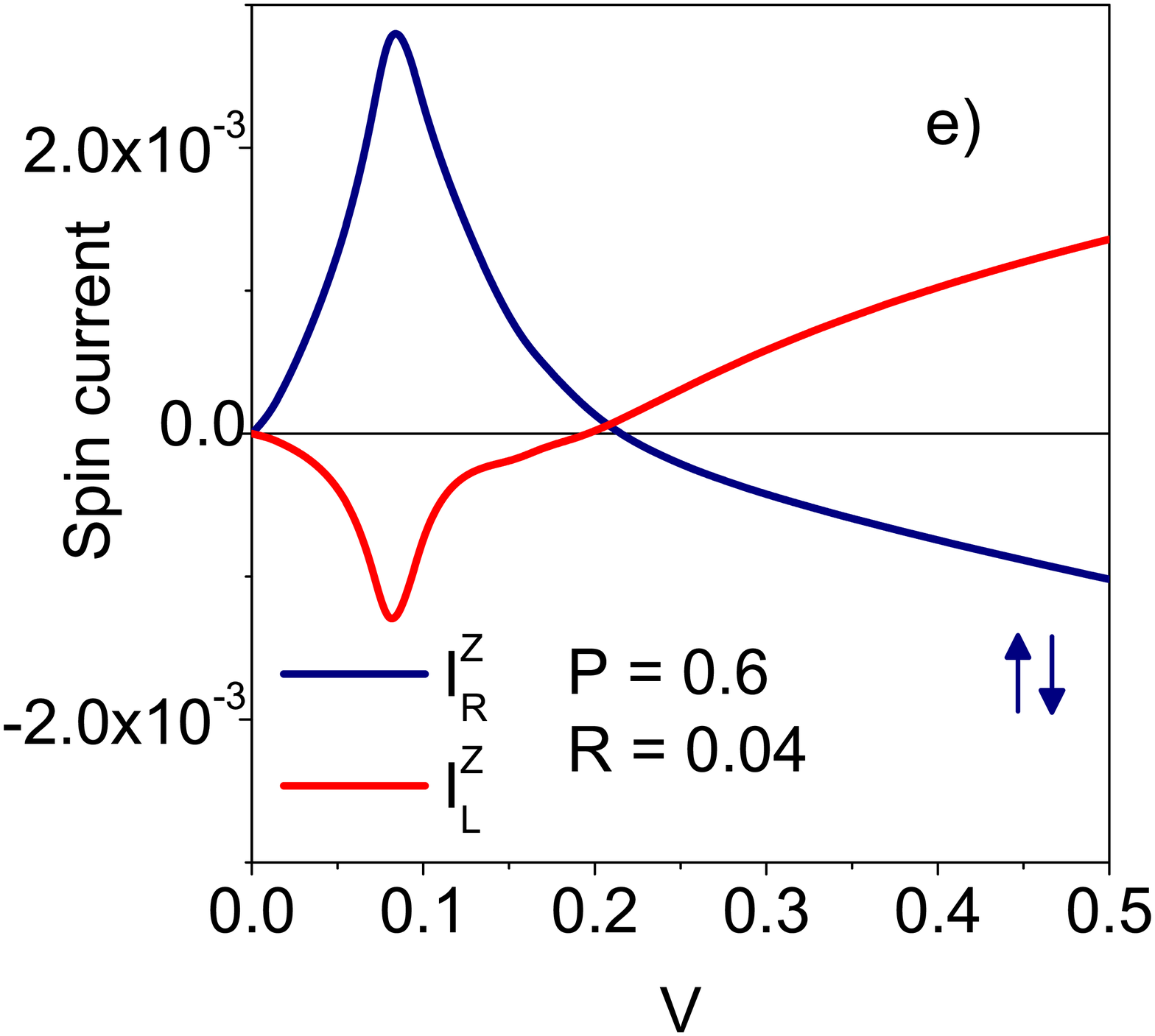}
\includegraphics[width=6 cm,bb=0 0 741 725,clip]{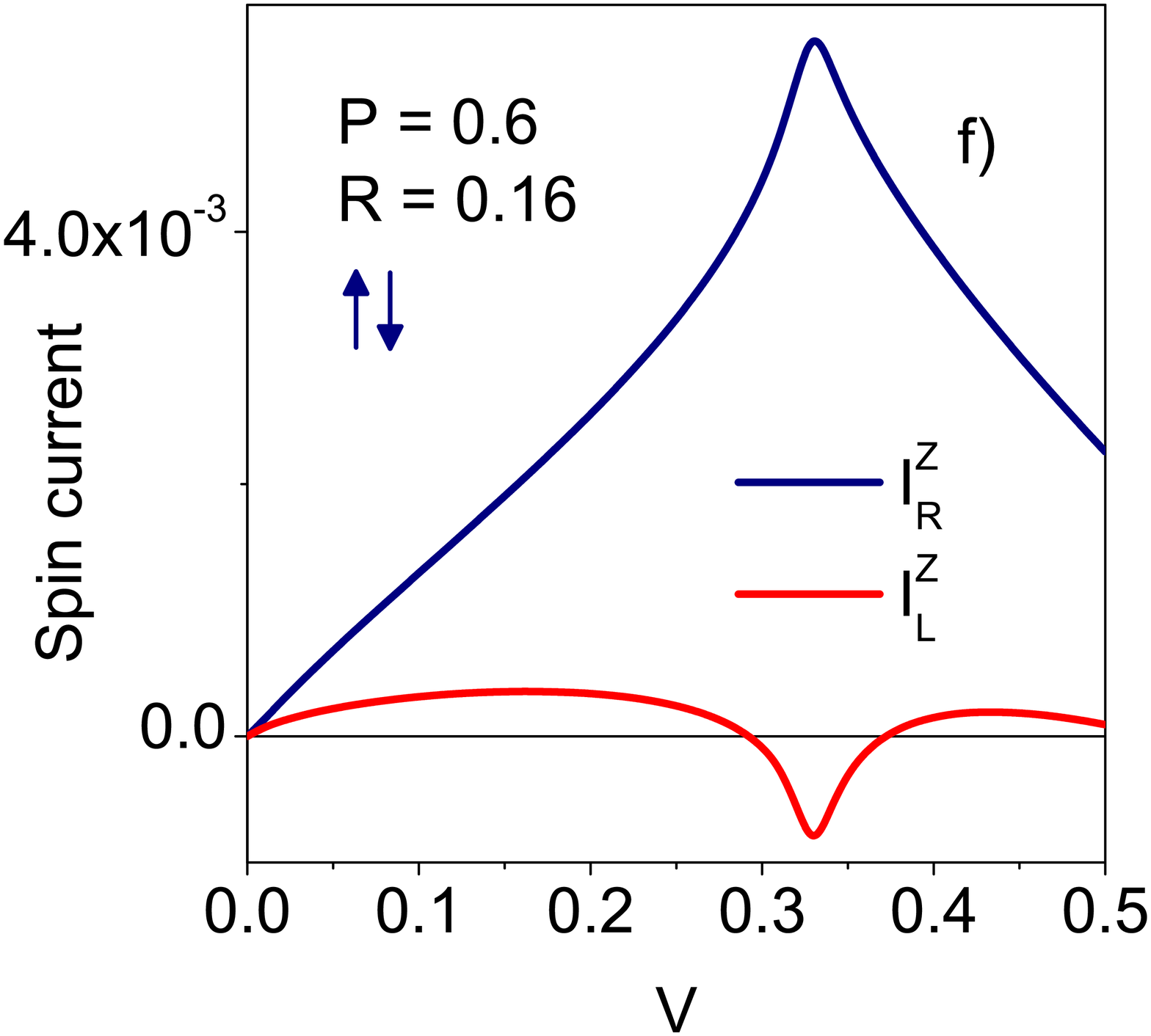}
\caption{\label{fig:epsart} (Color online) Spin currents of CNT-QD in the Kondo range ($\epsilon=-6$, ${\cal{U}} = 15$) in the presence of spin- flip scattering for parallel (a,b) and antiparallel (c-f) configuration of polarizations of the leads ($P=0.6$).}
\end{figure}
Fig. 13a-d present spin currents ${\cal{I}}^{x,y,z}=\frac{{\cal{I}}^{x,y,z}_{L}-{\cal{I}}^{x,y,z}_{R}}{2}$, where  ${\cal{I}}^{z}_{\alpha}={\cal{I}}^{z}_{\alpha+}-{\cal{I}}^{z}_{\alpha-}$ and spin-flip currents  ${\cal{I}}^{x}_{\alpha}=Re[\cal{I}^{+}_{\alpha}]$ and ${\cal{I}}^{y}_{\alpha}=Im[{\cal{I}}^{+}_{\alpha}]$ and ${\cal{I}}^{+}_{\alpha}$ can be expressed similarly as (11) by  ${\cal{I}}^{+}_{\alpha}(t)=2\sum_{km}t_{\alpha}[G^{<}_{m-,k\alpha m+}(t)-G^{<}_{k\alpha m-,m+}(t)]$.
 Minima  or maxima of spin currents observed for ${\cal{AP}}$ orientation occur for $V\simeq2{\cal{R}}$,  small finite bias    exchange field  is of minor importance in this case. For ${\cal{P}}$ configuration the characteristic energies of exchange or spin-flip splitting are less  clearly marked, but looking for example at  the ${\cal{I}}^{z}$ curve one recognizes that a local minimum and the discontinuity point of differential conductance roughly correspond to the positions of the peaks in diagonal transmission ${\cal{T}}^{\sigma\sigma}_{LR}$. The interesting effect of spin-flips is  negative differential conductance of spin currents (SCNDR).  In the case of ${\cal{I}}^{z}$ component for example SCNDR signals that the minority spin transmission entering the transport window changes for the corresponding energies much more rapidly than the majority transmission. The same mechanism can also result in the change of sign of spin current with voltage. Change of the sign of ${\cal{I}}^{z}$  means change of polarization of current.   Reversal  of polarization of current can occur on both electrodes (e.g.  Fig. 13e)  or on only one of them   (Fig. 13f). For some values of bias, current becomes unpolarized at one of the electrodes, but remains polarized for another.  Polarization of current can change across the system when spin-flip is present.  Interesting point visible on Figures 13 c,d is the occurrence of equilibrium spin current (ESC)  ${\cal{I}}^{y}$ for ${\cal{AP}}$ configuration of the magnetizations of the leads.  Nonzero spin current induced by spin-flip  processes can flow through QD with polarized electrodes  without bias. This phenomena is known in literature for systems with inhomogeneous magnetization and spin-orbit coupling ~\cite{124,125}. Comparing Figures 13c,d it is seen that the direction of flow of ESC for a given polarization of the leads might change with the spin-flip scattering strength. The occurrence of a given component of ESC can be inferred from symmetry of equilibrium state alone. In specific, for the case discussed, correlators $\langle c^{+}_{k\alpha m\sigma}d_{m\sigma'}\rangle$ and $\langle d^{+}_{m\sigma}c_{k\alpha m\sigma'}\rangle$ are equal for $\sigma=\sigma'$ and different for $\sigma\neq\sigma'$. This implies ${\cal{I}}^{z}_{\alpha}=0$ and ${\cal{I}}^{+(-)}_{\alpha}\neq0$. For parallel orientation ${\cal{I}}^{+(-)}_{L}={\cal{I}}^{+(-)}_{R}$ and for ${\cal{AP}}$ configuration ${\cal{I}}^{+(-)}_{L}=-{\cal{I}}^{+(-)}_{R}$. These relations together with the general property ${\cal{I}}^{+}=({\cal{I}}^{-})^{*}$ leads to the conclusion on the nonvanishing of ${\cal{I}}^{y}$ component of ESC.
 To clarify this point further let us focus on the simplest  case of ${\cal{AP}}$  configuration with $P =1$. The spin-flip scattering at the dot (3) induces in this case y component of the spin of the same absolute value but of opposite signs for electrons moving from the left electrode to the dot   ($\sigma_{z}=1$) and  for right moving electrons  ($\sigma_{z}=-1$). In consequence the electron flow in opposite directions is associated with opposite y-component of the spin. This happens in equilibrium, where charge current vanishes. In more general case ($P\neq1$) both $\sigma_{z}$ spin orientations play the role in the  flow in both directions and in addition the scattering processes are energy dependent (Fig.12b), what reflects in bias dependence of spin current with possibility of  change of its sign.

\begin{figure}
\includegraphics[width=6 cm,bb=0 0 741 725,clip]{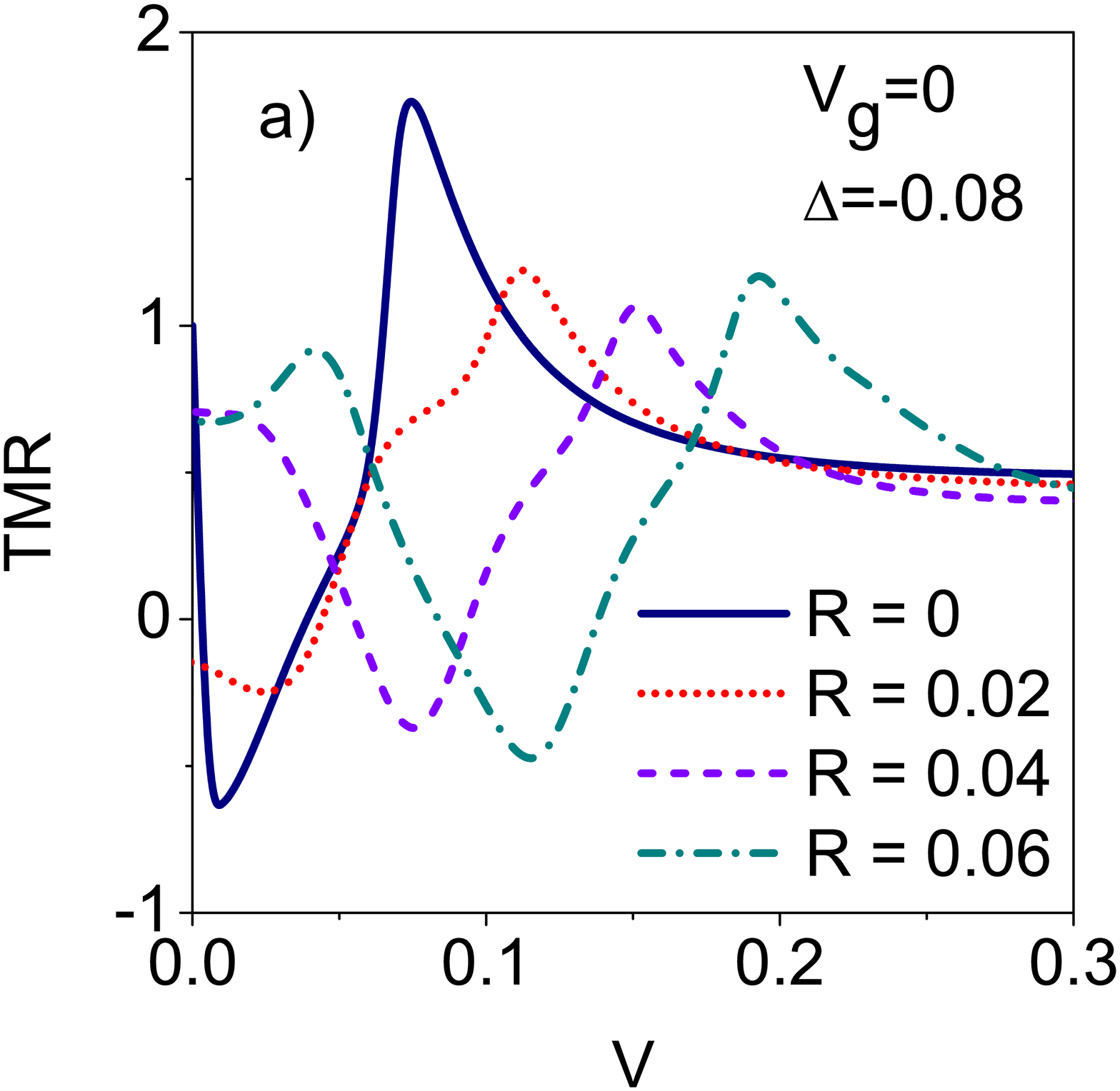}
\includegraphics[width=6 cm,bb=0 0 741 725,clip]{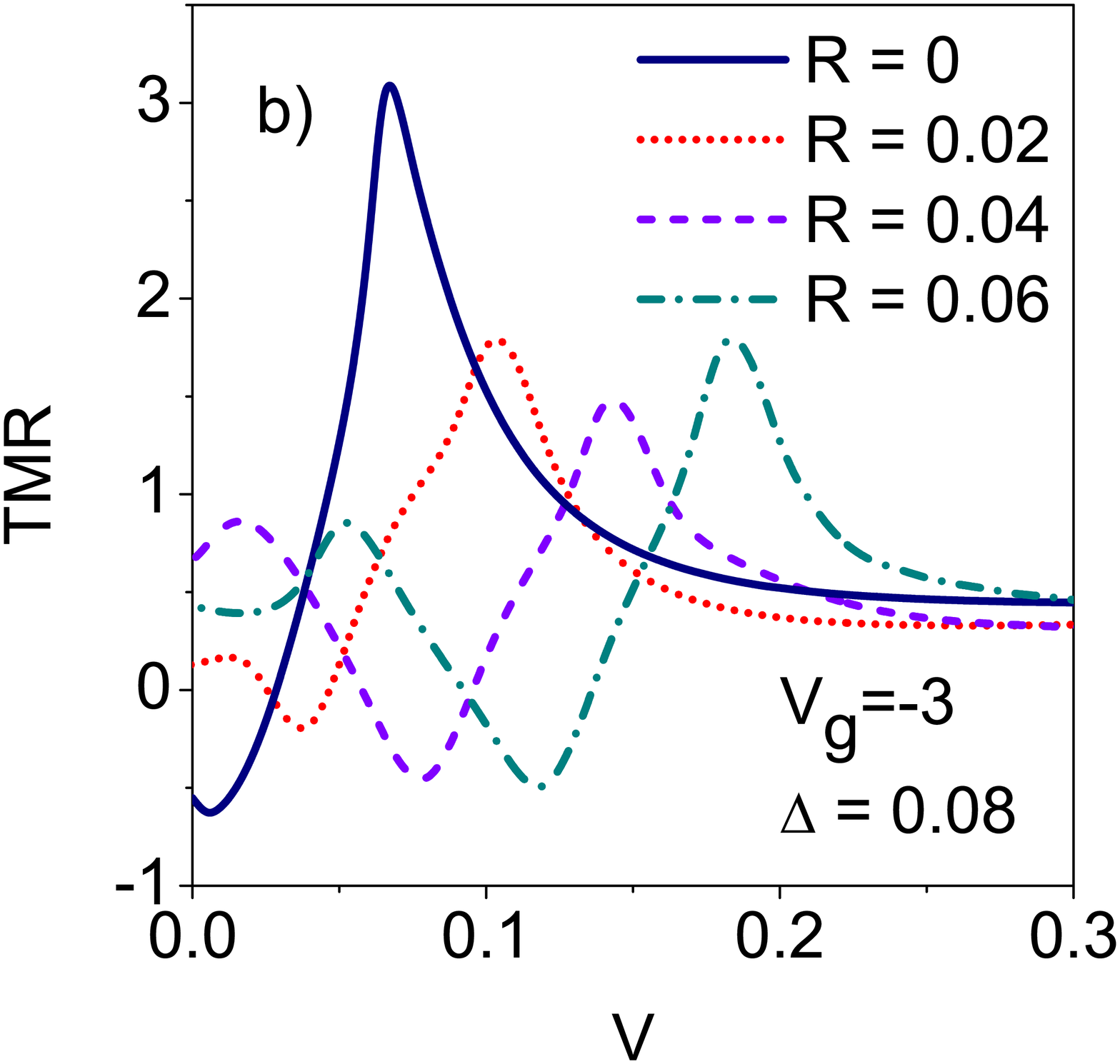}
\caption{\label{fig:epsart} (Color online) Bias dependencies of TMR of CNT-QD in the Kondo range ($\epsilon_{0}=-6$, ${\cal{U}} = 15$) for
               different values of spin-flip scattering amplitude and different gate voltages
                (a) $\Delta < 0$, (b) $\Delta > 0$.
}
\end{figure}

Fig 14 presents effect of spin-flips on ${\cal {TMR}}$. The decisive role in linear ${\cal {TMR}}$  plays a competition of the central  transmission peak  for ${\cal{P}}$ configuration and central peak for ${\cal{AP}}$ orientation (Fig.9c).  Spin flip scattering tends to change the sign of linear ${\cal {TMR}}$.  For positive exchange splitting ($\Delta>\Delta_{0}$)  the increase of ${\cal{R}}$ results in a decrease of the weight of central transmission  ${\cal{AP}}$ peak  at the Fermi level and the ${\cal{P}}$ conductance dominates in this case. ${\cal TMR}$ changes from negative to positive. For $\Delta<\Delta_{0}$   the opposite scenario is realized and  opposite   change of the sign of ${\cal TMR}$ is observed. The  oscillating character of ${\cal {TMR}}$ displayed on Fig. 14 results from   entering of the succeeding satellites  into the transport window. Whether the satellite marks on ${\cal {TMR}}$ curve as a distinct  maximum or minimum or only as an inflection point, or is not visible at all  depends on the height  of transmission peak and its separation from other peaks. For   example for ${\cal{R}}=0.04$ curve (Fig.14 a)  maximum at $V\simeq \widetilde{T_{K}}-\Delta+2{\cal{R}}$ reflects the dominant role played in this range by high energy satellite  for ${\cal{P}}$  configuration  and minimum at $V\sim2R$ in turn  exhibits   the leading role  in this range played by  spin-flip induced  ${\cal{AP}}$ satellite.  For ${\cal{R}} =0.06$ additional low voltage  peak of ${\cal {TMR}}$ is observed around $V\simeq \widetilde{T_{K}}+\Delta+2{\cal{R}}$ reflecting the influence of down spin satellite for ${\cal{P}}$ alignment, which in this case is  well separated from the central peak. The observed possibility of control of the sign of ${\cal {TMR}}$ both in linear and nonlinear regimes by the strength of spin-flip scattering  (e.g. by weak change of the transverse magnetic field) is interesting from application point of view.

\begin{figure}
\includegraphics[width=6 cm,bb=0 0 741 725,clip]{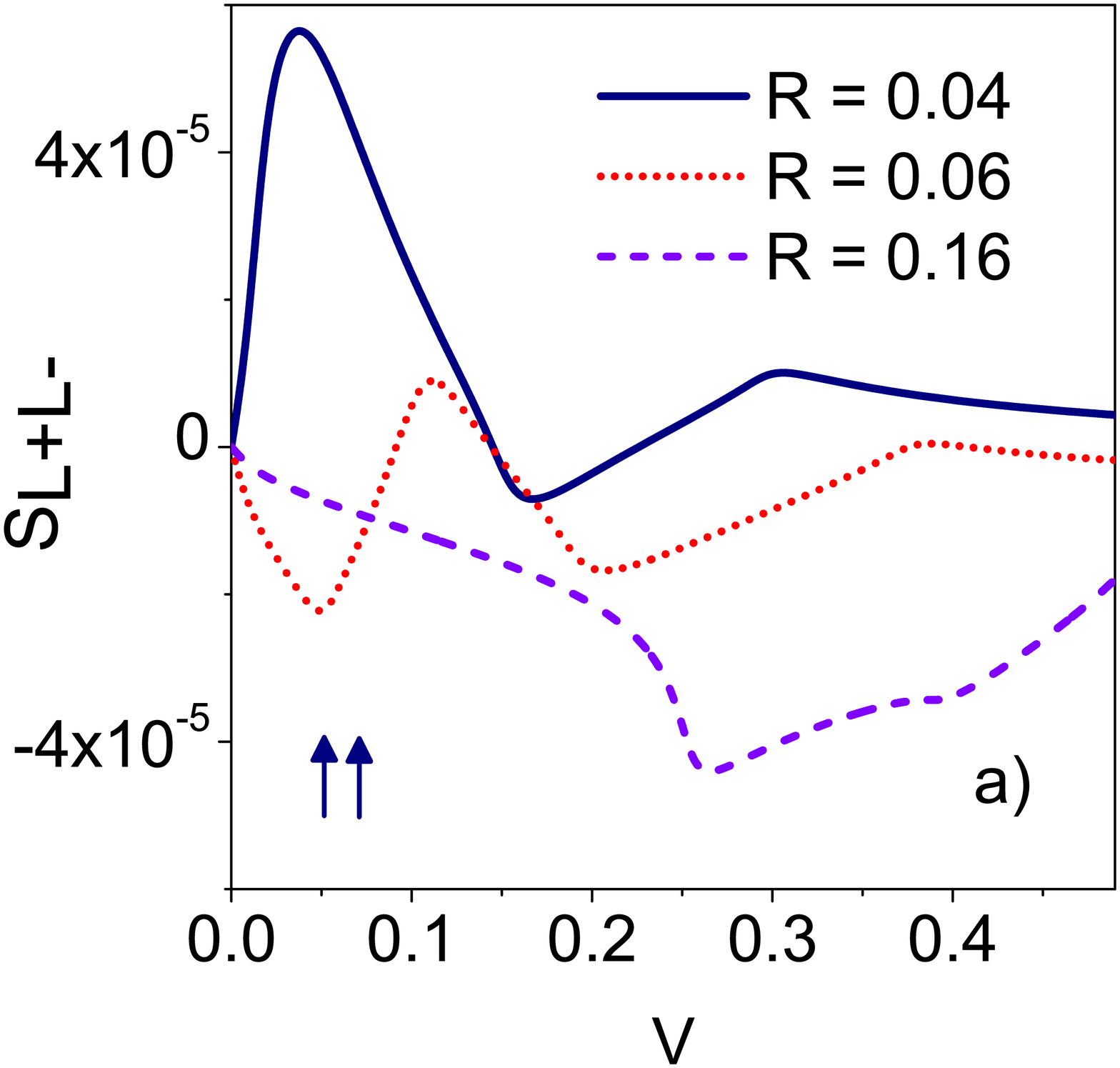}
\includegraphics[width=6 cm,bb=0 0 741 725,clip]{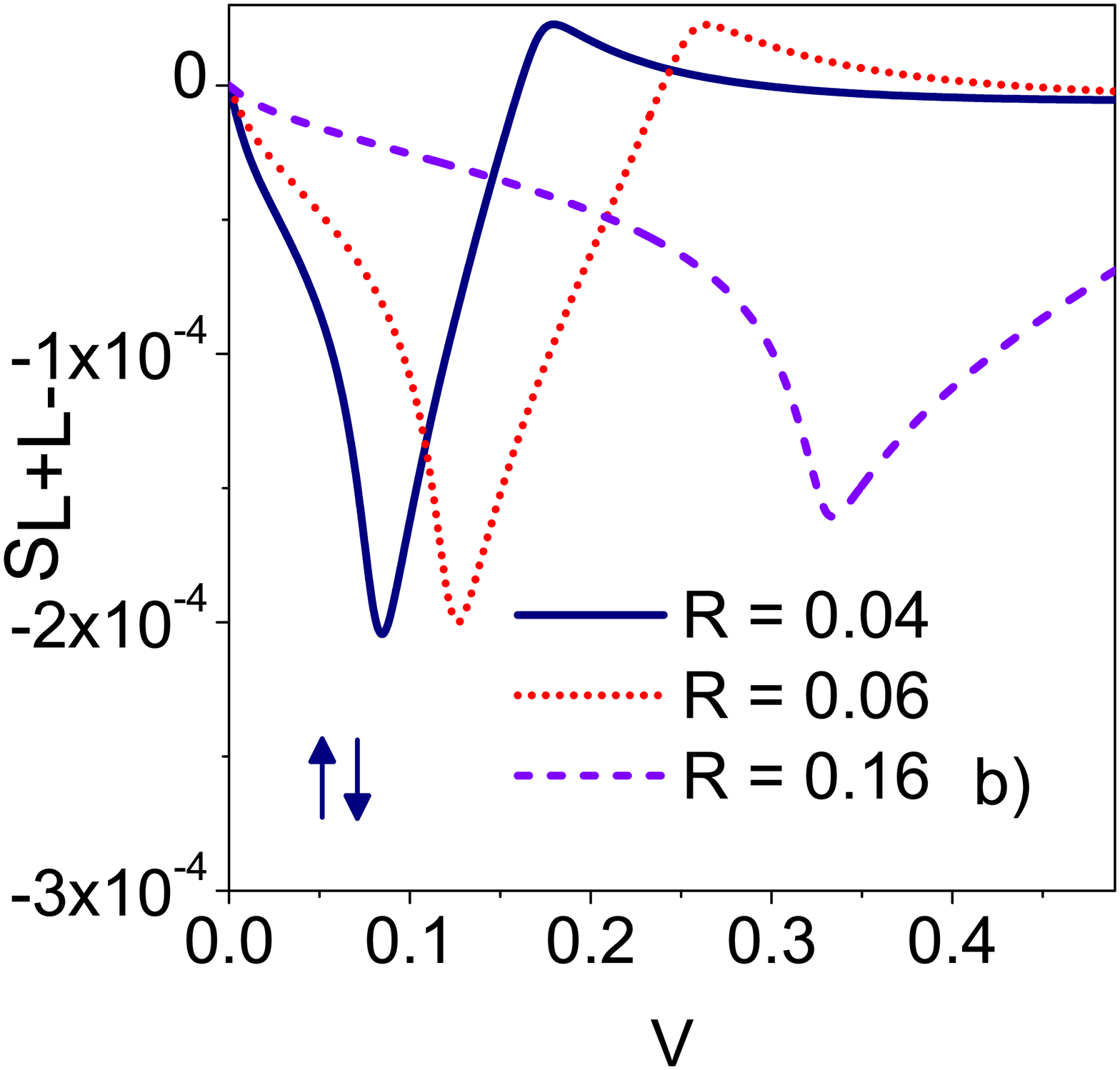}
\includegraphics[width=6 cm,bb=0 0 741 725,clip]{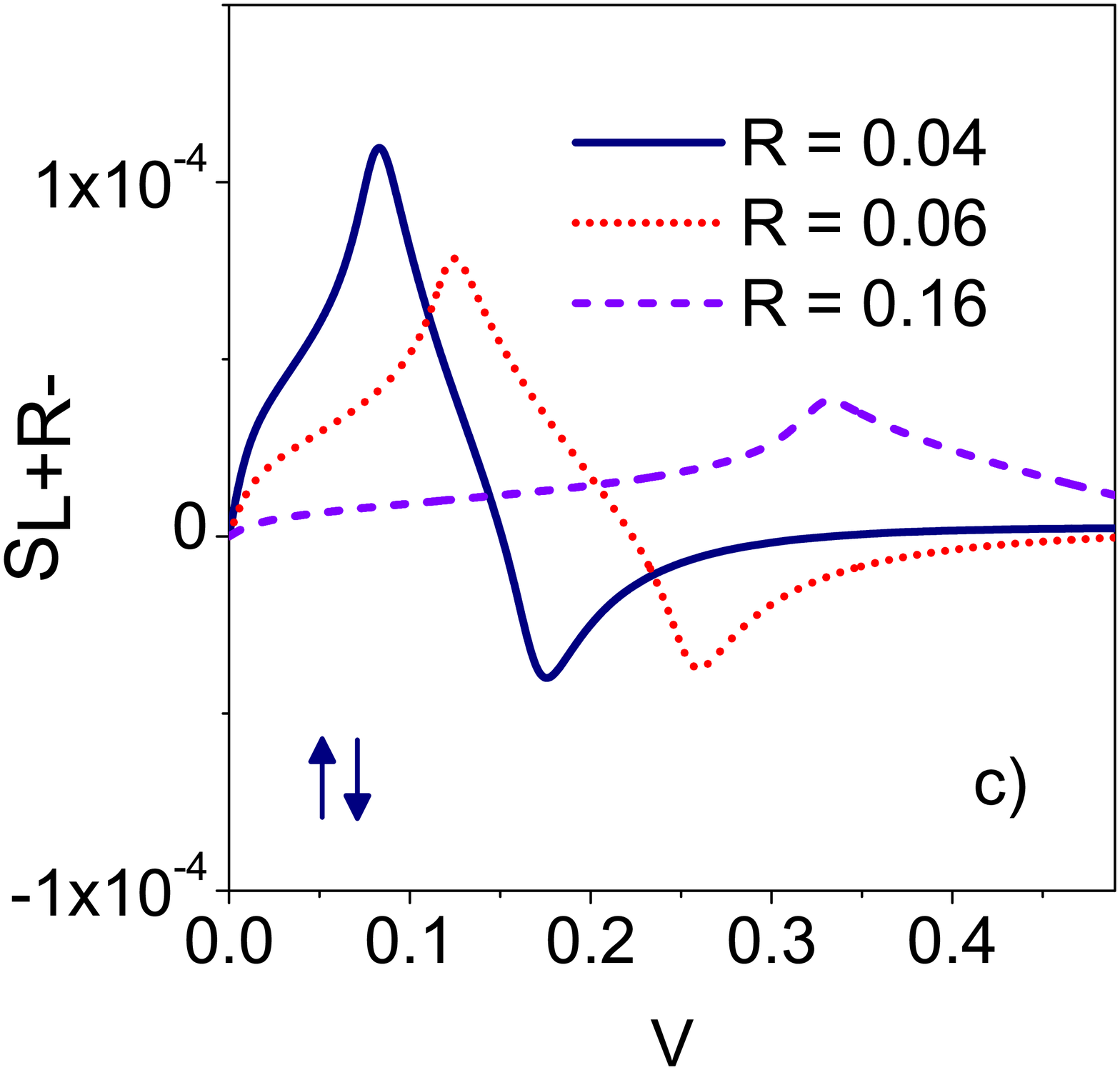}
\caption{\label{fig:epsart} (Color online) Spin-opposite shot noise of CNT-QD in the Kondo regime ($\epsilon=-6$, ${\cal{U}} = 15$) for
                parallel orientation of polarizations of the leads (a) and for antiparallel configuration (b,c) ($P=0.6$).
}
\end{figure}

Fig. 15 presents  the examples  of spin-resolved current noise. In general case of  nonvanishing spin-flip scattering the spin current is not conserved and therefore both the cross- and auto-correlations are needed  for characterization of the shot noise (sixteen noise components). On Fig. 15 we show only spin-opposite  noise, for ${\cal{P}}$ configuration and symmetric coupling case ($\gamma= 1$), it is characterized by only one element (${\cal{S}}_{L+L-}={\cal{S}}_{L-L+}={\cal{S}}_{R+R-}={\cal{S}}_{R-R+}=-{\cal{S}}_{L+R-}=-{\cal{S}}_{L-R+}=-{\cal{S}}_{R+L-}=-{\cal{S}}_{R-L+}$) and for ${\cal{AP}}$ orientation by two (${\cal{S}}_{L+L-}={\cal{S}}_{L-L+}=-{\cal{S}}_{R+L-}=-{\cal{S}}_{R-L+}$ and ${\cal{S}}_{R+R-}={\cal{S}}_{R-R+}=-{\cal{S}}_{L+R-}=-{\cal{S}}_{L-R+}$). Zero frequency shot noise  ${\cal{S}}_{+-}(V)$  can be expressed in terms of product of spin opposite transmissions (Eq.20).  The sign of spin-opposite  noise is determined by interference of spin-flip transmissions and the difference of Fermi distributions ensures that only transmissions in the range  between the Fermi levels of the leads contribute. Interesting observation is that depending on voltage the spin-opposite current noise  ${\cal{S}}_{\sigma-\sigma'}$   might be positive or negative indicating that  due to interference of spin raising  and lowering transmissions the fluctuation in the opposite spin channels mutually amplify or weaken.  Similarly  as for other discussed  transport characteristics the peaks in bias dependencies appear for voltages, for which new transmission peaks enter the transport window.  The exciting problem of fluctuations  of spin-opposite currents has been only announced here and we leave a more detailed analysis as an open question for future work.

\section{CONCLUSIONS}
We have investigated the effects of symmetry breaking perturbations on transport through CNT-QD in spin-orbital Kondo regime. Our study is addressed to spintronics and we have probed the symmetries examining the impact of magnetic field and polarization of electrodes. The conclusions drawn  in this paper can be easily adopted also  for the case of manipulating  of orbital degree  of freedom (orbitronics). The difference of orbital currents plays then  the role analogous to spin polarization of current, the torsional  strain inducing orbital level mismatch is analogue  of perpendicular magnetic field, the interorbital hopping corresponds to spin-flip scattering rate etc.. As we discussed, noise in the systems with strong interactions cannot be understood solely in terms of transmission, but still interpretation of symmetry breaking effects on the shot noise  based on Landauer-B\"{u}ttiker type form with interaction renormalized transmission is a  reasonable starting point. The linear conductance cannot reliably distinguish between SU(2) and SU(4) Kondo effects in the unitary limits ~\cite{4,70}. The shot noise distinction is evident. For SU(2) symmetry the shot noise vanishes in this limit and as has been first shown in ~\cite{82} and is confirmed by our calculations, SU(4) Kondo dot remains noisy (${\cal{F}}=1/2$). The background  for this difference lies in a remarkable difference in the structure of Kondo resonances for both symmetries,  the SU(2) resonance  is pinned to the Fermi level and  SU(4) peak is broader and  shifted from $E_{F}$ by $\omega\sim T^{SU(4)}_{K}$.  It also reflects in finite bias location of differential conductance maximum and minimum of the shot noise. Naturally the perturbation induced reconstruction of transmission close to $E_{F}$  is also quite different for both symmetries. In  case of SU(4) dot the DOS satellite of one of the spin orientations  moves with the increase of spin-dependent perturbation towards Fermi level and when reaches $E_{F}$  the minimal value of Fano factor is observed. In   CNT-QDs,  the values of the field, polarization or bias voltages, where  the shot noise for one of the spin directions is maximally suppressed  depend on the orientation of the  field and  gate voltage. The latter dependence  is especially important for dots coupled to ferromagnetic electrodes due to the gate dependence of the  exchange field. We have shown that for CNT-QDs with orbital level mismatch efficient spin filtering can be achieved in small magnetic fields. The giant values  of ${\cal {TMR}}$ have been predicted  in the Kondo range for negative exchange splitting.  We have also found,  that depending on the gate voltage both direct and inverse ${\cal {TMR}}$ can occur. Special attention in our discussion play spin currents, which have recently attracted wide interest due to possible applications in storage technology and quantum computing ~\cite{126,127}. Our calculations suggest the occurrence of equilibrium spin current in the presence of spin-flip scattering for  dots coupled to ferromagnetic electrodes in antiparallel configuration. Polarization of current can change across the system.  Spin-flips diminish ${\cal {TMR}}$ and might change its sign. The scattering processes converting spin up into spin down and vice versa induce spin-opposite correlations.  Correlations between currents of opposite spins are not necessarily negative and we have shown that cross-spin noise  oscillates with bias voltage, taking both positive and negative values indicating that  depending on the voltage the fluctuation in one of the spin channels prevents a fluctuation in another or enhances it.

	CNT-QDs provide interesting model to test the theory of exotic spin-orbital Kondo effect. From the experimental point of view CNTs are ideally suited for shot noise measurements due to the high Kondo temperatures, in which case relatively high currents can be applied. So far only one report has been published on the shot noise measurements in the spin-orbital Kondo regime ~\cite{82}. There is still a lack of spin-resolved noise measurements in this range. The technology of coupling of CNTs to ferromagnetic electrodes is well elaborated ~\cite{94} and a number of interesting transport results have been obtained in the spin-orbital Kondo range for CNT-QD attached to paramagnetic leads ~\cite{60,61,62,63}. Spin-resolved current noise measurements are within reach of present-day measuring techniques e.g. by spin filtering methods ~\cite{128}, or by detecting
magnetization fluctuation in the leads which senses the spin current noise via spin-transfer torque ~\cite{129}.  We believe that results presented in this paper will stimulate the experimental effort to use the noise measurement as a tool to probe the spin effects of SU(4)  Kondo correlations.

\begin{acknowledgments}
This work was supported by the EU grant CARDEQ under contract
IST-021285-2 and by Polish Ministry of Science and High Education through grant N N202 065636.
\end{acknowledgments}

\newpage 
\def\refname{References}


\begin{thebibliography}{99}

\bibitem{1}
Ya.M. Blanter, M. B\"{u}ttiker, Phys. Rep. \textbf{336}, 1 (2000).

\bibitem{2}
\emph{Quantum Noise in Mesoscopic Physics}, ed. by Yu. Nazarov and M. Blanter (Kluver, Dordrecht, 2003).

\bibitem{3}
C. Beenakker, C. Sch\"{o}nenberger, Phys. Today \textbf{56}, 37 (2003).

\bibitem{4}
R. Egger, Nature Physics \textbf{5}, 175 (2009).

\bibitem{5}
E. Onac, F.Balestro, B. Trauzettel, C.F.J. Lodewijk, and L.P. Kouvenhoven, Phys. Rev. Lett. \textbf{96}, 026803 (2006).

\bibitem{6}
A.C. Hewson, \emph{The Kondo Problem to Heavy Fermions}, Cambridge University Press, Cambridge, 1993.

\bibitem{7}
D. Goldhaber-Gordon, H. Shtrikman, D. Mahlul, D. Abuschmagder, U.Meiraev, and M.A. Kastner, Nature \textbf{391}, 156 (1998).

\bibitem{8}
S.M. Cronewett, T.H. Oesterkamp, and L.P. Kouwenhoven, Science \textbf{291}, 540 (1998).

\bibitem{9}
J. Schmid, J. Weis, K. Eberl, and K.v. Klitzing, Physica B \textbf{256-258}, 182 (1998).

\bibitem{10}
W.G. van der Wiel, S. De Franceschi, T. Fujisawa, J.M. Elzerman, S.Tarucha, and L.P. Kouvenhoven, Science \textbf{289}, 2105 (2000).

\bibitem{11}
J. Park, A.N. Pasupathy, J.I. Goldsmith, C. Chang, Y. Yaish, J.R. Petta, M. Rinkoski, J.P. Sethna, H.D. Abruna, P.L. McEuen, and D.C. Ralph, Nature \textbf{417}, 722 (2002).

\bibitem{12}
W.J. Liang, M.P. Shores, M. Bockrath, J.R. Long, and H. Park, Nature \textbf{417}, 725 (2002).

\bibitem{13}
J. Nyg{\aa}rd, D.H. Cobden, and P.E. Lindelof, Nature \textbf{408}, 342 (2000).

\bibitem{14}
M.R. Buitelaar, A. Bachtold, T. Nussbaumer, M. Iqbal, and C. Sch\"{o}nenberger, Phys. Rev. Lett. \textbf{88}, 156801 (2002).

\bibitem{15}
L.M. Glazman, and M.E. Raikh, JETP Lett. \textbf{47}, 452 (1998).

\bibitem{16}
T.K. Ng, and P.E. Lee, Phys. Lett. \textbf{61}, 1768 (1998).

\bibitem{17}
S. Hershfield, J.H. Davies, and J.W. Wilkins, Phys. Rev. Lett. \textbf{67}, 3720 (1991).

\bibitem{18}
Y. Meir, N.S. Wingreen, and P.A. Lee, Phys. Rev. Lett. \textbf{70}, 2601 (1993).

\bibitem{19}
A. Levy-Yeyati, A. Martin-Rodero, and F. Flores, Phys. Rev. Lett. \textbf{71}, 2991 (1993); A. Levy-Yeyati, F. Flores, and A. Martin-Rodero, Phys. Rev. Lett. \textbf{83}, 600 (1999).

\bibitem{20}
N.S. Wingreen, Y. Meir, Phys. Rev. B \textbf{49}, 11040 (1994).

\bibitem{21}
K. Kang, and L. Min, Phys. Rev. B \textbf{52}, 10689 (1995).

\bibitem{22}
J.J. Palacois, L. Liu, and D. Yoshioka, Phys. Rev. B \textbf{55}, 15735 (1997).

\bibitem{23}
H. Schoeller, and J. K\"{o}nig, Phys. Rev. Lett. \textbf{84}, 3686 (2000).

\bibitem{24}
A. Kami\'{n}ski, Yu.V. Nazarov, and L.I. Glazman, Phys. Rev. B \textbf{62}, 8154 (2000).

\bibitem{25}
M. Krawiec, and K.I. Wysoki\'{n}ski, Phys. Rev. B \textbf{66}, 165408 (2002).

\bibitem{26}
B.R. Bu{\l}ka, and P. Stefa\'{n}ski, Phys. Rev. Lett. \textbf{86}, 5128 (2001).

\bibitem{27}
R. \'{S}wirkowicz, J. Barna\'{s}, and M. Wilczy\'{n}ski, Phys. Rev. B \textbf{68}, 195318 (2003).

\bibitem{28}
Y. Meir, and A. Golub, Phys. Rev. Lett. \textbf{88}, 116802 (2002).

\bibitem{29}
B. Dong, and X.L. Lei, J. Phys.: Condens. Matter. \textbf{14}, 4963 (2002).

\bibitem{30}
R. Lopez, and D. Sanchez, Phys. Rev. Lett. \textbf{90}, 116602 (2003).

\bibitem{31}
Y. Avishai, A. Golub, and A.D. Zaikin, Phys. Rev. B \textbf{67}, 041301 (2003).

\bibitem{32}
R. Lopez, R. Aguado, and G. Platero, Phys. Rev. B \textbf{69}, 235305 (2004).

\bibitem{33}
D. Sanchez, and R. Lopez, Phys. Rev. B \textbf{71}, 035315 (2005).

\bibitem{34}
T.-F. Fang, and S.-J. Wang, J. Phys.: Condens. Matter. \textbf{19}, 026204 (2007).

\bibitem{35}
E. Sela, Y. Oreg, F. von Oppen, and J. Koch, Phys. Rev. Lett. \textbf{97}, 086601 (2005).

\bibitem{36}
A. Golub, Phys. Rev. B \textbf{73}, 233310 (2006).

\bibitem{37}
O. Zarchin, M. Zaffalon, M. Heilblum, D. Mahalu, and V. Umansky, Phys. Rev. B \textbf{77}, 241303(R) (2008).

\bibitem{38}
T.A. Costi, Phys. Rev. Lett. \textbf{85}, 1504 (2000); Phys. Rev. B \textbf{64}, 24310 (2001).

\bibitem{39}
J.E. Moore, X.-G. Wen, Phys. Rev. Lett. \textbf{85}, 1722 (2000).

\bibitem{40}
B. Dong, X.L. Lei, Phys. Rev. B \textbf{63}, 235306 (2001).

\bibitem{41}
A. Rosch, J. Paaske, J. Kroha, and P. W\"{o}lfe, Phys. Rev. Lett. \textbf{90}, 076804 (2003).

\bibitem{42}
A. Kogan, A. Amasha, D. Goldhaber-Gordon, G. Granger, M.A. Kastner, and H. Shtrikman, Phys. Rev. Lett. \textbf{93}, 166602 (2004).

\bibitem{43}
J. Paaske, A. Rosch, and P. W\"{o}lfe, Phys. Rev. B \textbf{69}, 155330 (2004).

\bibitem{44}
N. Sergueev, Q.F. Sun, H. Guo, B.G. Wang, and J. Wang, Phys. Rev. B \textbf{65}, 165303 (2002).

\bibitem{45}
B.R. Bu{\l}ka, and S. Lipi\'{n}ski, Phys. Rev. B \textbf{67}, 024404 (2003).

\bibitem{46}
R. Lopez, and D. Sanchez, Phys. Rev. Lett. \textbf{90}, 116602 (2003).

\bibitem{47}
J. Martinek, Y. Utsumi, H. Imamura, J. Barna\'{s}, S. Maekawa, J. K\"{o}nig, and G. Sch\"{o}n, Phys. Rev. Lett. \textbf{91}, 247202 (2003).

\bibitem{48}
M.S. Choi, D. Sanchez, and R. Lopez, Phys. Rev. Lett \textbf{92}, 056601 (2004).

\bibitem{49}
J. Martinek, M. Sindel, L. Borda, J. Barna\'{s}, R. Bulla, J. K\"{o}nig, G. Sch\"{o}n, S. Maekawa, and J. von Delft, Phys. Rev. B \textbf{72}, 121302 (2005).

\bibitem{50}
Y. Utsumi, J. Martinek, G. Sch\"{o}n, H. Imamura, and S. Maekawa, Phys. Rev. B \textbf{71}, 245116 (2005).

\bibitem{51}
R. \'{S}wirkowicz, M. Wilczy\'{n}ski, and J. Barna\'{s}, J. Phys.: Condens. Matter. \textbf{18}, 2291 (2006).

\bibitem{52}
M. Sindel, L. Borda, J. Martinek, R. Bulla, J. K\"{o}nig, G. Sch\"{o}n, S. Maekawa, and J. von Delft, Phys. Rev. B \textbf{76}, 045321 (2007).

\bibitem{53}
A.N. Papsupathy, R.C. Bialczak, J. Martinek, J.E. Donem, P.L. McEuen, and D.C. Ralph, Science \textbf{306}, 86 (2004).

\bibitem{54}
J.R. Hauptmann, J. Paaske, and P.E. Lindelof, Nature Phys. \textbf{4}, 373 (2008).

\bibitem{55}
K. Hamaya, M. Kitabatake, K. Shibata, M. Jung, M. Kowamura, K. Hirakawa, and T. Machida, App. Phys. Lett. \textbf{91}, 232105 (2007).

\bibitem{56}
M.R. Carlo, J. Fernandez-Rossier, J.J. Palacios, D. Jacob, D. Natelson, and C. Untiedt, Nature \textbf{458}, 1150 (2009).

\bibitem{57}
S. Sasaki, S. Amaha, N. Asakawa, M. Eto, and S. Tarucha, Phys. Rev. Lett. \textbf{93}, 017205 (2004).

\bibitem{58}
U. Wilhelm, J. Schmid, J. Weis, and K.v. Klitzing, Physica E (Amsterdam) \textbf{14}, 385 (2002).

\bibitem{59}
A. Holleitner, A. Chudnovskiy, D. Pfannkuche, K. Eberl, and R.H. Blick, Phys. Rev. B \textbf{70}, 075204 (2004).

\bibitem{60}
P. Jarillo-Herrero, J. Kong, H.S.J. Van der Zant, C. Dekker, L.P. Kouvenhoven, and S. De Franceschi, Nature (London) \textbf{434}, 484 (2005).

\bibitem{61}
A. Makarowski, J. Liu, and G. Finkelstein, Phys. Rev. Lett. \textbf{99}, 066801 (2007); A. Makarowski, A. Zhukov, J. Liu, and G. Finkelstein, Phys. Rev. B \textbf{75}, 241407 (2007).

\bibitem{62}
K. Grove-Rasmussen, H.J. J{\o}rgensen, and  P.E. Lindelof, Physica E \textbf{40}, 92 (2007).

\bibitem{63}
F. Wu, R. Danneau, P. Queipo, E. Kauppinen, T. Tsuneta, and P.J. Hakonen, Phys. Rev. B \textbf{79}, 073404 (2009).

\bibitem{64}
T. Pojhola, H. Schoeller, and G. Sch\"{o}n, Europhys. Lett. \textbf{54}, 241 (2001).

\bibitem{65}
L. Borda, G. Zarand, W. Hofsetter, B.I. Halperin, and J. von Delft, Phys. Rev. Lett. \textbf{90}, 026602 (2003).

\bibitem{66}
A.I. Chudnovskiy, Europhys. Lett. \textbf{71}, 672 (2005).

\bibitem{67}
M. Choi, R. Lopez, R. Aguado, Phys. Rev. Lett. \textbf{95}, 067204 (2005).

\bibitem{68}
S. Lipi\'{n}ski, and D. Krychowski, Phys. Status Solidi B \textbf{242}, 206 (2005); J. Alloys and Compounds \textbf{423}, 379-381 (2006).

\bibitem{69}
R. Lopez, D. Sanchez, M. Lee, M.-S. Choi, P. Simon, and K. Le Hur, Phys. Rev. B \textbf{71}, 115312 (2005).

\bibitem{70}
J.S. Lim, M. Choi, M.Y. Choi, R. Lopez, and R. Aguado, Phys. Rev. B \textbf{74}, 205119 (2006).

\bibitem{71}
G. Zarand, Philos. Mag. \textbf{86}, 2043 (2006).

\bibitem{72}
M.R. Galpin, D.E. Ogan, H.R. Krishnamurthy, J. Phys.: Condens. Matter \textbf{18}, 6571 (2006).

\bibitem{73}
R. Sakano, and N. Kawakami, Phys. Rev. B \textbf{73}, 155332 (2006).

\bibitem{74}
J. Mravlje, A. Ram\v{s}ak, and T. Rejec, Phys. Rev. B \textbf{73}, 241305 (2006).

\bibitem{75}
K. Le Hur, P. Simon, and  D. Loss, Phys. Rev. B \textbf{75}, 035332 (2007).

\bibitem{76}
C.A. B\"{u}sser, and G.B. Martins, Phys. Rev. B \textbf{75}, 045406 (2007).

\bibitem{77}
T.-F. Fang, W. Zuo, and H.-G. Luo, Phys. Rev. Lett. \textbf{101}, 246805 (2008).

\bibitem{78}
F.B. Anders, D.E. Logan, M.R. Galpin, and G. Finkelstein, Phys. Rev. Lett. \textbf{100}, 086809 (2008).

\bibitem{79}
M. Mizumo, E.H. Kim, and  G.B. Martins, J. Phys.: Condens. Matter \textbf{21}, 292208 (2009).

\bibitem{80}
P. Vitushinsky, A.A. Clerk, and K. Le Hur, Phys. Rev. Lett. \textbf{100}, 036603 (2008).

\bibitem{81}
C. Mora, X. Leyronas, and N. Regnault, Phys. Rev. Lett. \textbf{100}, 036604 (2008); Phys. Rev. Lett. \textbf{102}, 139902 (2009).

\bibitem{82}
T. Delattre, C. Feuillet-Palma, L.G. Herman, P.Morfin, J.-M. Berroir, G. F\`{e}ve, B. Pla\c{c}ais,D.C. Glattli, M.-S. Choi, C. Mora, and T. Kontos, Nature Physics \textbf{5}, 208 (2009).

\bibitem{83}
B.R. Bu{\l}ka, J. Martinek, G. Micha{\l}ek, and J. Barna\'{s}, Phys. Rev. B \textbf{60}, 12246 (1999).

\bibitem{84}
J. Barna\'{s}, J. Martinek, G. Micha{\l}ek, B.R. Bu{\l}ka, and A. Fert, Phys. Rev. B \textbf{62}, 12363 (2000).

\bibitem{85}
I. Weymann, J. Barna\'{s}, and S. Krompiewski, Phys. Rev. B \textbf{76}, 155408 (2007); Phys. Rev. B \textbf{78}, 035422 (2008).

\bibitem{86}
O. Saurent, and D.Feinberg, Phys. Rev. Lett. \textbf{92}, 106601 (2004).

\bibitem{87}
F.M. Souza, A.P. Jauho, and J.C. Egues, Phys. Rev. B \textbf{78}, 155393 (2008).

\bibitem{88}
B. Dong, and X.L.Lei, J. Phys.: Condens. Matter \textbf{14}, 4963 (2002).

\bibitem{89}
L. Rong, and L. Zhi-Rong, Chin. Phys. Lett. \textbf{24}, 195 (2007).

\bibitem{90}
J.Y. Luo, X.-Q. Li, J. Phys.: Condens. Matter \textbf{20}, 345215 (2008).

\bibitem{91}
C.P. Moca, I. Weymann, and G. Zarand, cond-mat/0907.0475 (2009).

\bibitem{92}
W. Liang, M. Bockrath, H. Park, Phys. Rev. Lett. \textbf{88}, 126801 (2002).

\bibitem{93}
M.S. Dresselhaus, G. Dresselhaus, and Ph. Avouris, \emph{Carbon nanotubes} (Springer, Berlin, 2000).

\bibitem{94}
A. Cottet, T. Kontos, S. Sahoo, H.T. Man, M.-S. Choi, W. Belzig, C. Bruder, A.F. Morpurgo, and C. Sch\"{o}nenberger, Semicond. Sci. Technol. \textbf{21}, S78 (2006).

\bibitem{95}
H. Haug, and A.-P. Jauho, \emph{Quantum Kinetics in Transport and Optics of Semiconductors} (Springer, Berlin, 1998).

\bibitem{96}
P. Coleman, Phys. Rev. B \textbf{29}, 3035 (1984); Phys. Rev. B \textbf{35}, 5072 (1987).

\bibitem{97}
G. Kotliar, and A.E. Ruckenstein, Phys. Rev. Lett. \textbf{57}, 1362 (1986).

\bibitem{98}
C. Lacroix, J. Phys. F: Metal Phys. \textbf{11}, 2389 (1998).

\bibitem{99}
O. Entin-Wohlman, A. Aharony, and Y. Meir, Phys. Rev. B \textbf{71}, 035333 (2005).

\bibitem{100}
V. Kashcheheyevs, A. Aharony, and O. Entin-Wohlman, Phys. Rev. B \textbf{73}, 125338 (2006).

\bibitem{101}
N.E. Bickers, Rev. Mod. Phys. \textbf{59}, 845 (1987).

\bibitem{102}
N.S. Wingreen, and Y. Meir, Phys. Rev. B \textbf{49}, 11040 (1999).

\bibitem{103}
L. DiCarlo, Y. Zhang, D.T. McClure, D.J. Reilly, C.M. Marcus, L.N. Pfeiffer, and K.W. West, Phys. Rev. Lett. \textbf{97}, 036810 (2008).

\bibitem{104}
X. Jehl, M. Sanquer, R. Calemczuk, and D. Mailly, Nature \textbf{405}, 50 (2000).

\bibitem{105}
P.-E. Roche, M. Kociak, S. Gueron, A. Kasumov, R. Reulet, and H. Bouchiat, Eur. Phys. J. B \textbf{28}, 217 (2002).

\bibitem{106}
F.Wu, L. Roschier, T. Tsuneta, M. Paalanen, T. Wang, and P. Hakonen, AIP Conf. Proc. \textbf{850}, 1482 (2008).

\bibitem{107}
W. Liang, M. Bockrath, H. Park, Phys. Rev. Lett. \textbf{88}, 126801 (2002).

\bibitem{108}
N. Hamada, S.I. Sawada, A. Oshiyama, Phys. Rev. Lett. \textbf{68}, 1579 (2000).

\bibitem{109}
D. Boese, W. Hofstetter, H. Schoeller, Phys. Rev. B \textbf{66}, 125315 (2002).

\bibitem{110}
K. Yamada, K. Yosida, K. Hanzawa, Proc. Theor. Phys. \textbf{71}, 450 (1984).

\bibitem{111}
F.D.M. Haldane, Phys. Rev. Lett. \textbf{40}, 416 (1978).

\bibitem{112}
T.-F. Fang, and S.-J. Wang, J. Phys.: Condens. Matter \textbf{19}, 026204 (2007).

\bibitem{113}
D.G. Langreth, in \emph{Linear and Nonlinear Electron Transport in Solids}, ed. by J.T. Devreese and V.E. Van Doren (Plenum, New York, 1976).

\bibitem{114}
T.K. Ng, Phys. Rev. Lett. \textbf{76}, 487 (1996).

\bibitem{115}
P. Jarillo-Herrero, S. Sapmaz, C. Dekker, L.P. Kouwenhoven, and H.S.J. Van der Zant, Nature \textbf{429}, 389 (2004).

\bibitem{116}
B. Babi\'{c}, T. Kontos, and C. Sch\"{o}nenberger, Phys. Rev. B \textbf{70}, 235419 (2004).

\bibitem{117}
L.P. Kouvenhoven, C.M. Marcus, P.L. McEuen, S. Tarucha, R.M. Westervelt, and N.S. Wingreen, in \emph{Proc. Advanced Study Institute on Mesoscopic Electron Transport}, ed. by L.L. Sohn, L.P. Kouvenhoven, and G. Sch\"{o}n (Kluver, Dordrecht, 1997).

\bibitem{118}
M.R. Graber, M. Weiss, S. Oberholzer, and C. Schonenberger, Semicond. Sci. Technol. \textbf{21}, S64 (2006).

\bibitem{119}
Estimation of Kondo temperature from the position of the center of Kondo resonance overestimates $T_{K}$ in comparision to the value obatained from temperature scaling of conductance.

\bibitem{120}
In contrast to SU(2) case no pathology of SBMFA solutions occurs for SU(4) for perpendicular field at $h = 2T_{K}$, because this field does not destroy the Kondo state, but only reduces its symmetry.

\bibitem{121}
D. Krychowski, S. Lipi\'{n}ski, and S. Krompiewski, J. Alloys and Compounds \textbf{442}, 379 (2007).

\bibitem{122}
M. Jullierre, Phys. Rev. Lett. \textbf{54A}, 225 (1975).

\bibitem{123}
A similar effect can be caused by spin-orbit interaction. Although it is widely believed that spin-orbit coupling is weak in CNTs, some recent papers call in question this statement due to curvature and cylindrical topology of these systems [see F. Kuemmeth, S. Ilani, D.C. Ralph, and P.L. McEuen, Nature \textbf{452}, 448 (2008)].  Spin-orbit interaction mixes both spin and orbital channels (${\cal {R}}^{so}(d_{m\uparrow}^{+}d_{-m\downarrow}-d_{-m\uparrow}^{+}d_{m\downarrow})+h.c.$), and as a result one can expect nonvanshing spin-opposite and orbital-opposite noise  ${\cal{S}}_{\alpha m\sigma,\alpha -m-\sigma}$.

\bibitem{124}
F. Liang, Y. Shen, and Y. Yang, Phys. Rev. Lett. A \textbf{372}, 4634 (2008).

\bibitem{125}
J. Wang, and K.S. Chan, Phys. Rev. B \textbf{74}, 035342 (2006).

\bibitem{126}
\emph{Semiconductor Spintronics and Quantum Computing}, ed. by D.D.Awschalom, D. Loss and N. Samarth (Springer, Berlin, 2002).

\bibitem{127}
I. Zutic, J. Fabian, and S. Das Sarma, Rev. Mod. Phys. \textbf{76}, 323 (2004).

\bibitem{128}
S.M. Frolov, A. Venkatesan, W. Yu, and J.A. Folk, Phys. Rev. Lett. \textbf{102}, 116802 (2009).

\bibitem{129}
J. Foros, A. Brataas, Y. Tsevkovnyak, and G.E. Bauer, Phys. Rev. Lett. \textbf{95}, 016601 (2005).

\end{thebibliography}
\end{document}